\documentclass[sigconf]{acmart}

\AtBeginDocument{%
  }

\newcommand{\beginsupplement}{%
        \setcounter{table}{0}
        \renewcommand{\thetable}{S\arabic{table}}%
        \setcounter{figure}{0}
        \renewcommand{\thefigure}{S\arabic{figure}}%
     }
\usepackage{xcolor}

\definecolor{brown}{rgb}{0.59, 0.29, 0.0}
\definecolor{darkblue}{rgb}{0.0, 0.0, 0.55}
\definecolor{darkgreen}{rgb}{0.0, 0.5, 0.0}

\usepackage{placeins}
\usepackage{multirow}
\usepackage{verbatim}
\usepackage{afterpage}
\usepackage{float}
\usepackage{graphicx}
\usepackage[table]{colortbl}
\usepackage{tikz}
\usepackage{nth}
\usetikzlibrary{calc}
\usepackage{zref-savepos}
\usepackage{xcolor}
\usepackage{booktabs,caption}
\usepackage{setspace}
\usepackage{placeins}
\usepackage{xcolor}
\usepackage{listings}
\usepackage{algorithm}
\usepackage{algpseudocode}
\usepackage[algo2e, ruled,vlined]{algorithm2e}
\usepackage{tikz}
\usepackage{cancel}
\usepackage{amsmath}

\usepackage{amsfonts}
\usepackage{amssymb}
\usepackage{enumitem}
\usepackage{subcaption}
\usepackage{todonotes}
\newcounter{cpModel}


\definecolor{tableauOrange}{HTML}{F28E2B}
\definecolor{tableauRed}{HTML}{E15759}
\definecolor{tableauBlue}{HTML}{4D79A7}
\definecolor{tableauGreen}{HTML}{77B7B3}
\definecolor{highlightPurple}{HTML}{b17aa1}

\copyrightyear{2025}
\acmYear{2025}
\setcopyright{cc}
\setcctype{by}
\acmConference[CHI '25]{CHI Conference on Human Factors in Computing Systems}{April 26-May 1, 2025}{Yokohama, Japan}
\acmBooktitle{CHI Conference on Human Factors in Computing Systems (CHI '25), April 26-May 1, 2025, Yokohama, Japan}\acmDOI{10.1145/3706598.3713962}
\acmISBN{979-8-4007-1394-1/25/04}

\begin{document}

\title[Characterizing Photorealism in AI-Generated Images]{Characterizing Photorealism and Artifacts in Diffusion Model-Generated Images}


\author{Negar Kamali}
\email{negar.kamali@u.northwestern.edu}
\orcid{0000-0002-1086-6735}
\affiliation{%
  \institution{Northwestern University}
  \city{Evanston}
  \state{Illinois}
  \country{USA}
}

\author{Karyn Nakamura}
\email{karynnakamura68@gmail.com}
\orcid{0009-0005-4419-0701}
\affiliation{%
  \institution{Northwestern University}
  \city{Evanston}
  \state{Illinois}
  \country{USA}
}

\author{Aakriti Kumar}
\email{aakriti.kumar@kellogg.northwestern.edu}
\orcid{0000-0002-9502-013X}
\affiliation{%
  \institution{Northwestern University}
  \city{Evanston}
  \state{Illinois}
  \country{USA}
}

\author{Angelos Chatzimparmpas}
\email{a.chatzimparmpas@uu.nl}
\orcid{0000-0002-9079-2376 }
\affiliation{%
  \institution{Utrecht University}
  \city{Utrecht}
  \state{}
  \country{Netherlands}
}

\author{Jessica Hullman}
\email{jhullman@northwestern.edu}
\orcid{0000-0001-6826-3550}
\affiliation{%
  \institution{Northwestern University}
  \city{Evanston}
  \state{Illinois}
  \country{USA}
}

\author{Matthew Groh}
\email{matthew.groh@kellogg.northwestern.edu}
\orcid{0000-0002-9029-0157}
\affiliation{%
  \institution{Northwestern University}
  \city{Evanston}
  \state{Illinois}
  \country{USA}
}
\renewcommand{\shortauthors}{Kamali et al.}

\renewcommand{\thesubfigure}{\textbf{\Alph{subfigure}}}
\captionsetup[sub]{labelformat=simple}  
\newcommand{\mybold}[1]{\textbf{#1}}


\begin{abstract}
Diffusion model-generated images can appear indistinguishable from authentic photographs, but these images often contain artifacts and implausibilities that reveal their AI-generated provenance. Given the challenge to public trust in media posed by photorealistic AI-generated images, we conducted a large-scale experiment measuring human detection accuracy on 450 diffusion-model generated images and 149 real images. Based on collecting 749,828 observations and 34,675 comments from 50,444 participants, we find that scene complexity of an image, artifact types within an image, display time of an image, and human curation of AI-generated images all play significant roles in how accurately people distinguish real from AI-generated images. Additionally, we propose a taxonomy characterizing artifacts often appearing in images generated by diffusion models. Our empirical observations and taxonomy offer nuanced insights into the capabilities and limitations of diffusion models to generate photorealistic images in 2024.
\end{abstract}



\maketitle

\section{Introduction}
The capabilities of diffusion models to generate photorealistic images of people are beginning to contribute to disinformation and erode trust in the media \cite{epstein2023art}. For example, in March 2023, realistic AI-generated images of world leaders went viral on social media, showing Pope Francis wearing what appeared to be a designer puffer jacket, Donald Trump getting arrested, and Vladimir Putin standing behind prison bars \cite{apnews2024}. These exemplar images may appear both provocative and realistic at first glance, but they are far from perfectly photorealistic; they contain distortions of hands and faces, implausible grasping of objects, and shadows that do not match the objects that appear to cast them. These distortions are not unique to these particular images but are pervasive in diffusion model-generated images produced by text--to--image tools such as Midjourney (the source of these fake images of world leaders), Stable Diffusion by Stability AI, and Firefly by Adobe~\cite{kamali2024distinguish}. While it is possible to generate images that seem indistinguishable from photographs, many diffusion model-generated images still leave behind human-identifiable artifacts. This raises an open research question for human--computer interaction: What drives human perception of photorealism in images generated by diffusion models?

We approach this question by conducting a large--scale, online experiment where we collect data on human participants' accuracy in identifying whether images are AI-generated or real. We measure photorealism following a psychophysics approach ~\cite{zhou2019hype} that defines photorealism as human discrimination performance. Accuracy scores are inversely associated with photorealism: a high accuracy score indicates low photorealism, whereas a low accuracy score indicates high photorealism. By defining photorealism based on discrimination performance, we avoid the speculation and subjectivity of asking participants questions like ``Is the image photorealistic?"~\cite{liang2024rich} and ``Could these images be taken with a camera?"~\cite{yan2024sanitycheckaigeneratedimage}. 

By comparing human detection accuracy across a diverse set of images, we can evaluate the contexts that influence the continuum of photorealism. Past research has demonstrated that GAN(Generative Adversarial Networks)-generated human portraits can be indistinguishable from real portrait images~\cite{nightingale2022ai}. However, open questions on context remain: How often do portrait images appear indistinguishable from real portraits? How does scene complexity across styles of photographic portraiture (e.g., single-subject close-up, single-subject full body, posed group, and candid group) influence aggregate measures of photorealism? How accurately can people identify real images across scene complexities? What kind of artifacts arise in diffusion model-generated images and how are the presence of those artifacts related to photorealism? How does display time of an image influence measures of photorealism? How does human curation of AI-generated image stimuli influence measures of photorealism?

We approach each of these questions in turn and then consider an interventional question: How should an AI literacy guide categorize artifacts and implausibilities that emerge in photorealistic AI-generated images to promote attention to and communicate these visual cues? We also consider a flipped version of that question: How do we help people avoid falsely identifying a real image as AI-generated? This question differs from the first because it focuses on how people can identify when to trust what they see. This is important because there has already been a case where a politician has incorrectly, publicly claimed that an authentic photograph of his opponent was AI-generated~\cite{apnews2024b, wired2024kamala}. Enhancing human skill at distinguish real from AI-generated remains important because technical platform--level solutions (e.g. watermarking and machine learning classification) lack robustness and are susceptible to error when images are slightly modified via cropping, compression, and other edits. AI literacy guides have the potential to help humans stay abreast of the capabilities and limitations of AI to better navigate assessing the authenticity of images.

Our contributions toward answering these research questions are fourfold: First, we contribute a taxonomy of artifacts and implausibilities in diffusion model-generated images of humans along five dimensions: (1) anatomical implausibilities: representations of human anatomy that deviate from realistic or common forms; (2) stylistic artifacts: visual elements that often appear in images generated by diffusion models like shiny or plastic textures; (3) functional implausibilities: design or structural flaws that would make an object or system unlikely to function as intended in the real world; (4) violations of physics: instances where an object or scenario defies the laws of physics; and (5) sociocultural implausibilities: representations of people that are unlikely to reflect the norms of a culture or society. Second, we conduct a large--scale digital experiment and present an empirical evaluation of photorealism – as measured by human detection accuracy – across images from three state-of-the-art diffusion models, varying photographic contexts, types of artifacts in an image, and randomized display time of an image. This evaluation offers insight into the capabilities and limitations of diffusion models to produce photorealistic images. Third, we provide empirical evidence revealing the influence of human curation on the level of photorealism in images generated by diffusion models. This finding reveals the importance of human curation and the limits of state-of-the-art diffusion models to producing photorealistic images, and create consistent and believable narratives via an automated deluge of fake images that are indistinguishable from real photographs. Fourth, we release a public dataset of 749,828 responses from 50,444 participants on 599 images to enable the replications of our results and further research on photorealism in diffusion models; replication data, code, and stimuli can be found at the following link: https://github.com/negarkamali/Replication-for-Characterizing-Photorealism-2025/.
\section{Background}\label{sec:relwork}

\subsection{Limitations of machine learning approaches to detect AI-generated images}

Machine learning models for detecting AI-generated images are brittle and lack robustness to simple data transformations. Corvi et al.~\cite{corvi2023intriguingpropertiessyntheticimages} compare four different machine learning approaches to deepfake detection and demonstrate that recropping and compression – simple modifications common on social media – lead to drops in accuracy such that the classifiers are nearly just as good as random guessing. Dong et al.~\cite{9879575}reveal the ease with which spectral artifacts used in the identification of GAN-generated images can be mitigated via blurring and resizing, demonstrating a noticeable decrease in accuracy under basic modifications. Cozzolino et al.~\cite{cozzolino2024raisingbaraigeneratedimage} demonstrate that post--processing images by random--cropping, resizing, and compression lead to a drop in AI-generated image detection from 90\% accuracy to 85\% accuracy. The fundamental problem is that machine learning models for deepfake detection lack robustness to context shift, out--of--distribution data and adversarial perturbations ~\cite{wang2023deepfakedetectioncomprehensivestudy, ha2024, groh2022identifying, hulzebosch2020detectingcnngeneratedfacialimages}. 

How an image is generated influences the ability of deepfake detection classifiers to accurately identify it as AI-generated. Classifiers trained to detect GAN-generated images tend to fail to detect diffusion model-generated images. For example, the approach to detecting GAN-generated images based on frequency spectra~\cite{marra2018gansleaveartificialfingerprints, yu2019attributingfakeimagesgans, 9035107, durall2020watchupconvolutioncnnbased, bi2023detectinggeneratedimagesreal, pmlr-v119-frank20a} and inconsistencies in head poses and facial landmark positions~\cite{yang2018exposingdeepfakesusing, yang2019exposinggansynthesizedfacesusing, Mundra_2023_CVPR}, do not generalize to detecting images generated by diffusion models~\cite{ojha2024universalfakeimagedetectors}. GAN-trained detection models miss these patterns because they have learned patterns for identifying GAN-generated images~\cite{wang2020cnn, ricker2024detectiondiffusionmodeldeepfakes}. Likewise, it is possible to learn the statistical regularities in diffusion model-generated images but these regularities are not invariant to image post-processing.~\cite{xi2023aigeneratedimagedetectionusing, 10334046, wang2023dirediffusiongeneratedimagedetection, ma2023exposingfakeeffectivediffusiongenerated, yang2023diffusion}. 

Moreover, machine learning models' lack of robustness for detection is exacerbated by the changing architectures of generative AI models~\cite{lin2024, Mirsky2021}. Vision transformers~\cite{radford2021learning, ojha2024universalfakeimagedetectors} and multi--architecture training~\cite{epstein2023onlinedetectionaigeneratedimages, porcile2024findingaigeneratedfaceswild, jia2024can} show promise for enhancing the detection of AI-generated images, but adversarial attacks and large architectural changes in generative models continue to affect robustness of detection.  

\begin{figure*}[h]
\centering
\captionsetup{justification=raggedright, singlelinecheck=false, skip=2pt}
\begin{subfigure}[t]{0.13\textwidth}
    \subcaption{}\vtop{\vskip0pt\hbox{\includegraphics[width=\linewidth]{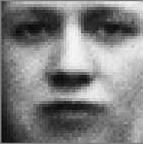}}}
    \caption*{\footnotesize GAN '14}
\end{subfigure}
\hfill
\begin{subfigure}[t]{0.13\textwidth}
    \subcaption{}\vtop{\vskip0pt\hbox{\includegraphics[width=\linewidth]{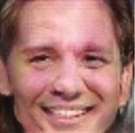}}}
    \caption*{\footnotesize DCGAN '15}
\end{subfigure}
\hfill
\begin{subfigure}[t]{0.13\textwidth}
    \subcaption{}\vtop{\vskip0pt\hbox{\includegraphics[width=\linewidth]{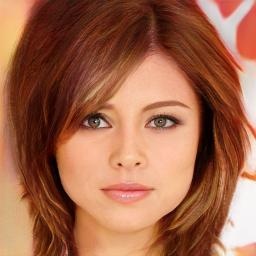}}}
    \caption*{\footnotesize PGGAN '18}
\end{subfigure}
\hfill
\begin{subfigure}[t]{0.13\textwidth}
    \subcaption{}\vtop{\vskip0pt\hbox{\includegraphics[width=\linewidth]{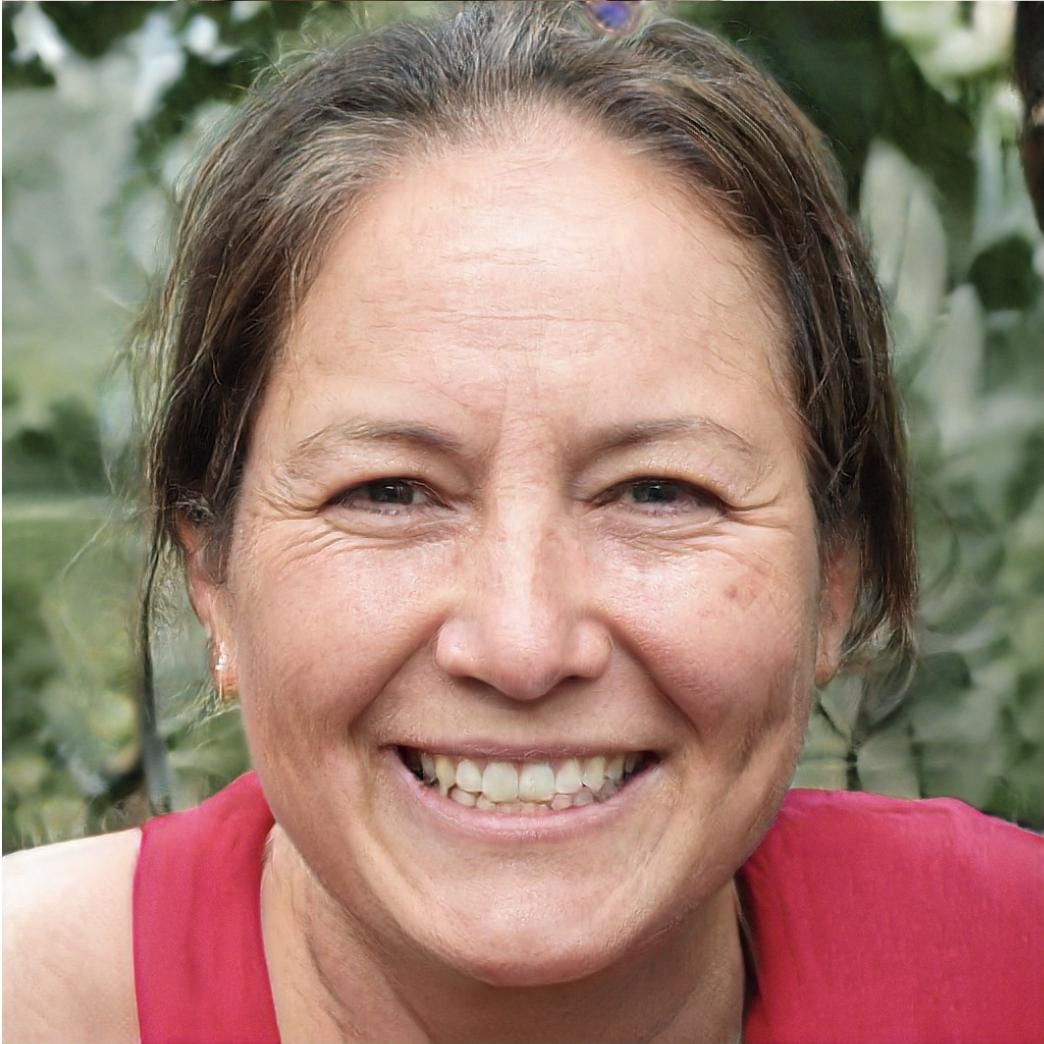}}}
    \caption*{\footnotesize StyleGAN '19}
\end{subfigure}
\hfill
\begin{subfigure}[t]{0.13\textwidth}
    \subcaption{}\vtop{\vskip0pt\hbox{\includegraphics[width=\linewidth]{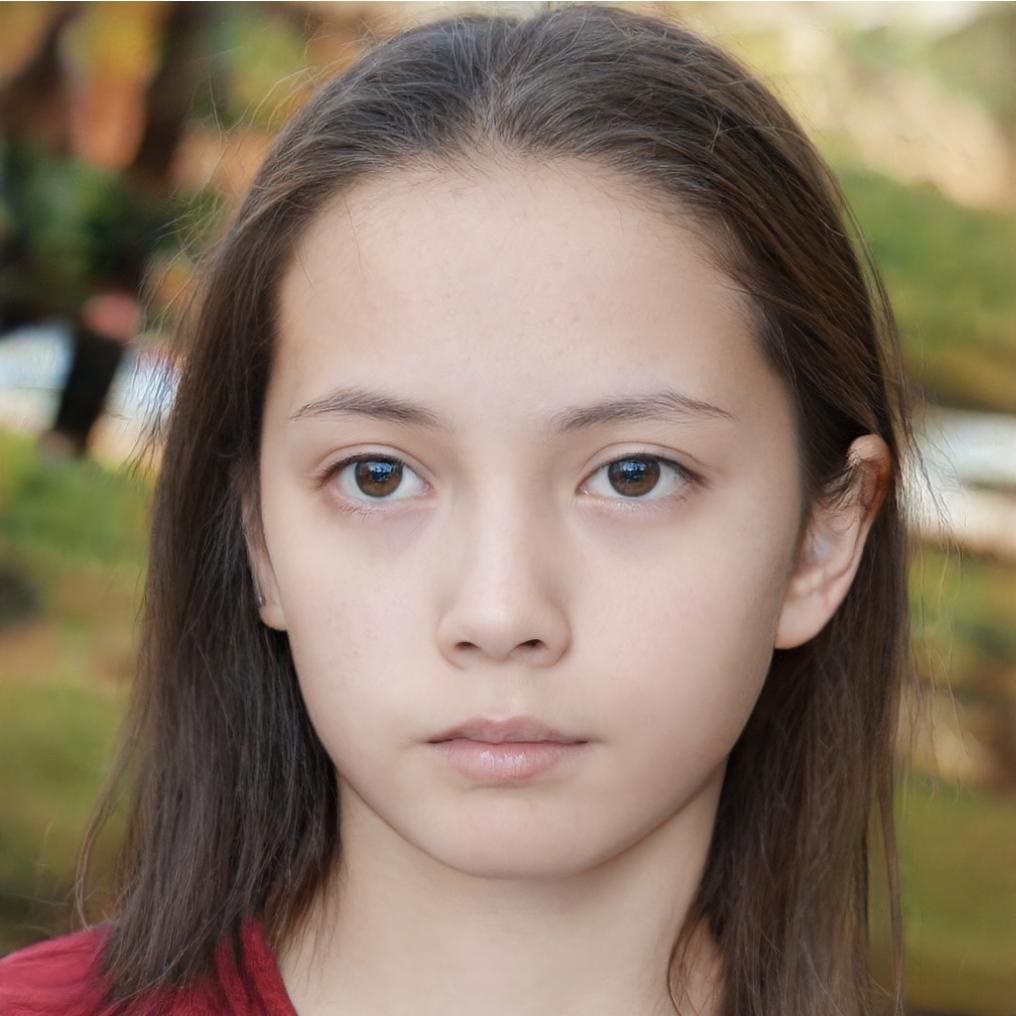}}}
    \caption*{\footnotesize StyleGAN2 '20}
\end{subfigure}
\hfill
\begin{subfigure}[t]{0.13\textwidth}
    \subcaption{}\vtop{\vskip0pt\hbox{\includegraphics[width=\linewidth]{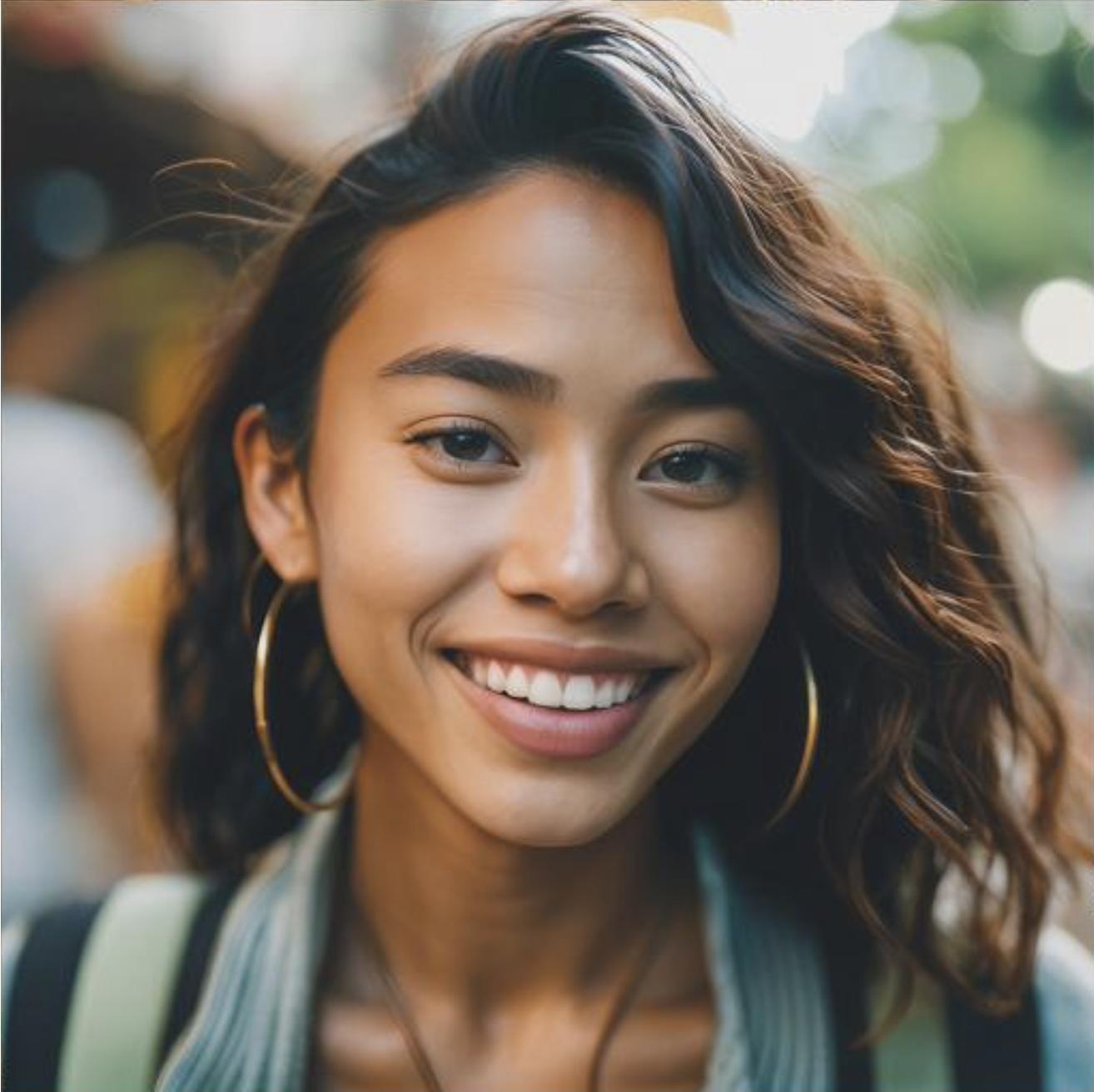}}}
    \caption*{\footnotesize SDXL '23}
\end{subfigure}
\hfill
\begin{subfigure}[t]{0.13\textwidth}
    \subcaption{}\vtop{\vskip0pt\hbox{\includegraphics[width=\linewidth]{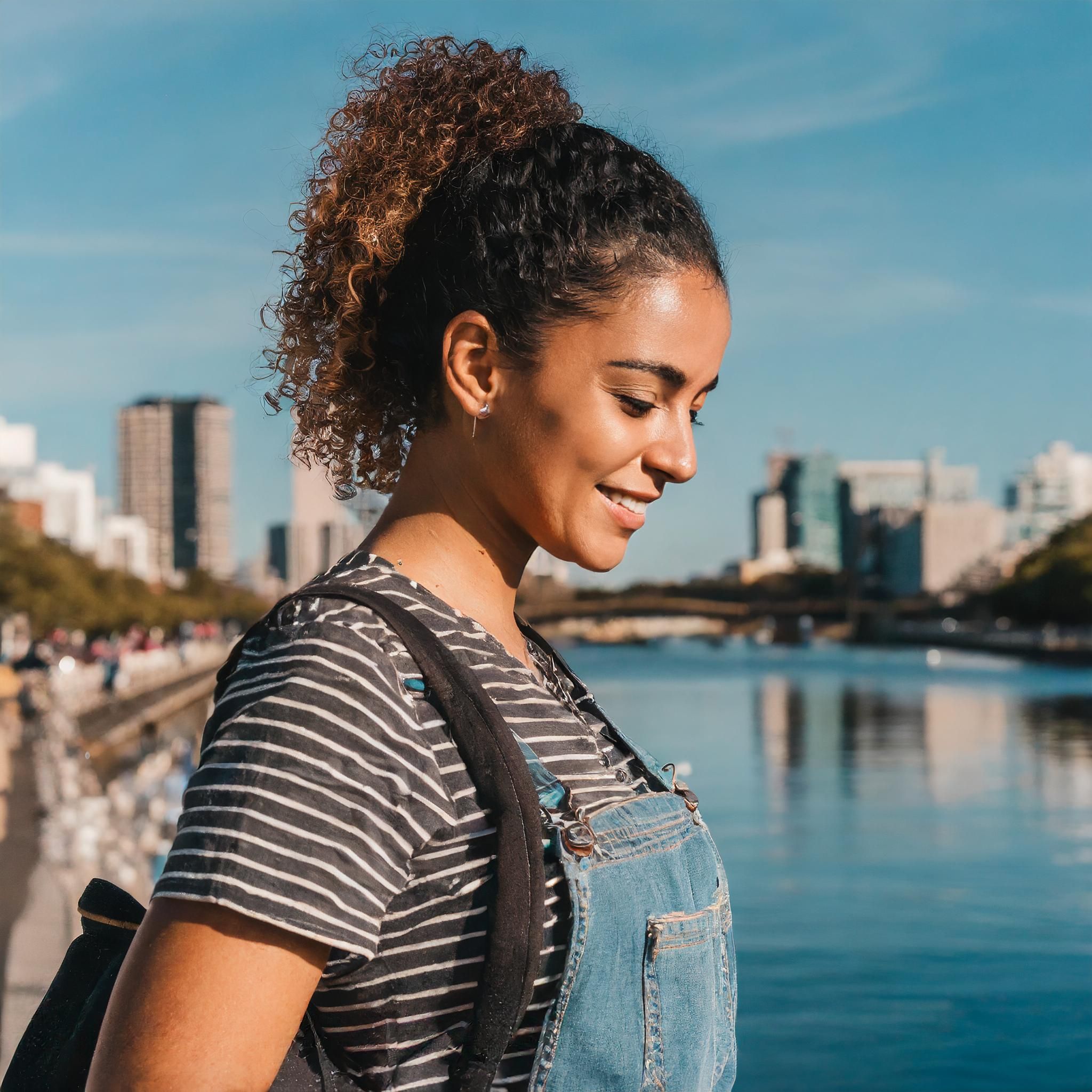}}}
    \caption*{\footnotesize Firefly '24}
\end{subfigure}
\caption{\mybold{Exemplar images of photorealism across a range of generative models.} \normalfont{Examples of AI-generated images from 2014 to 2024~\cite{goodfellow2014generativeadversarialnetworks,radford2015unsupervised, faceimageforgery,karras2018progressivegrowinggansimproved,karras2019style,karras2020analyzingimprovingimagequality, podell2024sdxl,adobe_firefly}.}}
\label{fig:AI-faces}
\Description{Portrait images from various image generation models that improve in quality and complexity over the years.}
\end{figure*}

Figure~\ref{fig:AI-faces} highlights the increasing complexity of AI-generated images over the past decade. The changing architectures and increasing photorealism pose a challenge for both humans and machines to distinguish real from AI-generated images. However, humans and machines are fundamentally different. For example, humans can critically reason about an image's elements and its context~\cite{wang2023context}. On the other hand, machine learning classifiers for detecting AI-generated images often oversimplify image authenticity as a question of real versus fake and ignore the critical reasoning about component parts and sub--questions that an ordinary person or digital forensics expert may consider when evaluating an image's authenticity~\cite{jacobsen2024deepfakes}.

\subsection{Human perception and evaluation of AI-generated media}

In response to the increasing realism of AI-generated media, researchers have been examining the degree to which humans can distinguish between authentic and AI-generated media. For example, researchers found that GAN-generated images of faces are indistinguishable from real face portraits~\cite{nightingale2022ai, Lago_2022}. However, for video deepfakes, humans are much better than random guessing~\cite{deepfakedetectionbyhumancrowds}, which may in part be due to humans' specialized ability to process the temporal elements of faces~\cite{deepfakedetectionbyhumancrowds, sinha2006face}. Researchers found that text--to--speech voices were rated as lower in quality and clarity than human voices in 2020~\cite{cambre2020choice} but have reached the point where research participants cannot tell the difference between short 20-second recordings of AI-generated voices and authentically recorded voices~\cite{barrington2024people}.

Recent research has identified specific cues and heuristics that people use to evaluate AI-generated media. For example, cues such as recording settings in the detection of text-to-speech audio~\cite{han2024uncovering} and speaking patterns in political deepfake videos~\cite{groh2024human}. However, two studies found that participants rarely attributed their judgments to specific visual features~\cite{hameleers2024they, wohler2021towards}, and in one of these deepfake studies, researchers found that participants are noticing the artifacts but rarely linking these to manipulation~\cite{wohler2021towards}. With respect to AI-generated text, research has highlighted that people tend to use flawed heuristics when attempting to distinguish AI-generated text from human--written text, like associating grammatical errors with AI-generation~\cite{Jakesch2022HumanHF}. 

Social context also plays a significant role in both what diffusion models generate~\cite{luccioni2024stable} and how people form beliefs about AI-generated images and their content. For example, researchers have found detection ability is influenced by shared identity between the viewer and subject of the content~\cite{mink2024s}. Furthermore, researchers have found that white AI-generated faces were disproportionately judged as human more frequently than their real counterparts~\cite{miller2023ai}. GAN-generated faces in portrait images were often perceived as more trustworthy than real faces~\cite{nightingale2022ai}, and as a result, people were less likely to question their authenticity~\cite{Lago_2022}. In instances where AI-generated images are linked to misinformation, researchers find that labeling AI-generated images and the associated content as ``potentially misleading" instead of simply ``AI-generated" had a stronger influence on curtailing participants' self--reported intentions to share misinformation~\cite{epstein2023label, wittenberg2024labeling}.

Researchers have approached a number of methods for measuring photorealism perceived by humans. For example, prior research has examined photorealism with carefully worded questions such as ``Is the image photorealistic?''~\cite{liang2024rich}, ``Does the image look like a real photo or an AI-generated image?''~\cite{lee2024holistic, otani2023toward} and ``Whether this image could be taken with a camera?''\cite{yan2024sanitycheckaigeneratedimage}. These questions are useful for assessing participants' subjective opinions but do not capture the human ability to distinguish real images from fake images and can potentially suffer from demand characteristic bias. Another approach has been to characterize photorealism by examining the features that can influence realism, such as aesthetics and semantically meaningful content of an image~\cite{peng2024crafting}. A third approach involves simply defining images as photorealistic if they are rendered with computer graphics software~\cite{lyu2005realistic}. In this paper, we approach photorealism from the psychophysics perspective, examining participants' objective performance at distinguishing real images from fake images~\cite{zhou2019hype}. 

\subsection{Categorizing artifacts and implausibilities in diffusion model-generated images} \label{sec:artifimpl}

Previous research on earlier versions of diffusion models categorized the kinds of qualitative failures of diffusion model-generated images as distorted body parts, impossible geometry, physics violations, illogical relationships in a scene, and noise~\cite{borji2023qualitative}. In addition to obvious issues with hands, feet, eyes, and teeth, research at the intersection of digital foresnics and AI-generated images shows details such as corneal reflections~\cite{hu2021exposing} and irregular pupil shapes~\cite{guo2022eyes} can also be artifacts.  Likewise, violations of physics like implausible shadows, lighting, and perspective errors~\cite{farid2022perspectiveinconsistencypainttext, farid2022lighting, sarkar2024shadows} often occur in diffusion model generated images that otherwise appear photorealistic. 
\section{Methods}\label{sec:method}
\begin{figure*}[!htb]
    \centering
    \includegraphics[width=\linewidth]{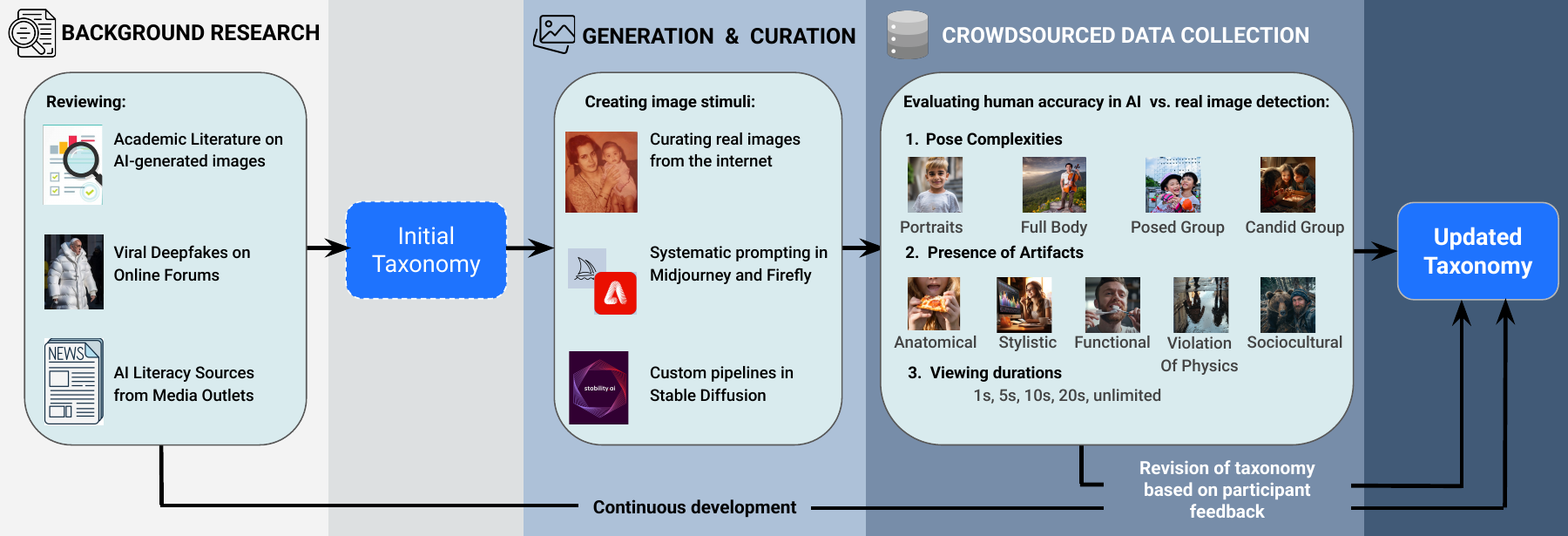}
    \caption{\mybold{Overview of the taxonomy development process.} \normalfont{In the background research stage, we reviewed existing literature on visible features of AI-generated images from a wide range of sources. This included academic literature, practitioner perspectives in AI literacy articles, and discussions on the photorealism of AI-generated images online. From these features, we developed an initial taxonomy of artifacts. In the Generation and Curation stage, we used our taxonomy of artifacts to create a dataset of 599 images. Of these images, 149 were real photographs curated from the internet, and 450 were generated in Midjourney, Firefly, and Stable Diffusion through extensive iteration with photorealistic image generation techniques. We used the dataset of images for an online crowdsourced experiment where we evaluated participant accuracy in identifying AI-generated images. We iteratively refined the taxonomy based on results from the experiment and continued monitoring new literature on AI-generated images as generative models evolved.}}
    \label{fig:overview}
    \Description{Overview of the Taxonomy Development Process. In the background research stage, we reviewed existing literature on visible features of AI-generated images from a wide range of sources. This included academic literature, practitioner perspectives in AI literacy articles, and discussions on the photorealism of AI-generated images online. From these features, we developed an initial taxonomy of artifacts. In the Generation and Curation stage, we used our taxonomy of artifacts to create a dataset of 599 images. Of these images, 149 were real photographs curated from the internet, and 450 were generated in Midjourney, Firefly, and Stable Diffusion through extensive iteration with photorealistic image generation techniques. We used the dataset of images for an online crowdsourced experiment where we evaluated participant accuracy in identifying AI-generated images. We iteratively refined the taxonomy based on results from the experiment and continued monitoring new literature on AI-generated images as generative models evolved.}
\end{figure*}

We develop a detailed taxonomy of visual features, qualities, and artifacts that offer cues that an image is AI-generated or not following a three-step process based on the taxonomy development method proposed by Nickerson et al.~\cite{nickerson2013method}. We began by drafting an initial version of the taxonomy based on a review of visual features previously identified in AI literacy resources, academic literature, and online discussions (\autoref{sec:initializing-taxonomy}). We then employed two parallel processes to develop the taxonomy: iteratively generating and curating a dataset of 599 images to showcase the taxonomy artifacts (\autoref{sec:curation}) and conducting an online, crowdsourced experiment using these curated images to assess human detection ability (\autoref{sec:onlexp}). Third, we integrated participant feedback---both accuracy metrics and thematic comments---back into the taxonomy, allowing real-world human detection behaviors to inform the final categorization.

While the taxonomy development was guided by data, we acknowledge that subjectivity is inherent in the categorization process. To mitigate this subjectivity and ensure methodological rigor, multiple team members independently identified recurring patterns and artifacts during image generation and curation, and we reconciled any differences through structured discussion until stable, consistently observed phenomena emerged. In Figure~\ref{fig:overview}, we show an overview of the taxonomy development process.

\subsection{Initializing the Taxonomy} \label{sec:initializing-taxonomy}
In addition to reviewing academic literature discussed in Section \ref{sec:relwork}, we surveyed traditional and social media discussions about distinguishing AI-generated images. These included AI literacy resources on how to identify AI-generated content in media (see~\autoref{fig:nytimes}), online discussions of AI-generated images in response to viral deepfakes, and popular posts discussing photorealism on online forums for AI image creators (Reddit channels such as r/Midjourney and r/StableDiffusion) to initialize the taxonomy. These sources highlighted several visual cues, including (1) anatomical implausibilities such as pupil dilation and misaligned eyes \cite{norton2024deepfakes, techtarget2024deepfakes, nytimes2024deepfake}, teeth \cite{realitydefender2024deepfake}, hair  \cite{norton2024deepfakes, realitydefender2024deepfake}, fingers, and alignment of body parts  \cite{norton2024deepfakes, realitydefender2024deepfake, nytimes2024deepfake}; (2) irregular reflections and shadows \cite{realitydefender2024deepfake, eweek2024deepfake, techtarget2024deepfakes}; (3) unnatural color balances  \cite{norton2024deepfakes, realitydefender2024deepfake}; (4) a mismatch in textures and styles within an image  \cite{norton2024deepfakes, realitydefender2024deepfake, eweek2024deepfake}; (5) garbled or nonsensical text \cite{nytimes2024deepfake}; 
(6) photoshoot-like perfection and overly cinematic scenarios \cite{reddit2023}.

While some prior research has suggested that AI-generated face images can be indistinguishable from real ones \cite{hulzebosch2020detectingcnngeneratedfacialimages, nightingale2022ai}, more complex scenes, such as group photos have not been thoroughly explored.  We address this by introducing a detailed categorization of \textit{scene complexity} across all images. We identified four distinct scene types that capture varying levels of detail within an image:
 \begin{itemize}[leftmargin=*, label=\tiny{$\bullet$}]
    \item \textbf{Portraits (Single-Subject Close-Up)}: An image featuring a single individual, typically focusing on the face and torso. The individual is the primary focus, often set against a blurred or minimal background. Portraits have relatively low scene complexity.
    
     \item \textbf{Full-Figure (Single-Subject Full Body)}: An image featuring a single individual whose entire body is visible along with the surrounding environment. These images exhibit moderate scene complexity, as they include more details than portraits, such as the person's posture and interaction with their setting.
         
     \item \textbf{Posed Group}: An image featuring multiple people posing for the camera in a structured manner. These images involve higher scene complexity due to the presence of multiple subjects, their interactions, and the added challenge of capturing each person accurately. 
    
     \item \textbf{Candid Group}: An image of multiple people captured in candid moments. These images often feature intricate interactions between people and their environments, representing the highest level of scene complexity.
 \end{itemize}

\subsection{Stimuli Generation and Curation} \label{sec:curation}
We created a dataset of 599 images. This image set included 149 real photographs curated from the internet, from which we derived the scenarios for 450 images that we generated using AI. Of the AI-generated images, 207 were generated in Midjourney, 133 in Firefly, and 110 in Stable Diffusion. 

Drawing on techniques shared on online forums and articles, we developed strategies to generate photorealistic images. We first curated real photographs and then experimented extensively with the three AI-generation tools to create over 3000 images that depict similar scenarios as the real images. The final dataset represents a selection of images from this larger set that we judged to be not immediately identifiable as AI-generated at first glance. This selection enabled us to focus on more challenging cases, better assess participants' ability to detect subtle artifacts, and enhance the relevance of our taxonomy in real-world scenarios. 

\subsubsection{Curating Real Photographs.}  
We sourced real photographs from online platforms, selecting them to represent diverse scenarios (e.g., diverse cultural settings with celebrity and non-celebrity figures engaging in common and uncommon activities) across the four dimensions of scene complexity \autoref{sec:initializing-taxonomy}. We established these categories to curate a diverse range of real photographs and ensure our dataset accurately captures how the features in our taxonomy may manifest and be perceived in both real and AI-generated images. 
We verified that these images were real photographs by confirming details like the creation date, photographer, and publisher. We include a complete list of image sources and verification details in the following link: https://github.com/negarkamali/Replication-for-Characterizing-Photorealism-2025/. We used these real images to inform the prompts to generate images using AI tools.

\begin{figure*}[htbp]
\centering
\captionsetup{justification=raggedright, singlelinecheck=false, skip=2pt, aboveskip=0pt, belowskip=2pt}

\begin{subfigure}[t]{0.24\linewidth}
    \subcaption{}\includegraphics[width=\linewidth]{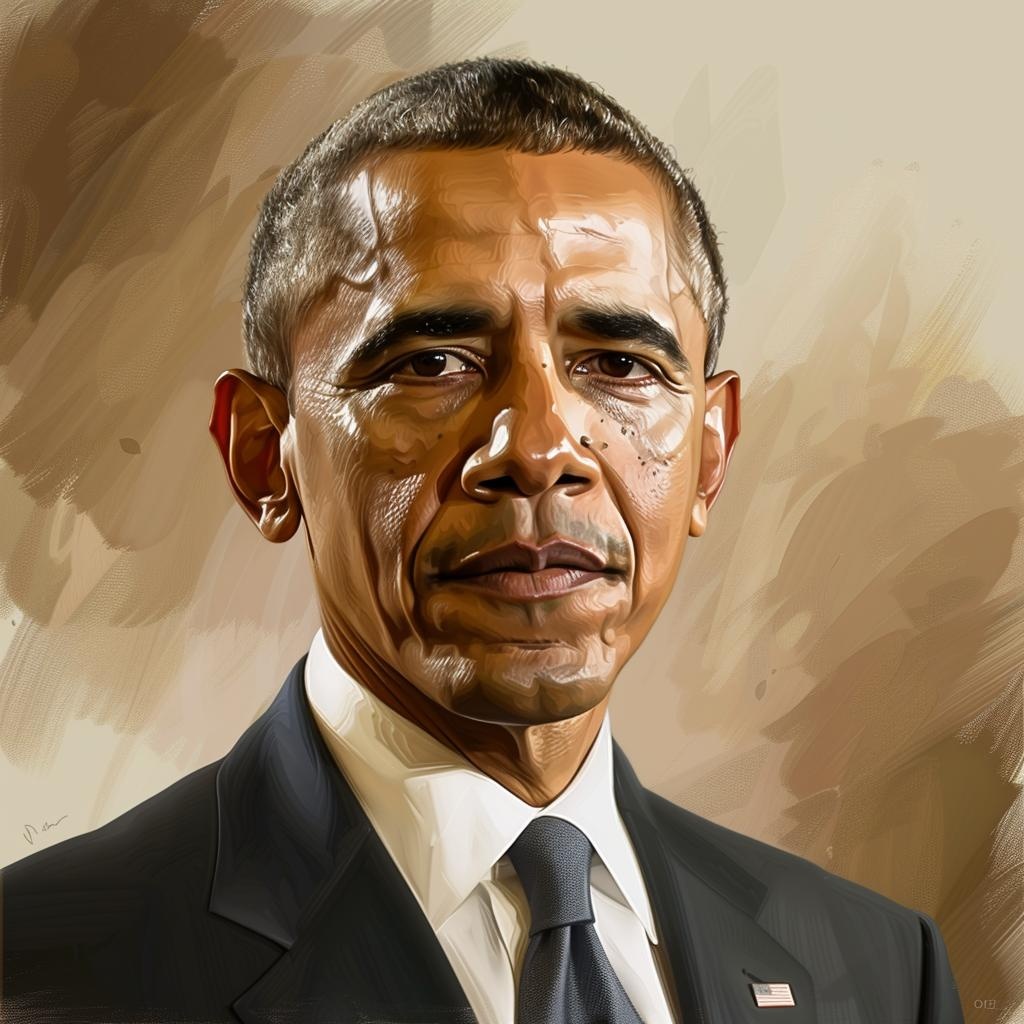}
\end{subfigure}
\hfill
\begin{subfigure}[t]{0.24\linewidth}
    \subcaption{}\includegraphics[width=\linewidth]{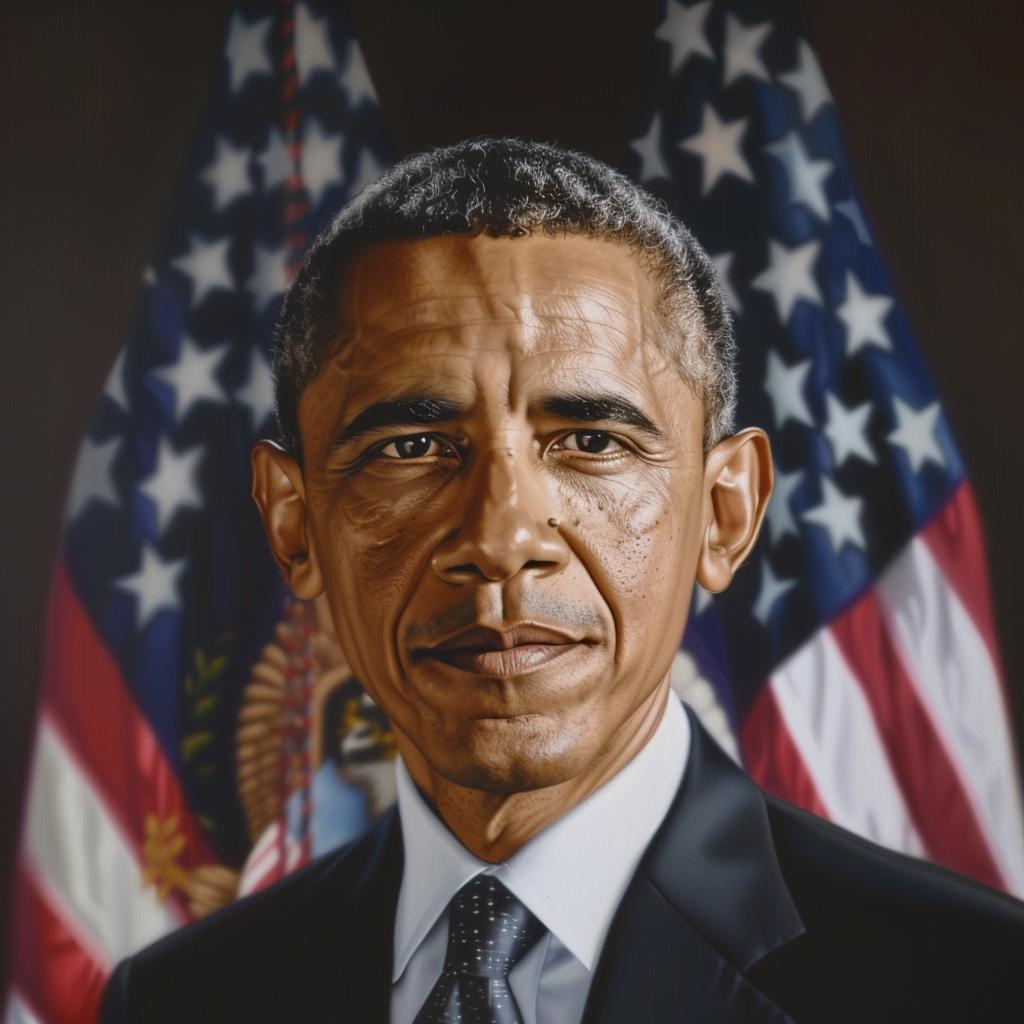}
\end{subfigure}
\hfill
\begin{subfigure}[t]{0.24\linewidth}
    \subcaption{}\includegraphics[width=\linewidth]{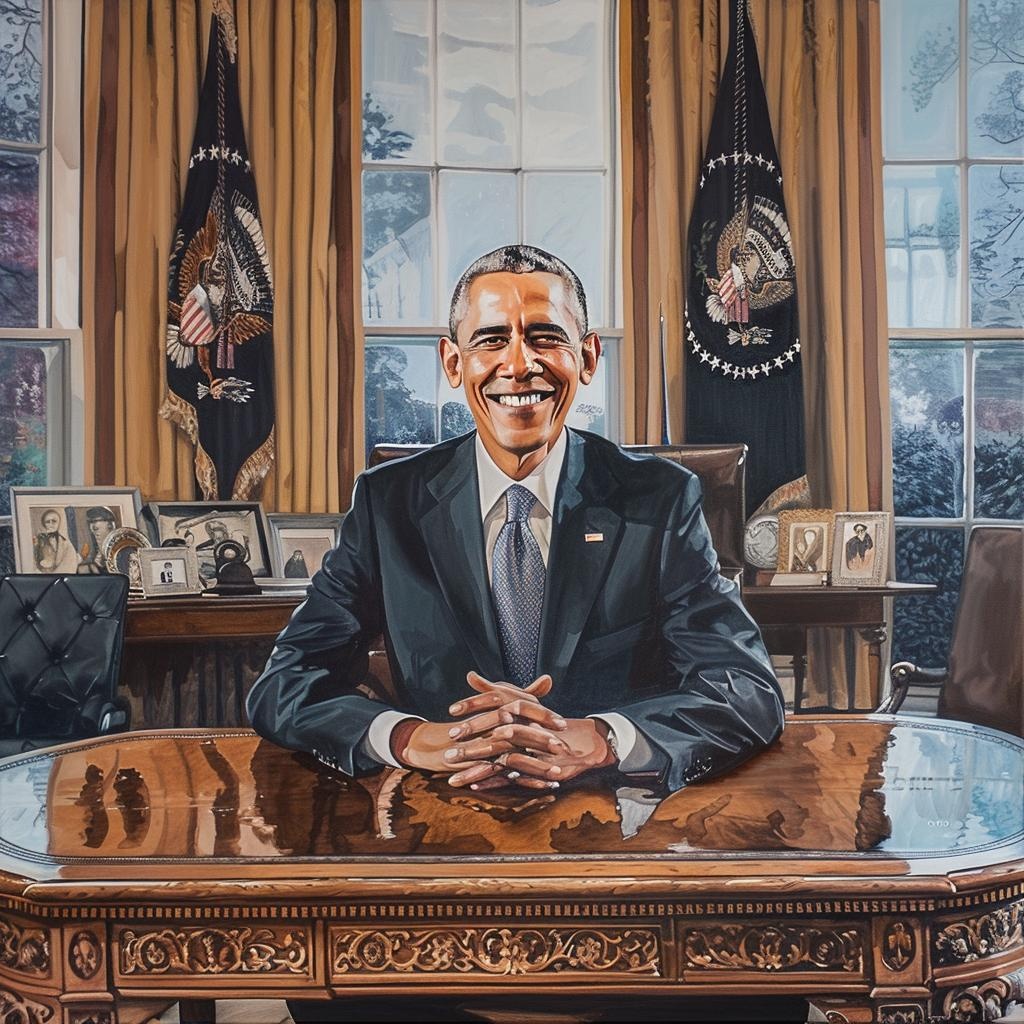}
\end{subfigure}
\hfill
\begin{subfigure}[t]{0.24\linewidth}
    \subcaption{}\includegraphics[width=\linewidth]{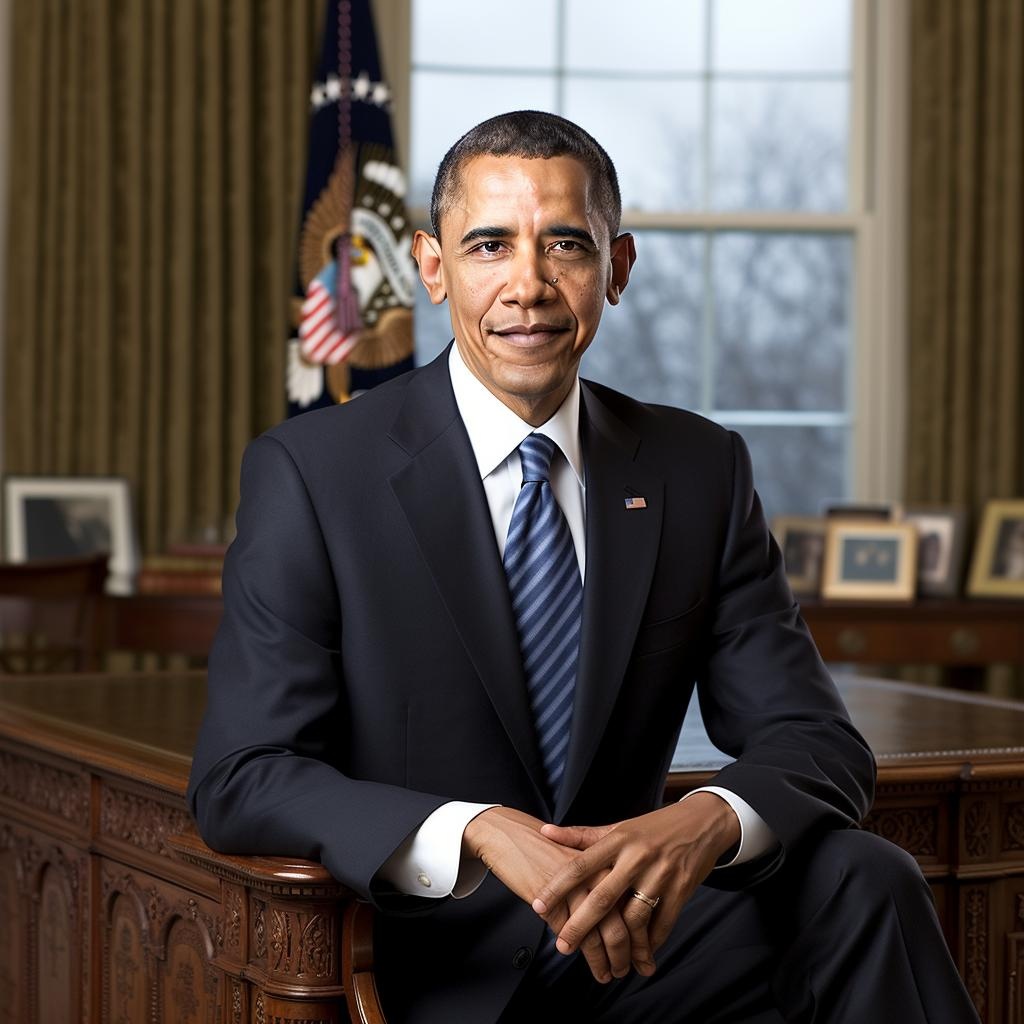}
\end{subfigure}

\caption{\mybold{Images of Barack Obama generated in Midjourney V5.} \normalfont{Images were created by progressively adding details to the prompt shown below each image: \textbf{A.} ``Portrait of Barack Obama." \textbf{B.} ``Portrait of Barack Obama, hyperrealistic, megapixel." \textbf{C.} ``Portrait of Barack Obama, sitting in his Oval Office, smiling, hyperrealistic, megapixel." \textbf{D.} ``A portrait of Barack Obama sitting in the Oval Office, smiling, wearing a suit and tie, shot on Kodak, hyperrealistic, grainy, official portrait."}}

\label{fig:obamaprompt}
\Description{Images of Barack Obama generated in Midjourney V5 by progressively adding more details to the prompt as such: \textbf{A.} "Portrait of Barack Obama". \textbf{B.} "Portrait of Barack Obama, Hyper realistic, Megapixel." \textbf{C.} "Portrait of Barack Obama, sitting in his oval office, smiling, hyperrealistic, megapixel". \textbf{D.} "A portrait of Barack Obama sitting in the Oval Office, smiling, wearing a suit and tie, shot on kodak, hyperrealistic, grainy, Official Portrait".} 
\end{figure*}

\subsubsection{Generating Images using AI tools.} Based on our curated set of real images, we generated images in Midjourney V5 and V6, Adobe Firefly Image 2, and Stable Diffusion to depict similar scenarios. In Midjourney and Firefly, we started the image generation process by creating a simple prompt describing the scenario. We then progressively refined the prompts by adding details about the quality of the image using keywords known to enhance image quality and resolution from the sources mentioned in~\autoref{sec:initializing-taxonomy}. Our prompts followed the basic structure of: ``[Subject description] [action or pose], [context or setting description], [clothing or appearance details], [image quality and style attributes], [camera or film type if applicable]." If the images were insufficiently realistic, then further details were added to the end of the above prompt such as: [specific details unique to the scenario], [`high resolution', `hyper-realistic', `megapixel', etc.]. The sequence of images in Figure \ref{fig:obamaprompt} shows the progression of a prompt and the resulting image qualities as more details and keywords are added in Midjourney V6. 

We also generated images inspired by real scenes of human interactions found in publicly available news sources and online media, ensuring they maintained a similar context and zoom level to real photographs. For example, we used a real reference image of a Ukrainian soldier getting married from New York Magazine~\cite{nymag2024ukraine}.

\begin{figure*}[htb]
\centering
\captionsetup{justification=raggedright, singlelinecheck=false, skip=2pt}
\begin{subfigure}[t]{0.45\textwidth}
    \subcaption{}\vtop{\vskip0pt\hbox{\includegraphics[width=\linewidth]{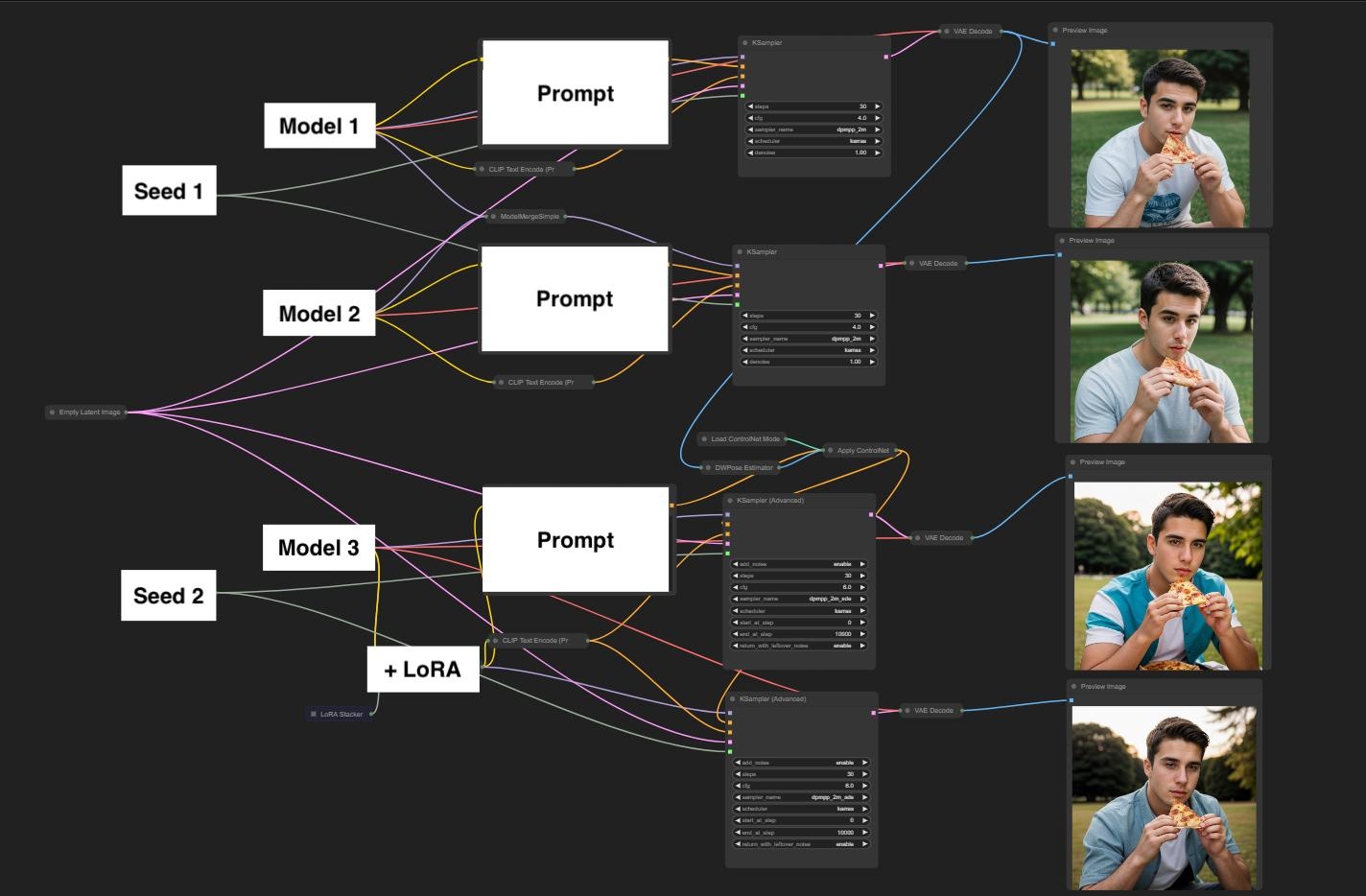}}}
\end{subfigure}
\begin{subfigure}[t]{0.5\textwidth}
    \subcaption{}\vtop{\vskip0pt\hbox{\includegraphics[width=\linewidth]{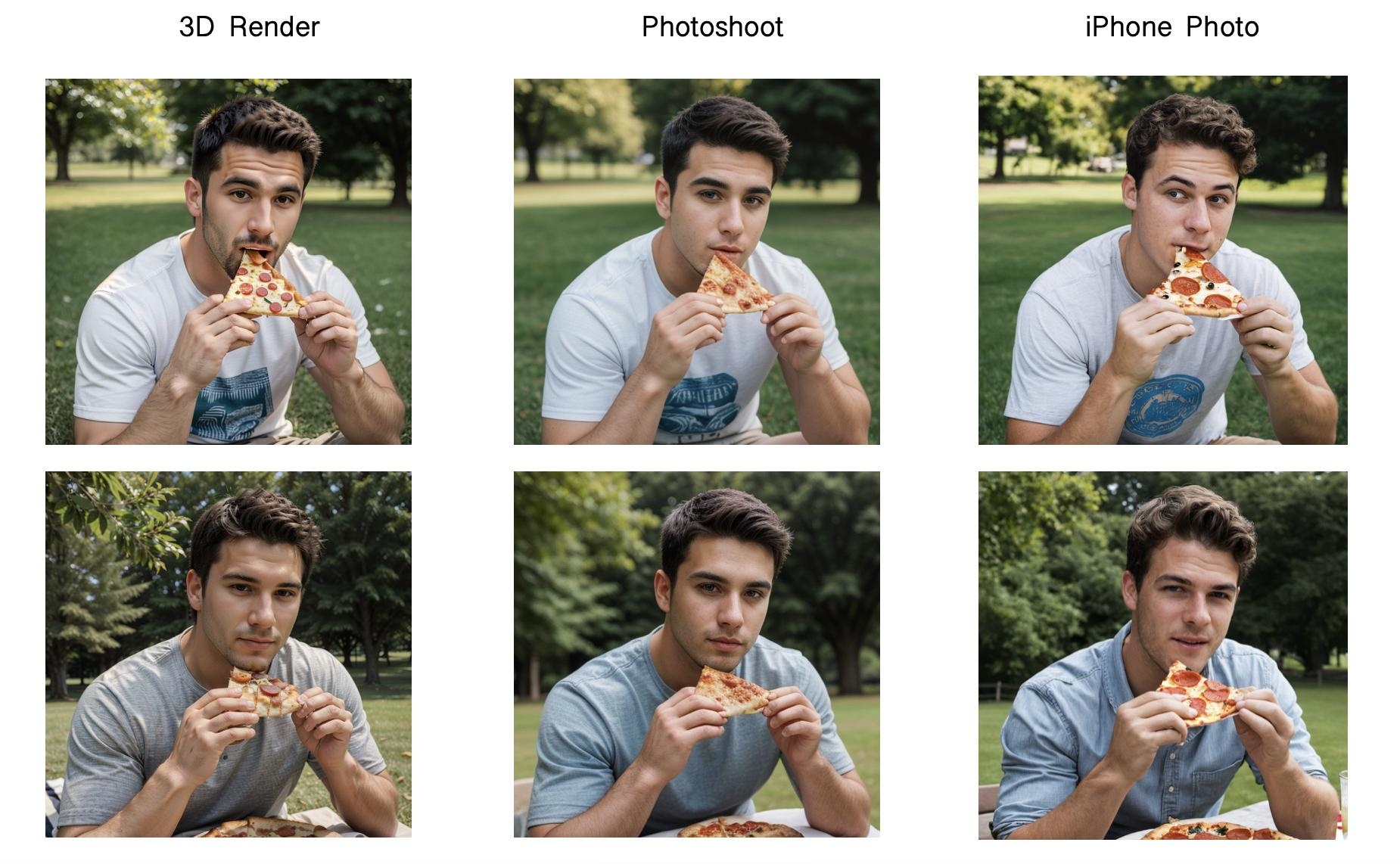}}}
\end{subfigure}
\caption{\mybold{Stable Diffusion pipeline and outputs of varied styles from the same pose and prompt.} \normalfont{\textbf{A.} Four pipelines for generating four variations of the prompt ``photo of a 25 year old man eating a slice of pizza, outside on the grass in a park, sunny, plain clothes." \textbf{B.} A sample of the variations that we labeled as having the style of a ``3D Render", ``Photoshoot", or ``iPhone photo." }}
\label{fig:stylespipe}
\Description{Two sets of images generated by Stable Diffusion showing style variations. A. Four variations of a man eating pizza in a park, generated with different pipelines. B. Samples labeled as "3D Render", "Photoshoot", or "iPhone photo" styles.}
\end{figure*}

In Stable Diffusion, we developed custom pipelines in order to generate images that were more realistic than the outputs of the original models. Using SD1.5 \cite{rombach2021highresolution} and SDXL \cite{podell2024sdxl} as the base models, we used techniques such as merging fine-tuned portrait models and combining outputs of different models to reduce obvious artifacts and generate highly photorealistic images, particularly for portraits.  We also experimented with generating the same poses in various styles in order to isolate the impact of certain categories of artifacts, as shown in Figure \ref{fig:stylespipe}. We used ControlNets \cite{zhang2023adding} to maintain consistent poses while altering other elements such as models, seed, and prompt scheduling. Additionally, we used Low-Rank Adaptation (LoRAs) \cite{hu2021loralowrankadaptationlarge} to introduce realistic imperfections like wrinkles and shorten the depth of field to produce iPhone-style images. We further refined images by implementing pipelines that regenerate artifacts in the hands and faces. A refining pipeline is shown in Appendix \ref{fig:refiningpipe}. 

\subsection{Crowdsourced Experiment} \label{sec:onlexp}

\begin{figure}[htb]
    \centering
    \includegraphics[width=\linewidth]{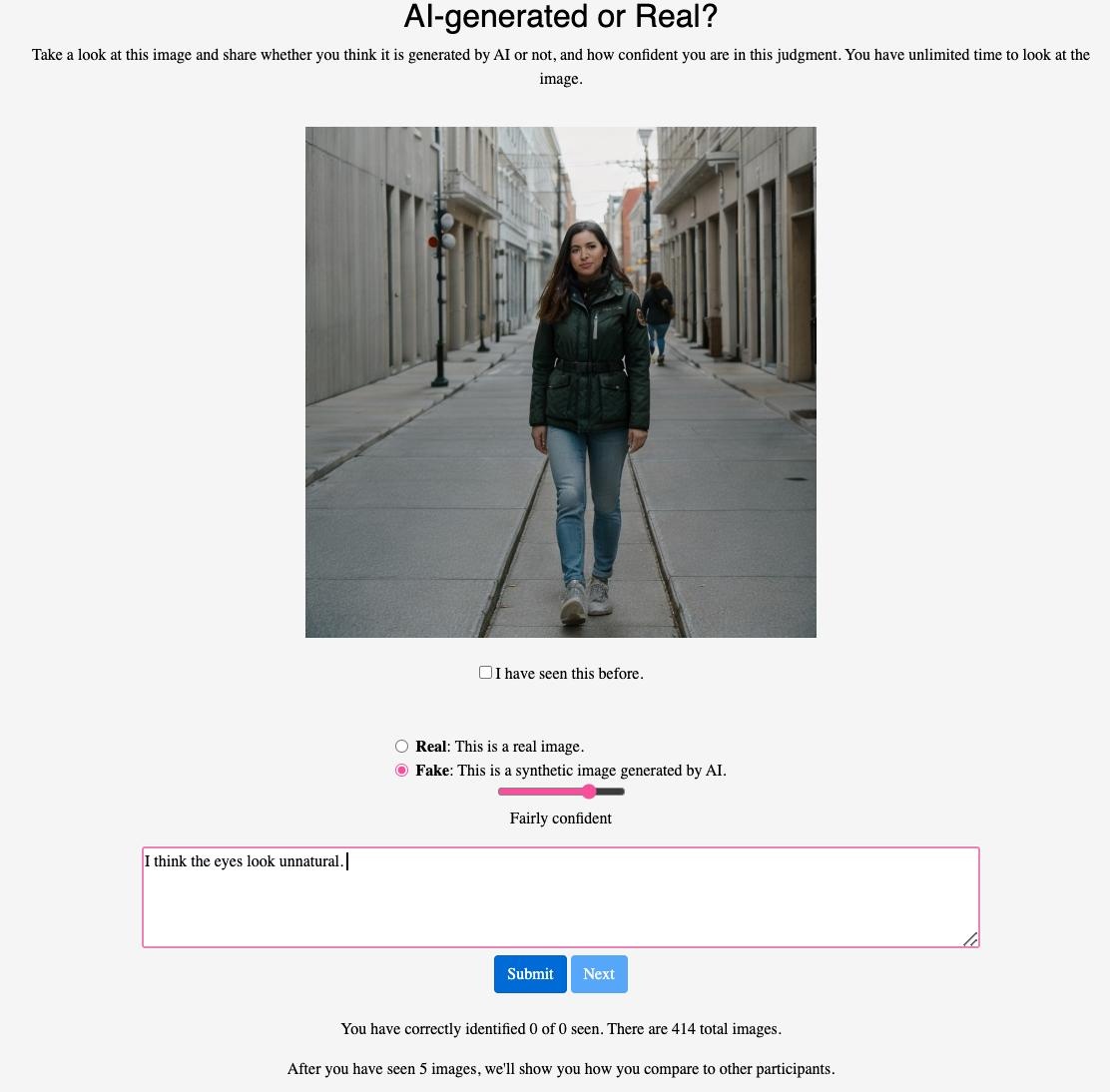}  
    \caption{A screenshot of the experiment website interface. }
    \label{fig:website}
    \Description{A screenshot of the experiment website interface}
\end{figure}

\subsubsection{Image Stimuli}
\label{sec:imagestimuli}

Across the entire experiment timeline, we collected 749,828 responses to whether 1083 images are AI-generated or real from 50,444 participants. Across the experiment timeline, we added and removed stimuli for two reasons. First, we included higher quality and more diverse images over the course of the experiment as new tools for controlling diffusion models became available (e.g., ControlNets and LORAs), and we identified prompt engineering techniques for producing more photorealistic images. Second, we split the experiment into two phases based on how we selected the diffusion model-generated images. In the first and main phase of the experiment, the stimuli were 149 real photographs and 450 most photorealistic images that our research team could generate with diffusion models. By comparison, the 482 stimuli in the second phase were based on generating 11 or more images for each of the 39 text prompts without curation. This second phase enables us to identify the effect of human curation (the selection bias involved in our research team selecting the most photorealistic AI-generated images) relative to no human curation on how accurately participants can distinguish AI-generated images.

\subsubsection{Experimental Design}\label{exp-design}

To ensure the quality of results, we implemented two measures. First, participants were shown an attention check image that was clearly AI-generated. Those who failed to identify this image correctly were excluded from the analysis. Second, we included an optional checkbox allowing participants to indicate if they recognized an image from outside the experiment, allowing us to filter out responses influenced by prior familiarity with the image. 

In the initial version of the experiment from February to May 2024, we prioritized presenting unseen images to participants rather than maintaining a balanced ratio of real and AI images. In the next version of the experiment, from May to June 2024, we used stratified random sampling to select stimuli, ensuring participants saw real images 50\% of the time and AI-generated images 50\% of the time. We repeated the analysis both with and without the data from the initial version of the experiment and did not find significant changes in the accuracy distributions. Details of this comparison are provided in Appendix \ref{tab:accuracy-comparison-dataset} and \ref{fig:accuracy-comparison-dstaset}. To ensure that newly added images to the stimuli set as described in \autoref{sec:imagestimuli} were adequately represented in participant responses, we implemented an up-sampling strategy that prioritized showing images that were labeled fewer than 100 times. 

After participants responded to five images, we randomized the display time of each subsequent image to one of the following conditions: unlimited time, 20 seconds, 10 seconds, 5 seconds, and 1 second. Participants were informed of the time limit at the start of each time-restricted trial by an on-screen message (e.g., "You have 20 seconds to view this image") and were instructed to click a button to reveal the image and begin the countdown.

\subsubsection{Participants}

We collected data through a public website (detectfakes.kellogg.northwestern.edu) where people could test their ability to detect AI-generated images. The website remained accessible throughout our taxonomy development, allowing us to gather responses as we updated image stimuli to reflect improvements in generation models. In total, 50,444 unique participants contributed 749,828 observations. According to Google Analytics, participants who visited our website came from 165 countries; the five countries with the most participants were United States, South Korea, United Kingdom, India, and Germany. We did not collect additional demographic data or other data on participants.

\subsubsection{Image-level and Participant Level Analyses}\label{data-analysis}
We define accuracy as a binary measure of whether a participant selected the correct label (Real/AI-generated) for an image. We aggregated accuracy at two levels: image-level accuracy and participant-level accuracy. For image-level accuracy measurements described in Sections~\ref{sec:acc-general},~\ref{sec:acc-scene-complexity},~\ref{sec:acc-time},~\ref{sec:acc-presence-artifacts} and~\ref{sec:human-curation}, we aggregated and averaged the binary responses (0 for "real" and 1 for "AI-generated") provided by participants for each image. Image-level accuracy was calculated as the mean of correct identifications across various factors contributing to photorealism, which are described in each section. For participant-level accuracy measurements described in Section \ref{sec:indiv-acc}, we calculated each participant's accuracy by averaging their correct identifications across all viewed images. 

We present descriptive statistics to summarize our findings, focusing on mean accuracies and their associated 95\% confidence intervals (CIs) obtained through non-parametric bootstrapping~\cite{TransparentStatsJun2019}. We use these measures to describe trends and patterns in the data without using statistical significance to dichotomize effects.
However, readers can apply that interpretation to the CIs if they desire.

For a qualitative analysis of the 34,675 optional comments provided on our website, we utilized GPT-3.5-turbo to extract ten recurring themes and map the comments to our taxonomy categories based on the types of artifacts and patterns reported by participants. 

\subsubsection{Ethics} \leavevmode
This research complied with all relevant ethical regulations. The Northwestern University Institutional Review Board (IRB) determined that it met the criteria for exemption from further review. The study's IRB identification number is STU00220627.

\section{Taxonomy of Artifacts in AI-generated Images}
\label{sec:taxonomy}
 \begin{figure*}[h!]
\centering
\includegraphics[width=0.9\linewidth]{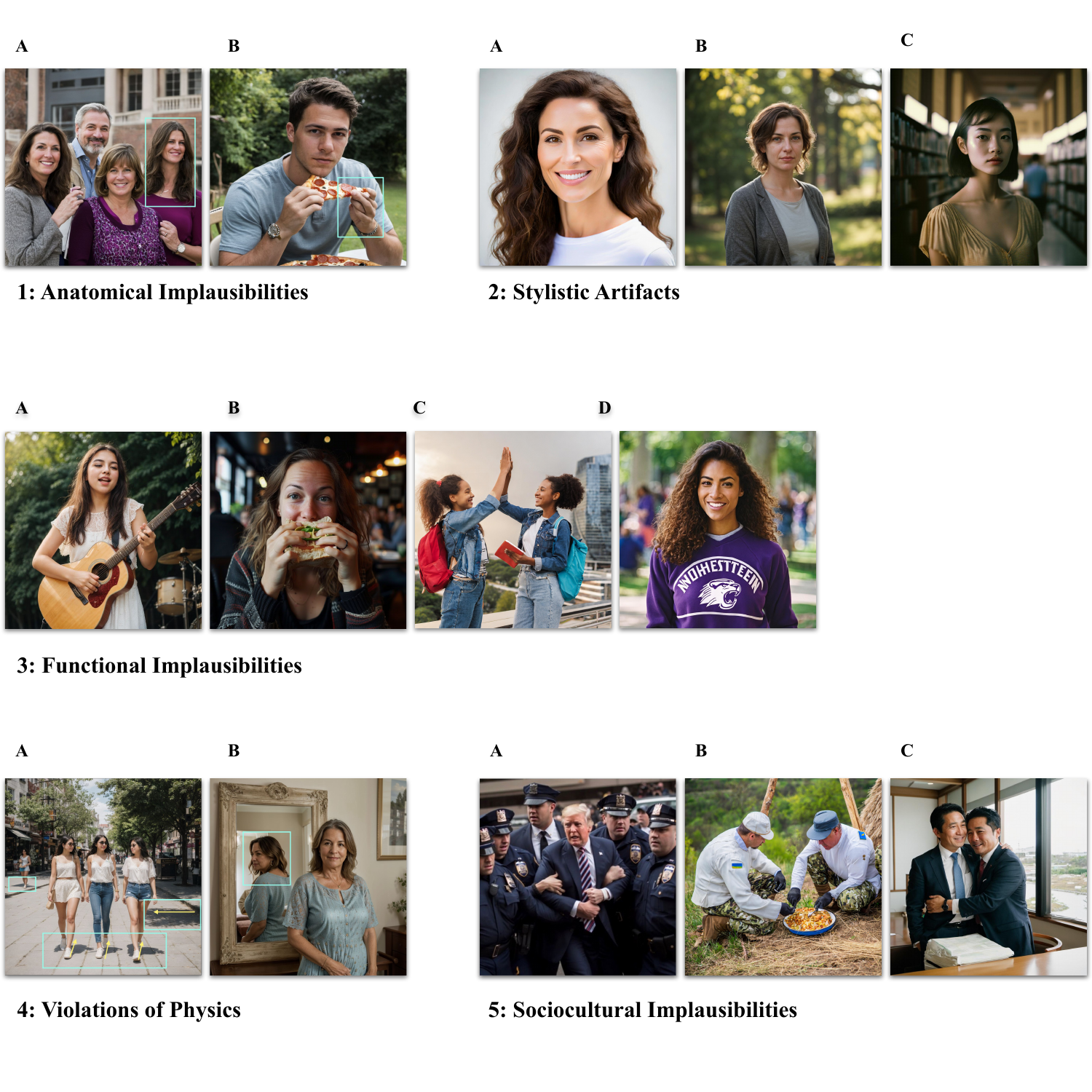}
\caption{\mybold{Categories of artifacts in AI-generated images.}  
\normalfont{This figure presents representative examples of common artifacts found in AI-generated images across five categories: 
\textbf{1. Anatomical Implausibilities:}  
\textbf{A.} Stable Diffusion image of a group of people where one woman has an abnormally long neck.  
\textbf{B.} Stable Diffusion image of a man eating pizza where his left fingers appear anatomically implausible. 
\textbf{2. Stylistic Artifacts:}  
\textbf{A.} Firefly image of a woman with a waxy texture.  
\textbf{B.} Stable Diffusion image with a cinematized style.  
\textbf{C.} Midjourney image of a woman with glossy skin. 
\textbf{3. Functional Implausibilities:}  
\textbf{A.} Stable Diffusion image where the guitar strings are not taut.  
\textbf{B.} AI-generated image of a woman holding a sandwich in an unlikely way.  
\textbf{C.} Firefly image where the strap on the red backpack merges into the denim jacket.  
\textbf{D.} AI-generated image of a woman wearing a shirt with incomprehensible text. 
\textbf{4. Violations of Physics:}  
\textbf{A.} Stable Diffusion image where the shadows fall in inconsistent directions.  
\textbf{B.} Stable Diffusion image of a woman standing in front of a mirror in which her reflection is inconsistent with the direction of her face. 
\textbf{5. Sociocultural Implausibilities:}  
\textbf{A.} Midjourney image of Donald Trump being restrained \cite{higgins2023tweet}.  
\textbf{B.} Firefly image depicting Ukrainian servicemen dressed in white shirts and hats that are not commonly part of the uniform. The flags on their shirts are different, and on the right serviceman, the flag is positioned awkwardly on their back and not their arm.  
\textbf{C.} Stable Diffusion image of an unlikely scenario of two Japanese businessmen hugging in a professional setting.}}

\label{fig:artifacts}

\Description{A composite figure illustrating five categories of AI-generated image artifacts:

1. Anatomical Implausibilities:  
   A. AI-generated image of a group of people, where one woman has an unnaturally long neck.  
   B. AI-generated image of a man eating pizza, where his left fingers appear anatomically incorrect, with extra or missing segments.  

2. Stylistic Artifacts:  
   A. AI-generated image of a woman with an unnaturally smooth and waxy skin texture.  
   B. AI-generated image with an exaggerated cinematic style, creating an overly dramatic effect.  
   C. AI-generated image of a woman with a glossy, plastic-like skin appearance.  

3. Functional Implausibilities:  
   A. AI-generated image of a guitar where the strings are not properly aligned or taut.  
   B. AI-generated image of a woman holding a sandwich in an awkward and impractical way.  
   C. AI-generated image where a red backpack strap appears to merge into the denim jacket, lacking clear separation.  
   D. AI-generated image of a woman wearing a shirt with distorted, incomprehensible text.  

4. Violations of Physics:  
   A. AI-generated image where shadows are inconsistent, falling in conflicting directions.  
   B. AI-generated image of a woman standing in front of a mirror, where her reflection does not match her actual pose.  

5. Sociocultural Implausibilities:  
   A. AI-generated image of Donald Trump being restrained in an unlikely or fabricated scenario.  
   B. AI-generated image depicting Ukrainian servicemen wearing incorrect uniforms with misplaced flags.  
   C. AI-generated image of two Japanese businessmen engaging in an uncharacteristically intimate hug in a formal setting.}
\end{figure*}

Our taxonomy organizes artifacts and implausibilities that may appear in AI-generated images into five high-level categories that are described in further detail in a how-to guide for identifying diffusion model-generated images\cite{kamali2024distinguish}. 

\textbf{1. Anatomical Implausibilities:} This category refers to artifacts that appear in the depiction of people within an image. These include unlikely artifacts in individual body parts, like hands with extra or missing fingers as shown in Figure \ref{fig:artifacts}--1B, or the disproportionately long woman's neck in Figure \ref{fig:artifacts}--1A. They also include artifacts in facial features such as an unnaturally empty gaze, overly shiny eyes, overlapping of the teeth and mouth, and unlikely proportions or configurations of limbs. In images of multiple people, this includes merged body parts and inconsistent proportions of body parts across different people. Anatomical implausibilities also include biometric artifacts such as size, shape, contours, and proportions of specific facial features if the person in the image is known. These biometric features include eyes, nose structure, mouth edges, interpupillary distance, ear shape and positioning, as well as distinctive markers like moles, dimples, and scars~\cite{Lakshmiprabha2011}

\textbf{2. Stylistic Artifacts}: This category refers to qualities of entire images or inconsistencies of those qualities within an image. This includes images of people that are waxy (Figure \ref{fig:artifacts}--2A), glossy (Figure \ref{fig:artifacts}-- 2C), shiny, and appear perfect like a model doing a photoshoot. These characteristics often appear in plastic--like skin and excessively soft hair. Additionally, this category includes noticeably cinematic, picturesque, and dramatic images that often appear in artistic photographs like Figure \ref{fig:artifacts}--2B. Stylistic artifacts also include inconsistencies between different subjects or parts of an image. This may appear as smudge-like distortions at the edges of different components or differences in resolution that make these parts look like they are cut out from different scenes.

\textbf{3. Functional Implausibilities}: Functional implausibilities result from a lack of understanding of the fundamental logic of real--world mechanical principles. This includes implausibilities in the objects themselves, their placement within the environment, and how the people in the image may be holding or using these objects, such as the woman holding a sandwich sideways in Figure \ref{fig:artifacts}--3B. Objects may also appear unable to function, like the loose strings of the guitar in Figure \ref{fig:artifacts}--3A, or placed in a way that they cannot function. Functional implausibilities also include distortion in fine details of the image. The image may present atypical designs in details like the print, buttons, and buckles on pieces of clothing, as seen in a backpack strap merging into a denim jacket in Figure \ref{fig:artifacts}--3C. Functional implausibilities also include errors in text, such as distorted or unconventional glyphs and odd spelling errors as seen in Figure \ref{fig:artifacts}--3D.

\textbf{4. Violations of Physics}: This category addresses inconsistencies in the image content that violate the expected logic of physical reality. Examples include shadows pointing in diverging directions, as shown in Figure \ref{fig:artifacts}--4A, or shadows that do not correspond to their light sources. Additionally, reflections on surfaces like water, mirrors, or shiny objects may appear misaligned with their surroundings, as illustrated in Figure \ref{fig:artifacts}--4B. Violations of physics also include depth and perspective issues, like warping and trajectories that do not align with the rest of the image. These distortions can also occur in real photographs, for example as seen with fish-eye lens distortions.

\textbf{5. Sociocultural Implausibilities}: This category includes scenarios that are socially inappropriate and unlikely to be seen in the real world, such as people wearing bathing suits at a funeral and a selfie with a bear. Violations of social and cultural norms could also be more subtle, present in details specific to certain cultures like Figure \ref{fig:artifacts}-- 5B and 5C attempting to depict Ukrainian and Japanese cultures, respectively. Historical inaccuracies and fake images of public figures in unlikely settings like 5A of Figure\ref{fig:artifacts} are also examples of sociocultural implausibilities.

\section{Accuracy in Distinguishing AI-generated Images from Real Photographs}\label{sec:results}

In the main phase of the experiment, we collected 539,749 responses on 599 images from 37,568 participants from February 5, 2024 to June 22, 2024. Sections~\ref{sec:acc-general} through \ref{sec:acc-model} focus on data from the main phase of the experiment. The second phase of the experiment started on June 22 and ended on August 30, with 83,577 responses on 482 images from 3,787 participants. Sections~\ref{sec:imagestimuli} and~\ref{sec:human-curation} describe the influence of human curation of the stimuli on how accurately participants identify the stimuli as AI-generated or real. 

The design of our experiment involves several important design choices. First, we selected the three models
of Midjourney, Firefly, and Stable Diffusion as the diffusion models. Second, we crafted prompts to produce realistic
outputs across various pose categories and content types. Third, we curated 450 images from over 3000 images generated
to use as image stimuli in the experiment. These images were selected to maximize realism while also representing
different visual artifacts and implausibilities. Inevitably, these design choices on models, prompts, and stimuli introduce some selection bias  into the experiment.

Additionally, we implemented two exclusion criteria that should be considered when interpreting our results. First, for all the analyses in Section ~\ref{sec:results}, we excluded observations where participants checked the box on the website ``I have seen this before''. These observations, which account for 2\% of the total observations, were excluded because of the strong possibility that participants who had previously seen the images were already aware of whether they were fake or real.
For these observations marked as having been seen before, 38\% of these observations were on AI-generated stimuli and 62\% were on real images. The image most frequently reported as 'seen before' is a real portrait of Martin Luther King Jr, which was one of the few real images of a well-known celebrity included in the experiment. 

Second, in line with our goals of studying detection ability on images for which there was some ambiguity, we excluded all images where participants' accuracy suggested very little ambiguity. We operationalized this as accuracy above 90\%. 

These exclusion criteria remove all observations on 68 fake images and 4 real images, which represent 14\% of observations from the entire experiment. 

In the human-coded analysis of artifacts discussed in Section~\ref{sec:acc-presence-artifacts}, we apply an additional exclusion criterion to make the coding tractable. Specifically, we exclude all images accurately identified in more than 80\% of observations. This exclusion criterion focuses the analysis on the most challenging images by excluding the most egregious distortions that lead to low photorealism (i.e., high participant accuracy).

\subsection{Overall Accuracy} \label{sec:acc-general}

\begin{figure}[H]
    \centering
    \includegraphics[width=\linewidth]{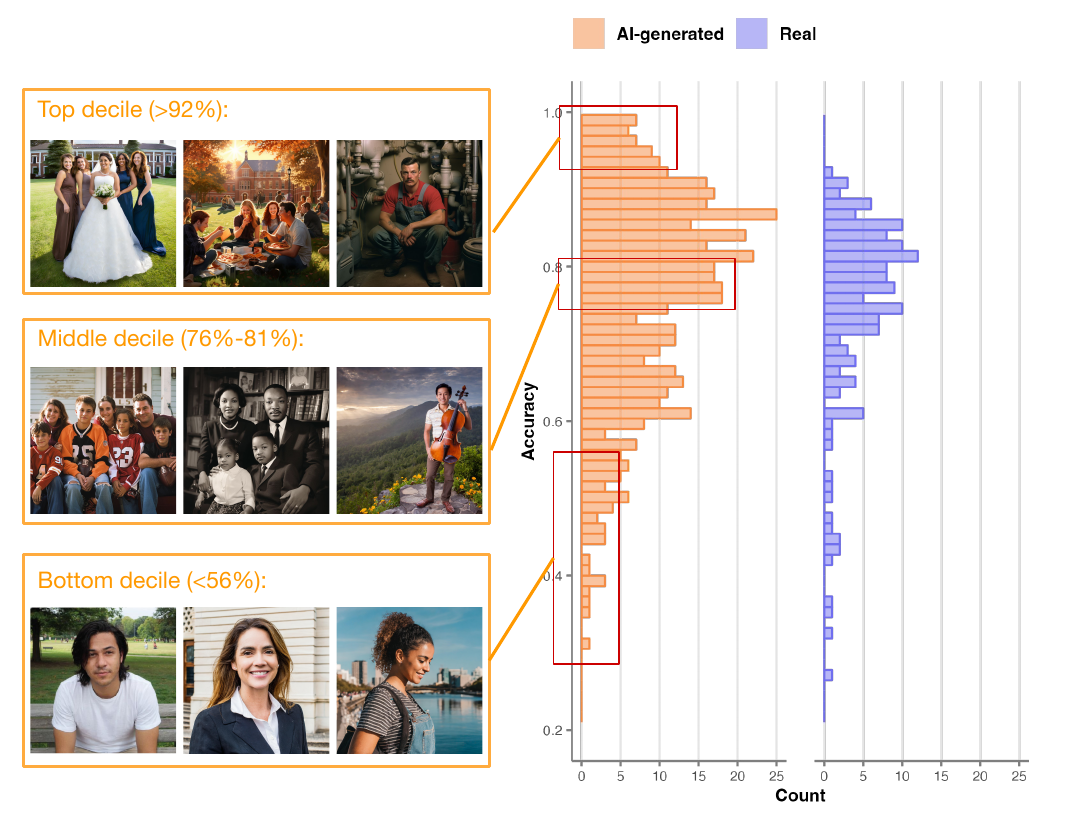}
    \caption{Distribution of accuracy scores for real and AI-generated images with example images representing different accuracy levels.}
    \label{fig:accuracy_real_fake}
    \Description{Histograms showing the distribution of accuracy scores for real and AI-generated images, accompanied by example images representing various accuracy levels.}
\end{figure}

In the main study, participants correctly identified AI-generated images and authentic photographs in 76\% and 74\% of observations, respectively. Accuracy varied substantially across images. Prior to implementing our accuracy-based exclusion described above, we found that for AI-generated images, accuracy ranged from 32\% to 99\%. Similarly, accuracy on real photographs ranged from 28\% to 92\%. Figure~\ref{fig:accuracy_real_fake} shows the distribution of accuracy in both AI-generated and real images with example images selected from the top, bottom, and middle deciles of each distribution. At the image level, the mean accuracy for identifying AI-generated and real images was 76\% (95\% CI:[74,77]) and 74\% (95\% CI:[72,76]), respectively. 

\begin{figure}[H]
\centering
\captionsetup{justification=raggedright, singlelinecheck=false, skip=2pt, font=small}
\begin{subfigure}[t]{0.3\linewidth}
    \subcaption{}\vtop{\vskip0pt\hbox{\includegraphics[width=\linewidth]{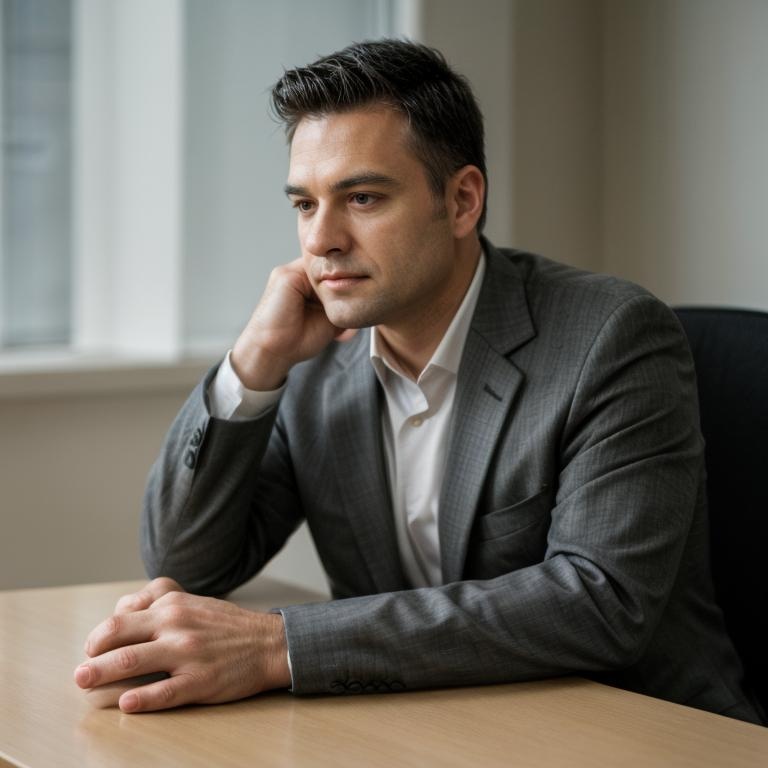}}}
\end{subfigure}
\hfill
\begin{subfigure}[t]{0.3\linewidth}
    \subcaption{}\vtop{\vskip0pt\hbox{\includegraphics[width=\linewidth]{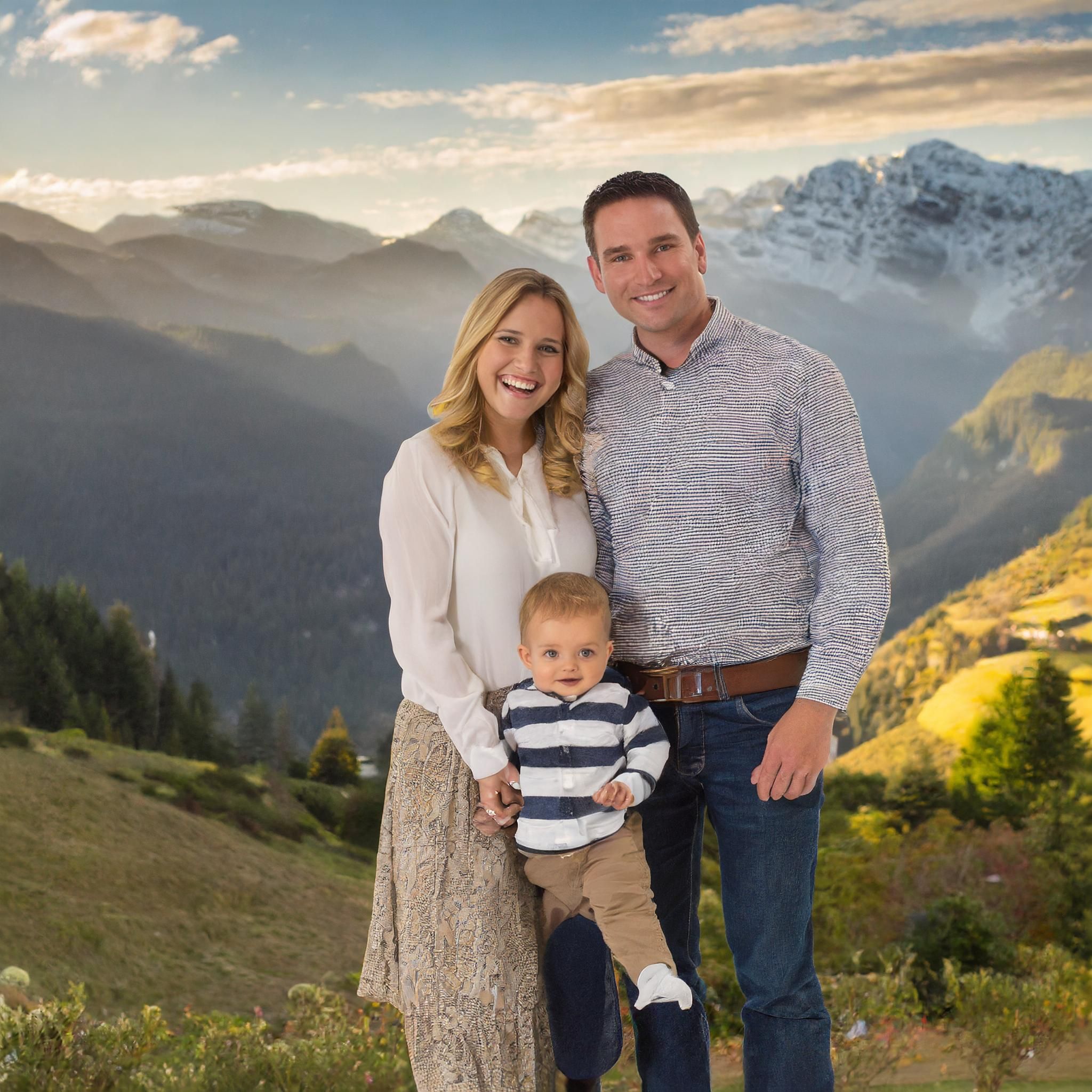}}}
\end{subfigure}
\hfill
\begin{subfigure}[t]{0.3\linewidth}
    \subcaption{}\vtop{\vskip0pt\hbox{\includegraphics[width=\linewidth]{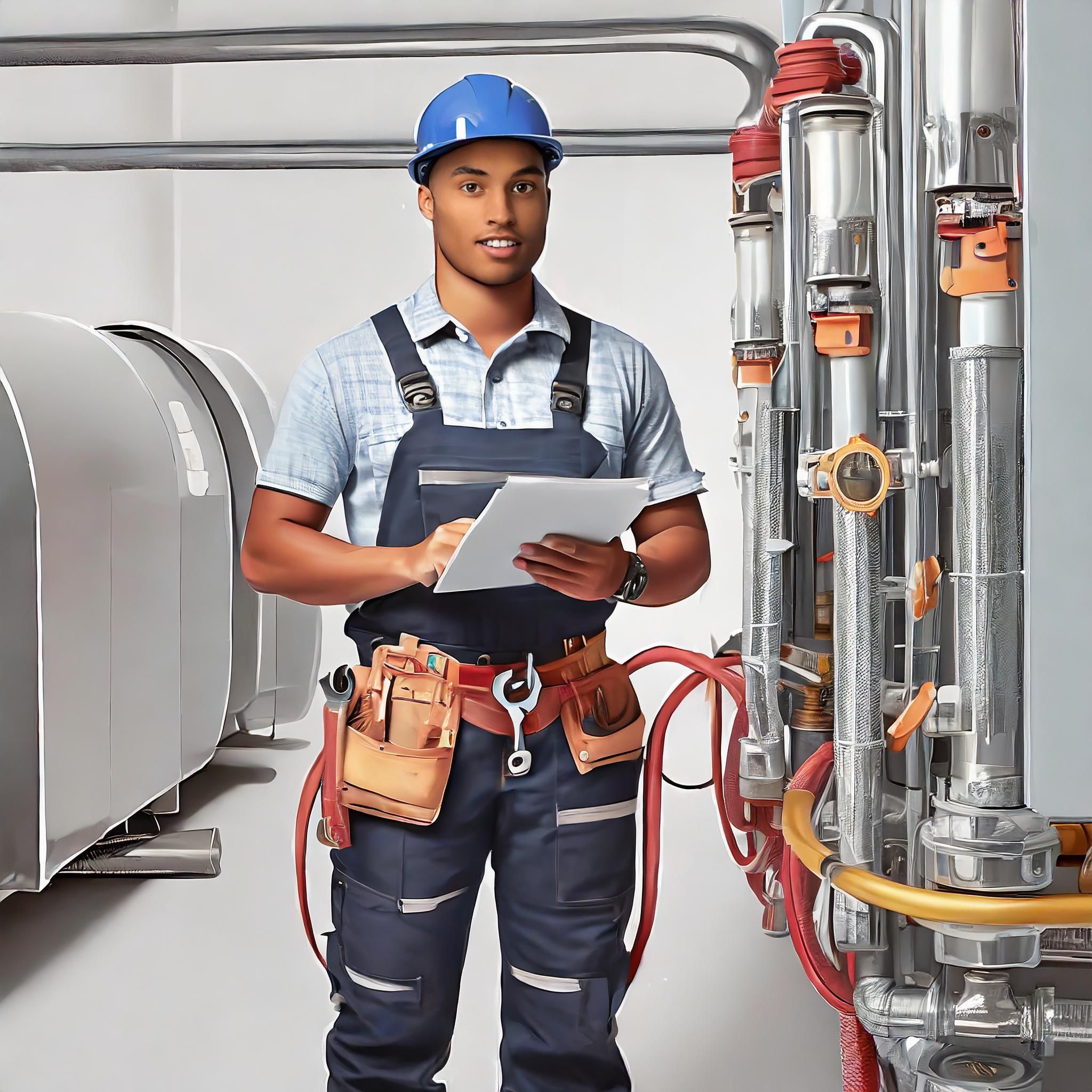}}}
\end{subfigure}
\caption{\mybold{Examples of obviously AI-generated images and their corresponding accuracy.} \normalfont{\textbf{A.} AI-generated portrait with 92\% accuracy. \textbf{B.} AI-generated posed group image with 95\% accuracy. \textbf{C.} AI-generated full-body image with 99\% accuracy.}}
\label{fig:three-fake-images}
\Description{Three examples of obviously AI-generated images with corresponding accuracy scores: A. Portrait with 92\% accuracy, B. Posed group image with 95\% accuracy, C. Full-body image with 99\% accuracy. }
\end{figure}

Despite our efforts to minimize obvious artifacts, some images - particularly non-portraits - were challenging to generate without noticeable artifacts. As a result, participants achieved nearly 100\% accuracy on a few AI-generated images with obvious features. We present examples of these images in Figure~\ref{fig:three-fake-images}.  
In contrast to AI-generated images, real photographs rarely contain definitive artifacts and visual cues often seen in AI-generated images, which limits participants from achieving near-perfect accuracy on real photographs.

\subsection{Participant Level Accuracy}\label{sec:indiv-acc}

\begin{figure*}[h]
    \centering
    \captionsetup{justification=raggedright, singlelinecheck=false, skip=2pt, font=small}

    \begin{subfigure}[t]{0.4\textwidth}  
        \centering
        \subcaption[]{}  
        \vspace{-3pt}  
        \includegraphics[width=\linewidth]{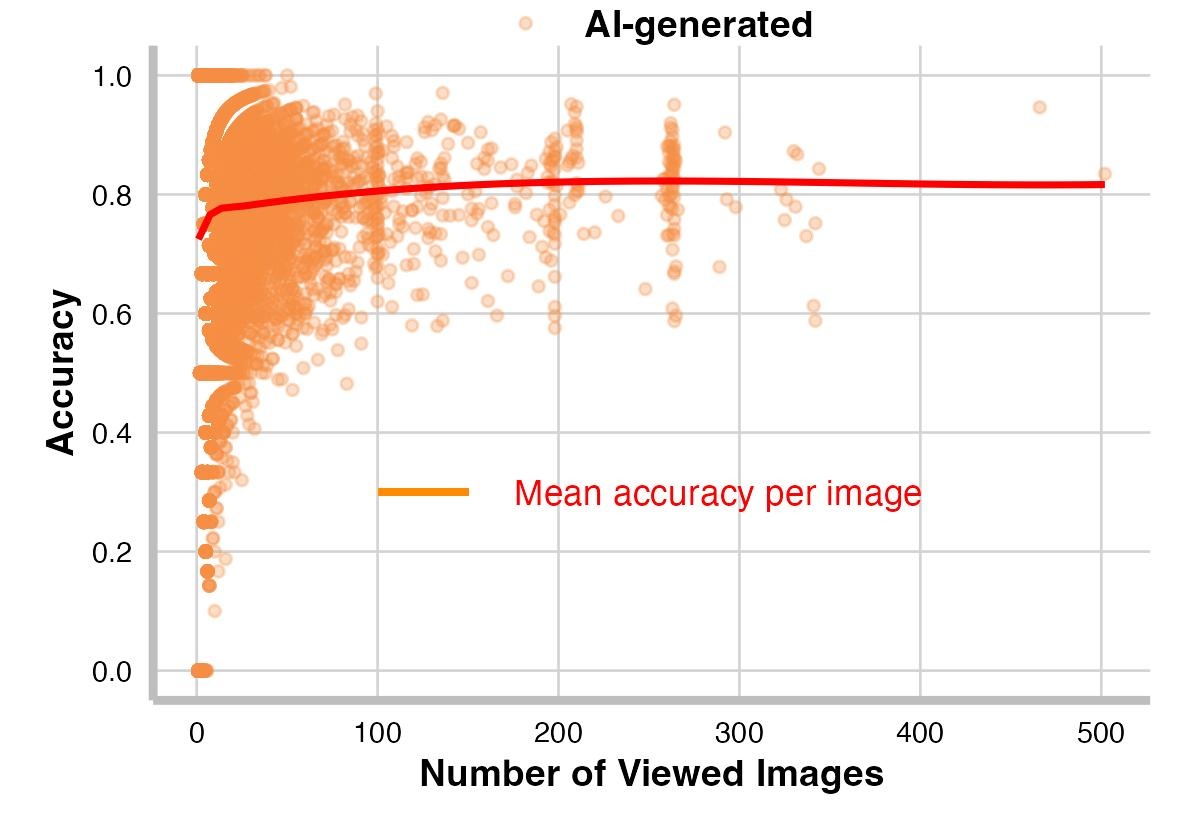} 
    \end{subfigure}
    \hspace{0.05\textwidth} 
    \begin{subfigure}[t]{0.38\textwidth}  
        \centering
        \subcaption[]{}  
        \vspace{-3pt}
        \includegraphics[width=\linewidth]{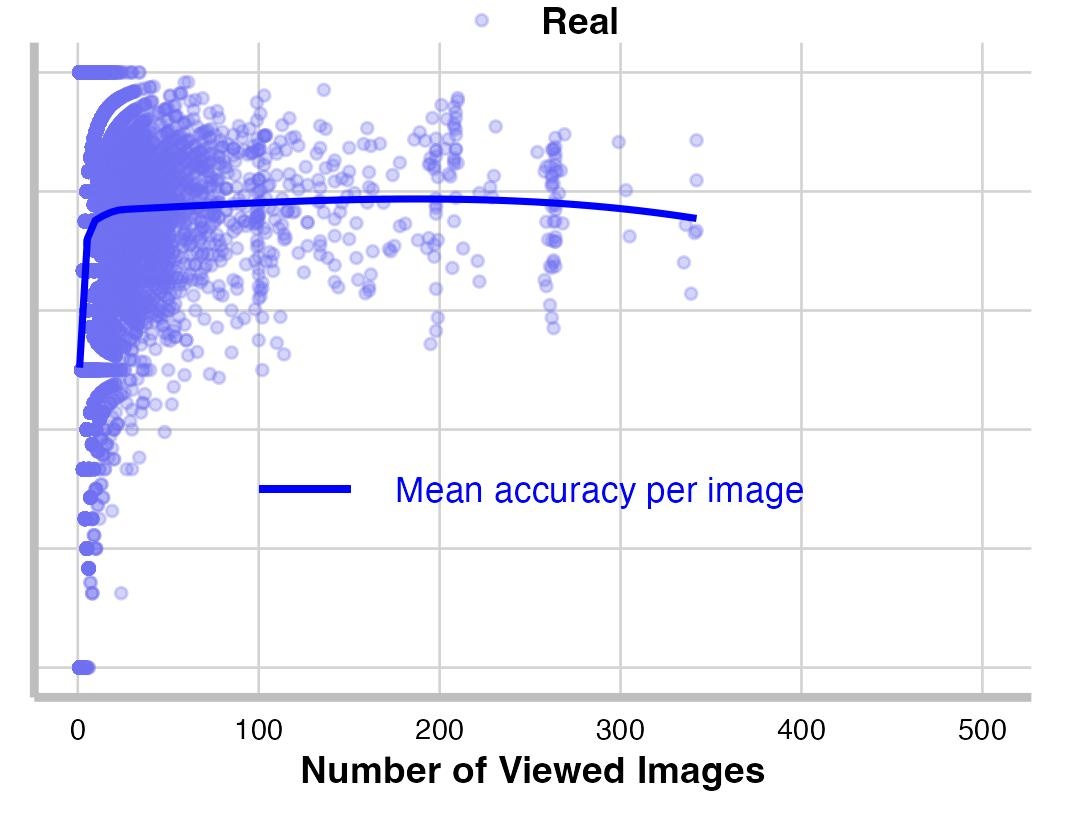}
    \end{subfigure}

    \begin{subfigure}[t]{0.36\textwidth}  
        \centering
        \subcaption[]{}  
        \vspace{-3pt}
        \includegraphics[width=\linewidth]{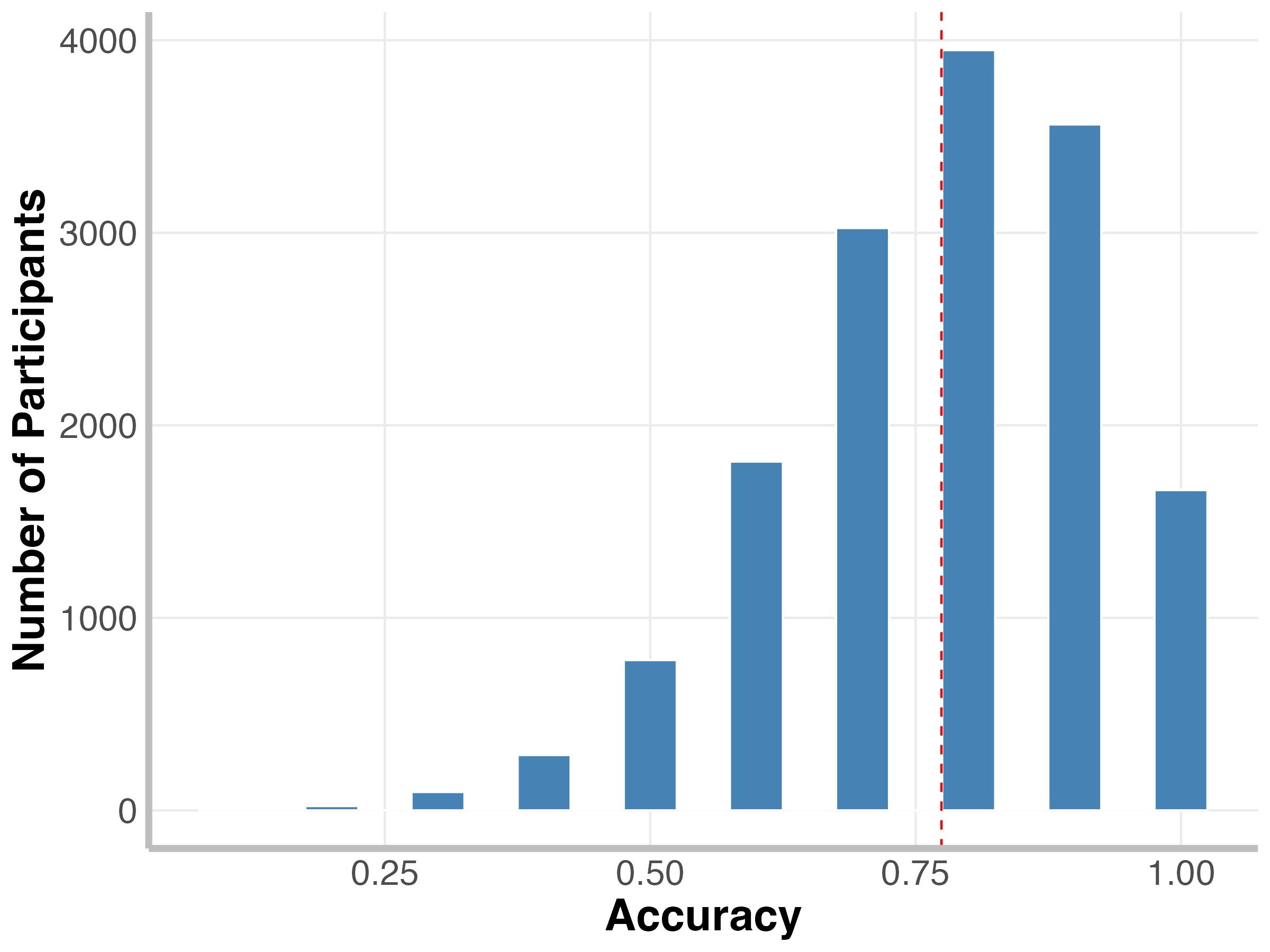}
    \end{subfigure}
    \hspace{0.05\textwidth}  
    \begin{subfigure}[t]{0.36\textwidth}  
        \centering
        \subcaption[]{}  
        \vspace{-3pt}
        \includegraphics[width=\linewidth]{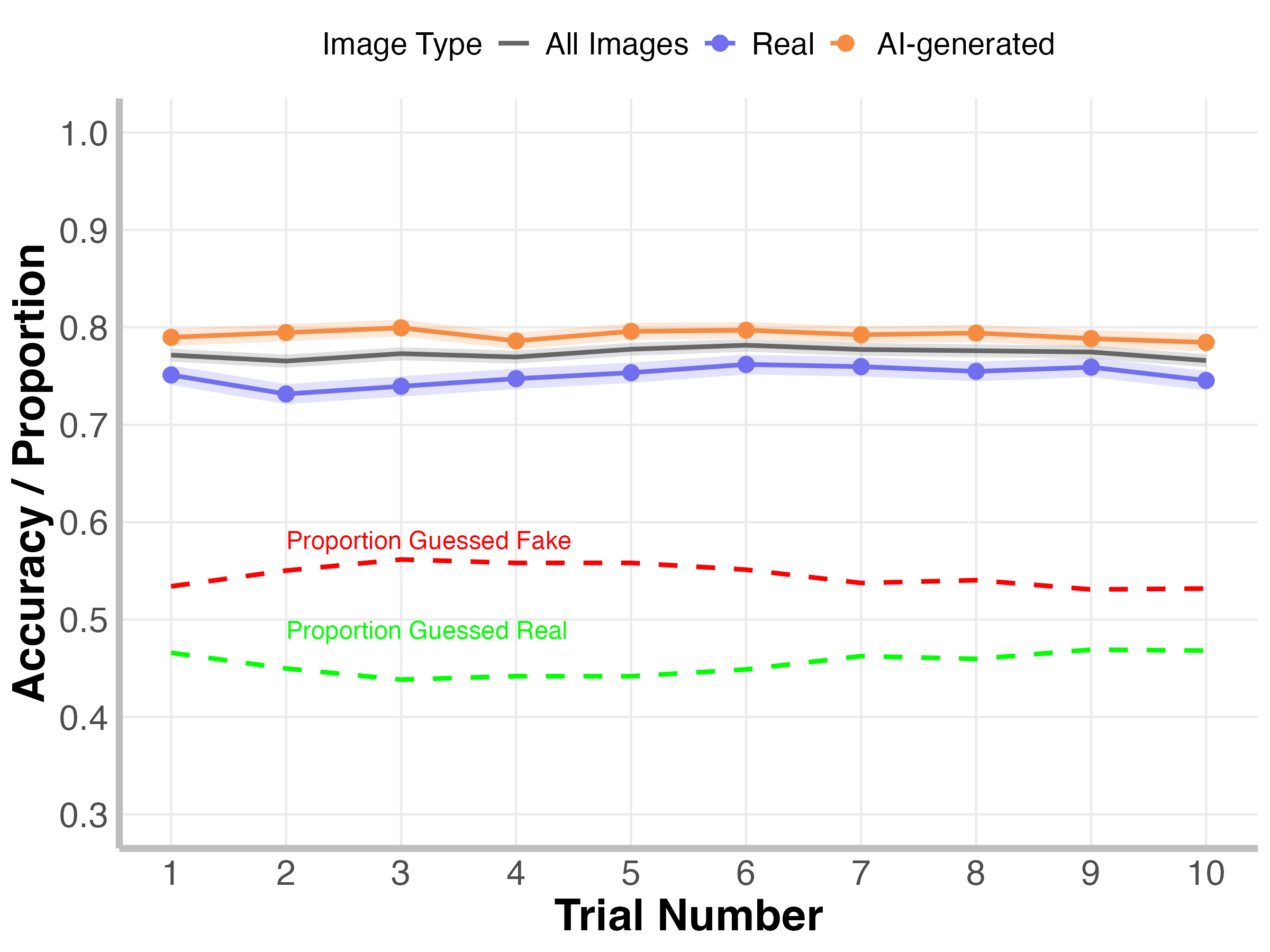}
    \end{subfigure}

    \caption{\textbf{Participant-level accuracy and learning trends.}  
    \normalfont{ \textbf{A.} Scatterplot of participant-level accuracy for AI-generated images. \textbf{B.} Scatterplot of participant-level accuracy for real images. \textbf{C.} Histogram showing the distribution of accuracy across the first ten images seen by participants who viewed at least 10 images. \textbf{D.} Learning curve illustrating accuracy trends and classification biases when detecting AI-generated and real images.}}
    
    \label{fig:combined-participant-accuracy}

    \Description{A composite figure showing participant accuracy trends:  
    A. Scatterplot displaying accuracy levels for detecting AI-generated images, with points representing individual participants.  
    B. Scatterplot for real images, structured similarly to A.  
    C. Histogram showing the distribution of accuracy for participants' first 10 images.  
    D. Learning curve tracking accuracy trends over time, highlighting biases in AI-generated and real image classification.}
\end{figure*}

Given the organic nature of participants' engagement with this experiment, we did not impose restrictions on the number of images a participant saw. Most participants in this study provided responses to at least seven images, but some participants only provided a single response, and one participant provided 502 responses. 

The vast majority of participants (75\%) saw 16 or fewer images. Figure~\ref{fig:combined-participant-accuracy}A and B present the distribution of participant--level accuracy by number of viewed images. 

In order to compare participant performance and avoid issues that arise with differential attrition, we focus on the first ten images seen by participants who saw at least 10 images, which includes 152,050 observations from 15,205 participants. First, we note that 34\% of these participants achieved 90\% accuracy or higher on the first ten images seen. If the AI-generated images were perfectly photorealistic such that the human ability to distinguish is no higher than random guessing, then we would have expected only 1\% of participants to achieve this threshold of accuracy (assuming random guessing at 50\% accuracy, with participants evaluating 10 images each, achieving at least 9 out of 10 correct responses would occur with a probability of approximately 1.07\%, based on the binomial probability distribution). Figure~\ref{fig:combined-participant-accuracy}C shows the distribution of accuracy across the first ten images seen by participants who saw at least 10 images.

In Figure~\ref{fig:combined-participant-accuracy}D, we present accuracy rates by the number of images seen. We find that on average, participants begin the experiment by disproportionally identifying images as fake in 63\% of observations. Notably, this bias is reduced after only a few images.

\subsection{Accuracy by Scene Complexity} \label{sec:acc-scene-complexity}

\begin{figure}[h]
    \captionsetup{justification=raggedright, singlelinecheck=false, skip=2pt, font=small}
    \centering

    \begin{subfigure}[t]{\linewidth}
        \subcaption{}
        \vspace{-12pt}  
        \includegraphics[width=\linewidth]{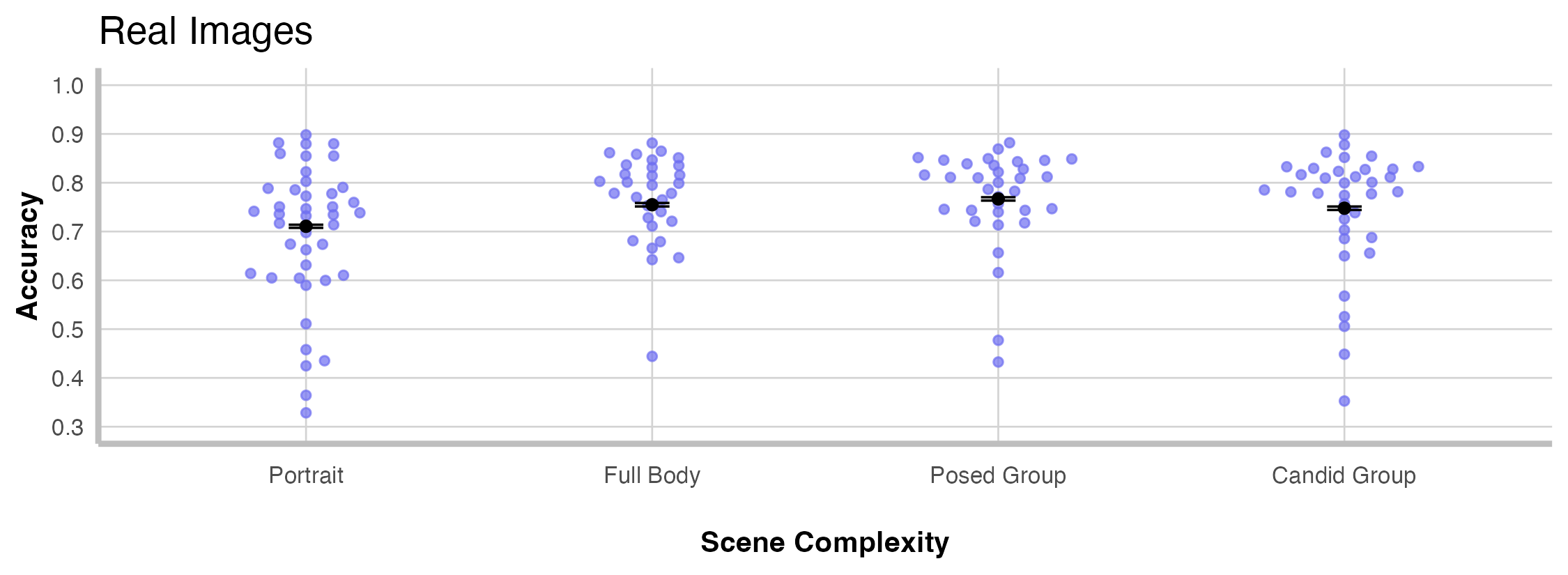}
    \end{subfigure}
    
    \begin{subfigure}[t]{\linewidth}
        \subcaption{}
        \vspace{-12pt}  
        \includegraphics[width=\linewidth]{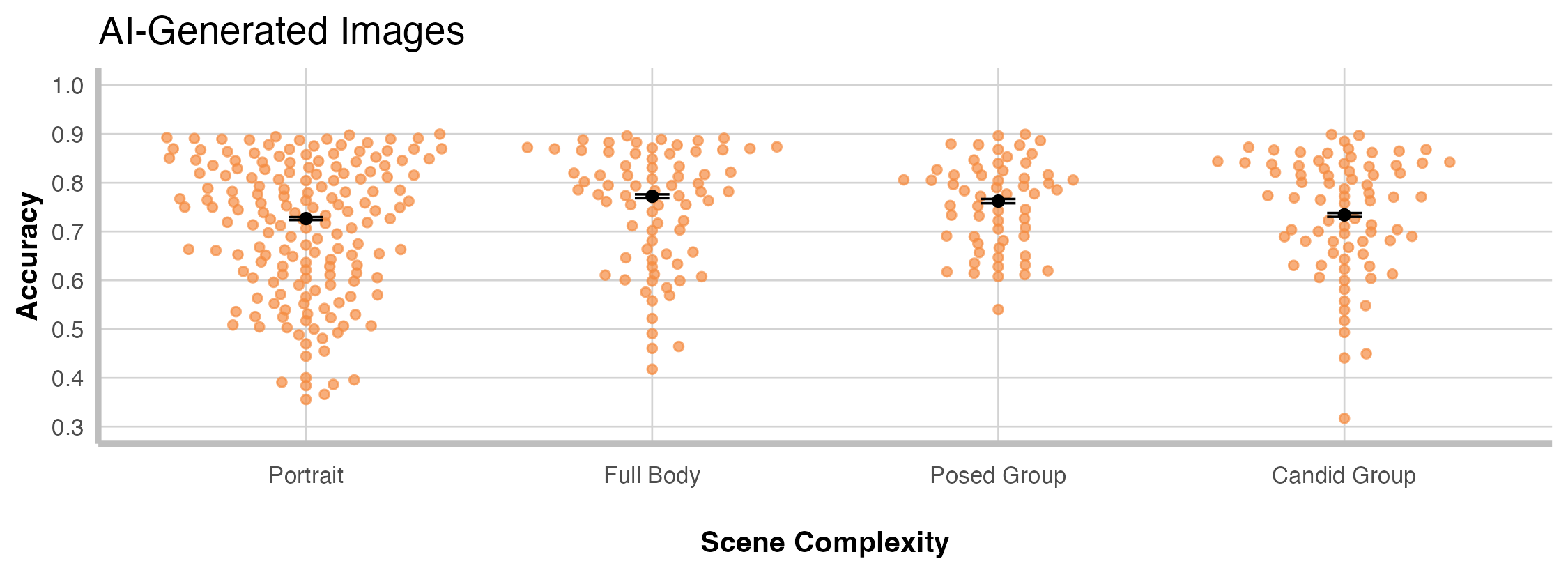}
    \end{subfigure}
    
    \begin{subfigure}[t]{\linewidth}
    \centering
        \subcaption{}
        \vspace{-12pt}  
        \includegraphics[width=0.9\linewidth]{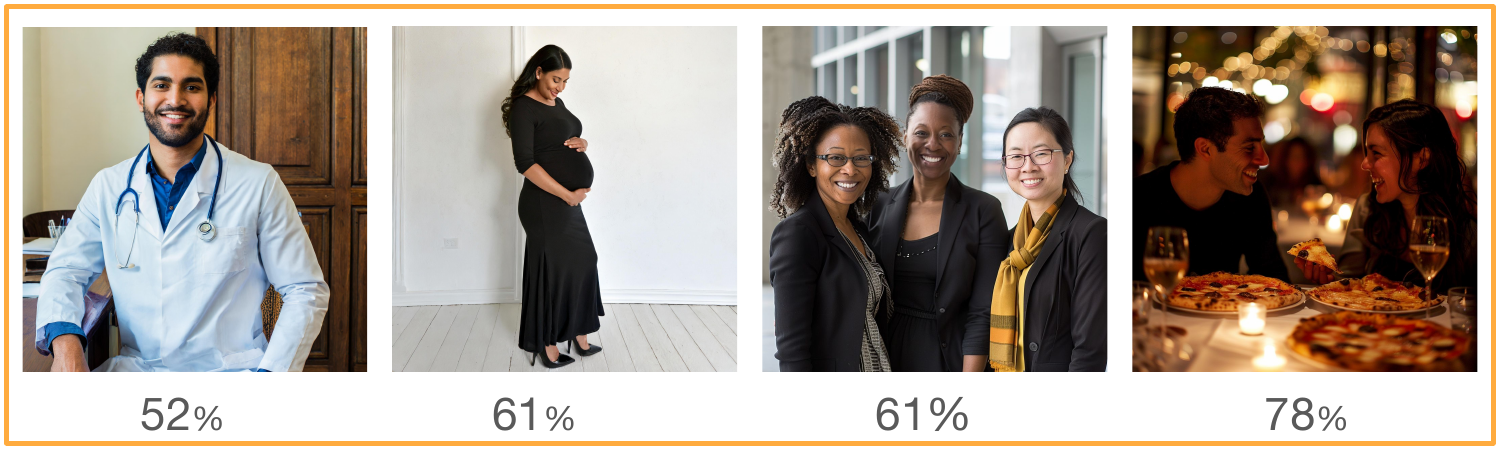}
    \end{subfigure}
    
    \caption{\textbf{Scene complexity}: Accuracy of real and AI-generated images by scene complexity levels. 
    \normalfont{Beeswarm plots of image-level accuracy for each dimension of scene complexity with bootstrapped 95\% confidence intervals. We exclude images identified with above 90\% accuracy in this analysis. \textbf{A.} Real images \textbf{B.} AI-generated images \textbf{C.} AI-generated images across scene complexities.}}
    
    \label{fig:pose-complexity}
    
    \Description{Beeswarm plots showing the accuracy of real and AI-generated images by pose complexity. Each dot represents an individual image, with error bars indicating the bootstrapped 95\% confidence interval around the mean.}
\end{figure}

We find that on average, participants' accuracy increases as scene complexity increases. For example, we find that 16\% of portraits appear in the bottom decile of accuracy scores (representing the highest level of photorealism), whereas only 3\% of AI-generated posed group images appear in the bottom decile. Figure~\ref{fig:pose-complexity} presents the distribution of accuracy for each category, separately for real and AI-generated images. For AI-generated images, the mean accuracy was 72.7\% (95\% CI: [72.4, 72.9]) for portraits, 77.2\% (95\% CI: [76.8, 78.6]) for full body, 76.2\% (95\% CI: [75.8, 76.7]) for posed groups, and 73.4\% (95\% CI: [73, 73.8]) for candid groups. For real images, the accuracy was 71.1\% (95\% CI: [70, 71.4]) for portraits, 75.5\% (95\% CI: [75.1, 75.8]) for full body, 76.7\% (95\% CI: [76.3, 77]) for posed groups, and 74.8\% (95\% CI: [74.4, 75.1]) for candid groups. 

As exemplified in Figure~\ref{fig:pose-complexity}C, we note that portraits, relative to the other levels of scene complexity, typically have less detail, simpler and more standardized poses, more blurred backgrounds, and fewer available cues than full-body or group images. 

\subsection{Accuracy by Presence of Artifacts}\label{sec:acc-presence-artifacts}
\begin{figure*}[h]
\centering
\captionsetup{justification=raggedright, singlelinecheck=false, skip=2pt, font=small}

\includegraphics[width=\linewidth]{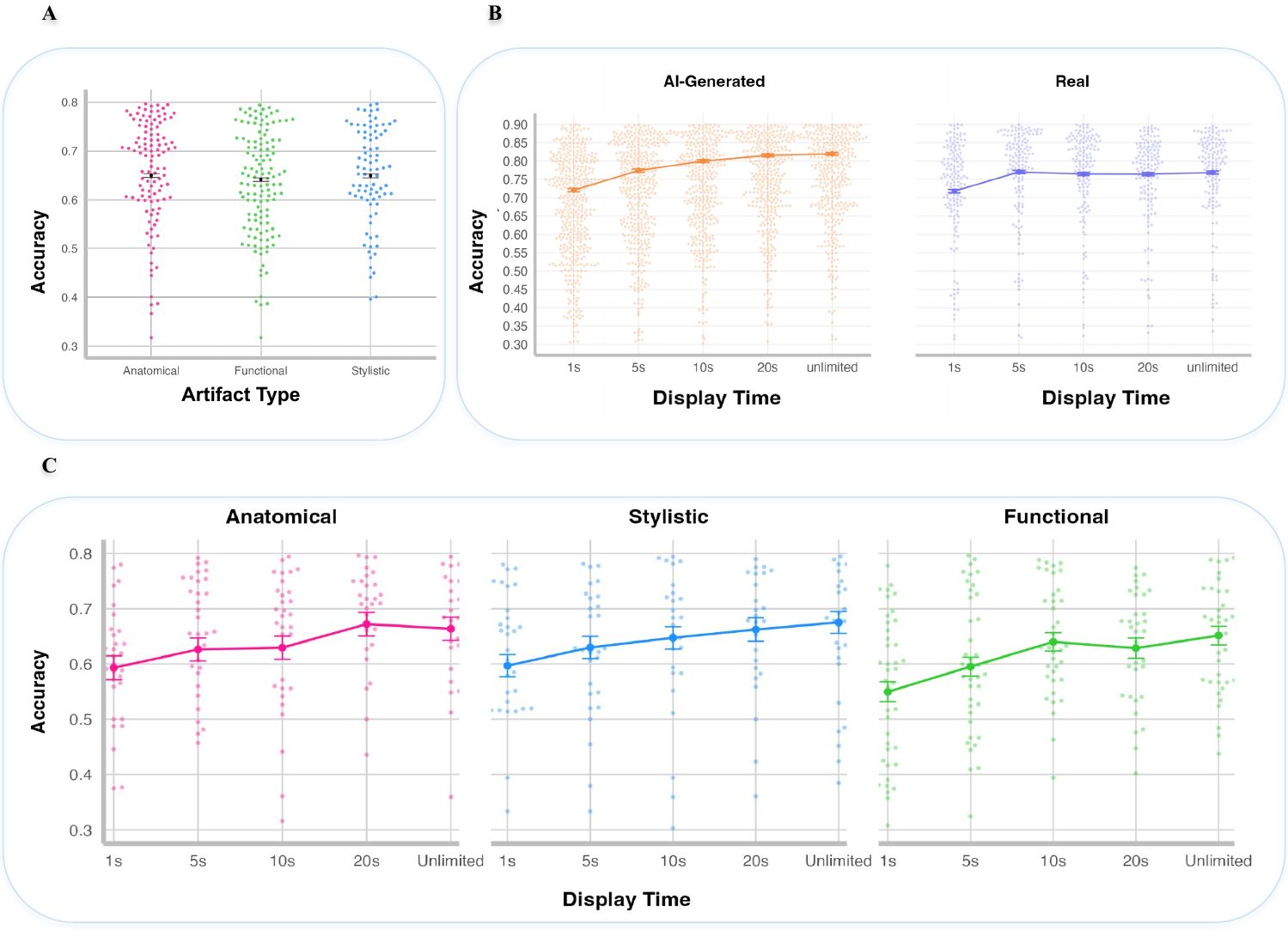}

\caption{\textbf{Accuracy by artifact types and display times} \normalfont{\textbf{A. Mean accuracy for different artifact types.} Distribution of accuracy scores by artifact type
for images with at least one artifact. \textbf{B. Mean accuracy over display time.} Change in mean accuracy across different display time assignments (1 second, 5 seconds, 10 seconds, 20 seconds, and unlimited) with 95\% confidence intervals and bee swarm plots of image accuracy for AI-generated and real images. \textbf{C. Mean accuracy over time for different artifact types.} Change in mean accuracy across different time assignments (1 second, 5 seconds, 10 seconds, 20 seconds, and unlimited) with 95\% confidence intervals and bee swarm plots of image accuracy for
images with anatomical (pink), functional (green), and stylistic
(blue) artifacts. The x–axis shows the display time intervals, and the
y–axis shows accuracy.}}

\label{fig:combined-artifact-trends}

\Description{A composite figure showing accuracy-related analyses:  
(A) A beeswarm plot displaying the distribution of accuracy scores for images containing at least one artifact, categorized by artifact type.  
(B) A line plot illustrating mean accuracy across different display time conditions (1 second, 5 seconds, 10 seconds, 20 seconds, and unlimited), with 95\% confidence intervals. Overlaid bee swarm plots represent individual accuracy scores for AI-generated and real images.  
(C) A line plot showing mean accuracy over time for different artifact types (Anatomical, Functional, and Stylistic). Each artifact type is color-coded (pink for Anatomical, green for Functional, and blue for Stylistic). Bee swarm plots depict individual accuracy scores for images within each artifact category. The x-axis represents display time intervals, and the y-axis represents accuracy.}

\end{figure*}

\begin{figure*}[h!]
\centering
\captionsetup{justification=raggedright, singlelinecheck=false, skip=2pt, font=small}

\begin{subfigure}[t]{0.27\linewidth}
\centering
    \subcaption{}
    \includegraphics[width=\linewidth]{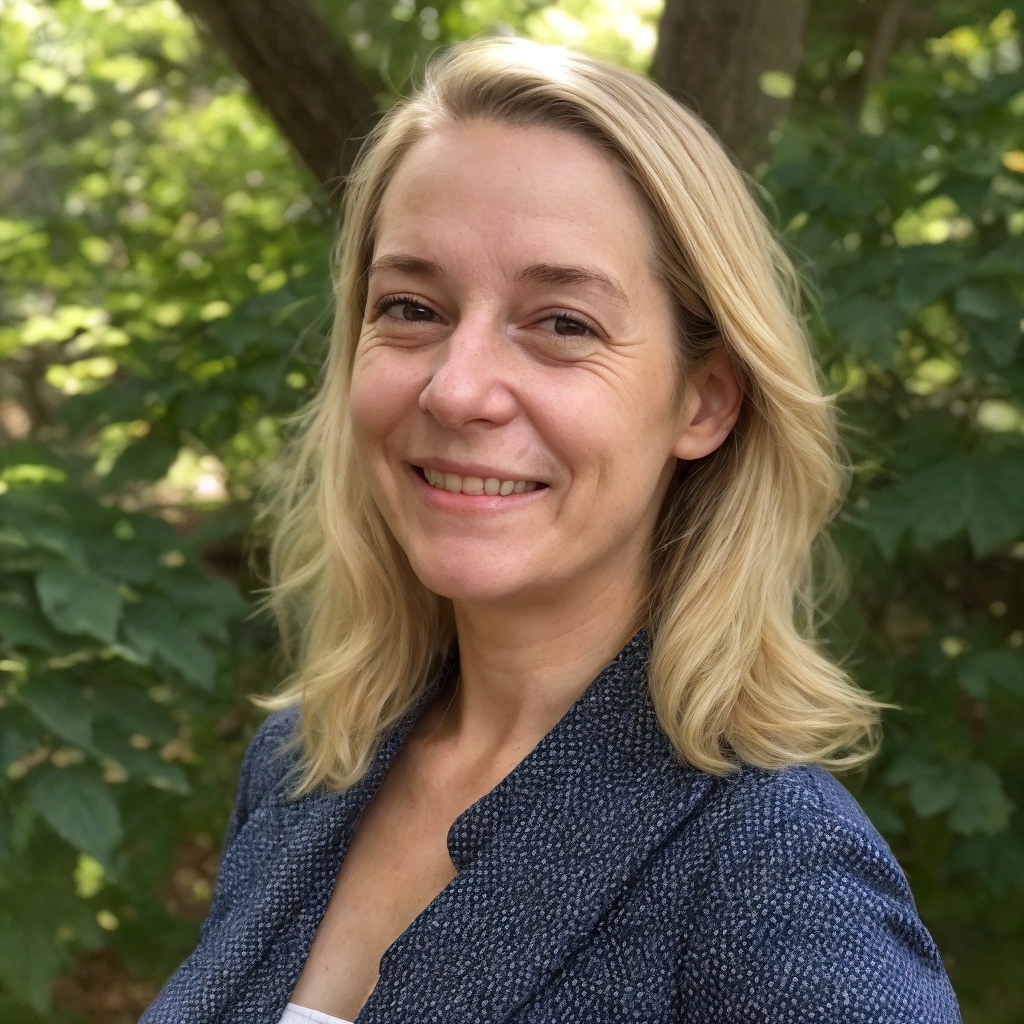}
\end{subfigure}
\hspace{1cm}
\begin{subfigure}[t]{0.27\linewidth}
\centering
    \subcaption{}
    \includegraphics[width=\linewidth]{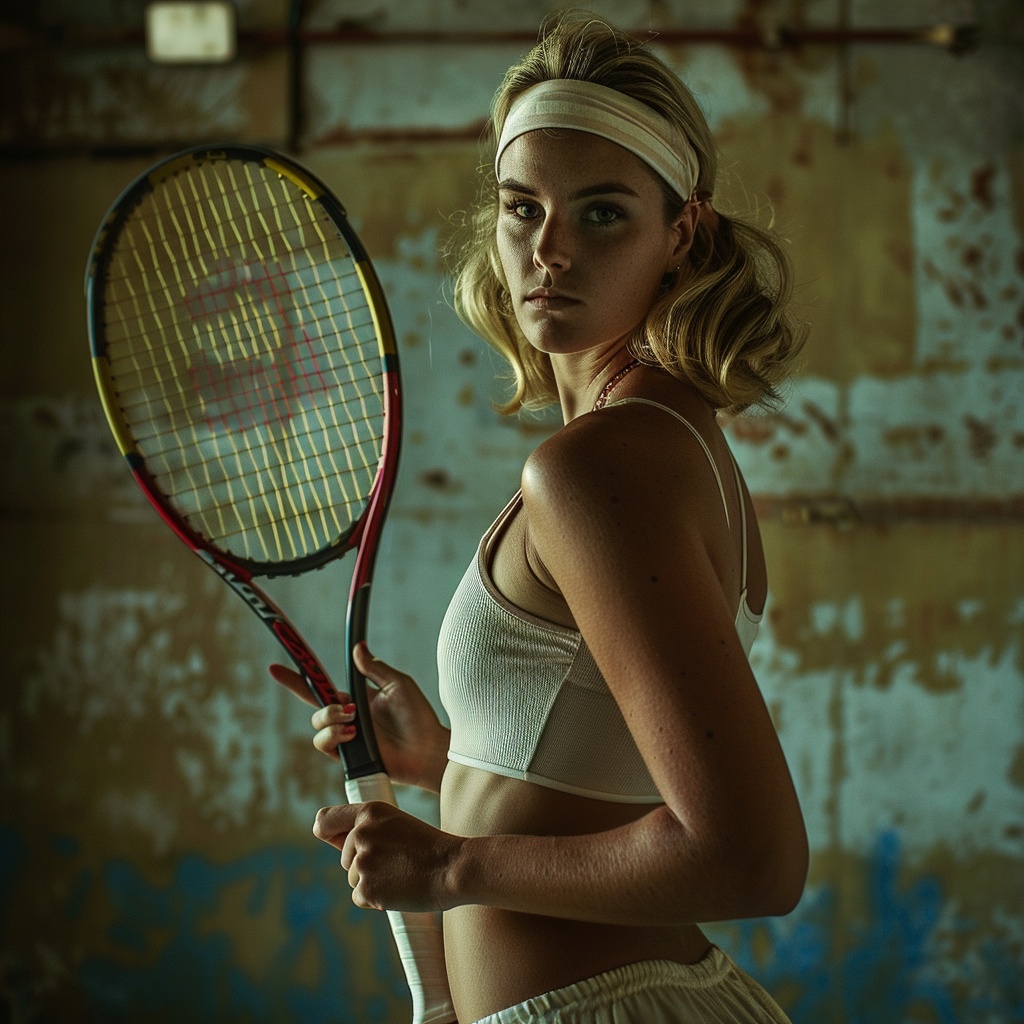}
\end{subfigure}
\hspace{1cm}
\begin{subfigure}[t]{0.27\linewidth}
\centering
    \subcaption{}
    \includegraphics[width=\linewidth]{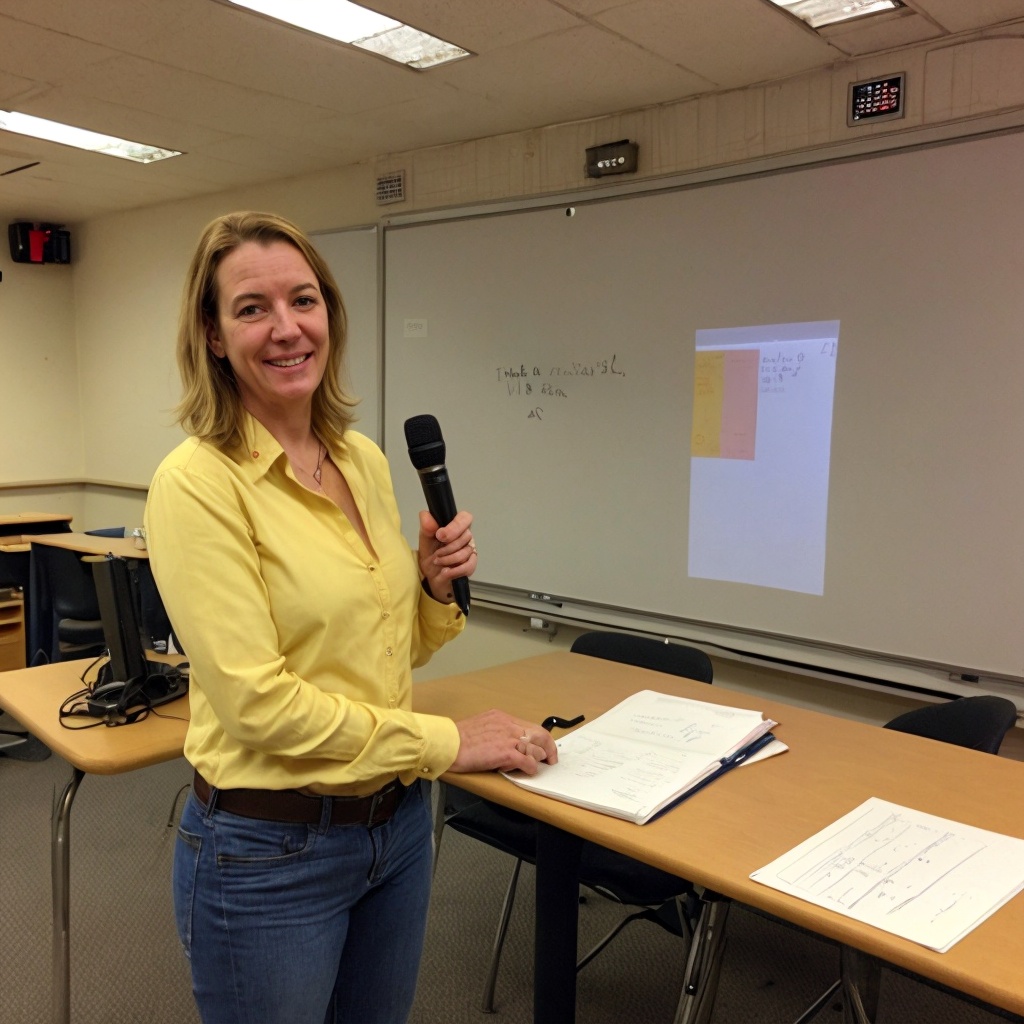}
\end{subfigure}

\vskip 5mm 

\begin{subfigure}[t]{0.27\linewidth}
\centering
    \subcaption{}
    \includegraphics[width=\linewidth]{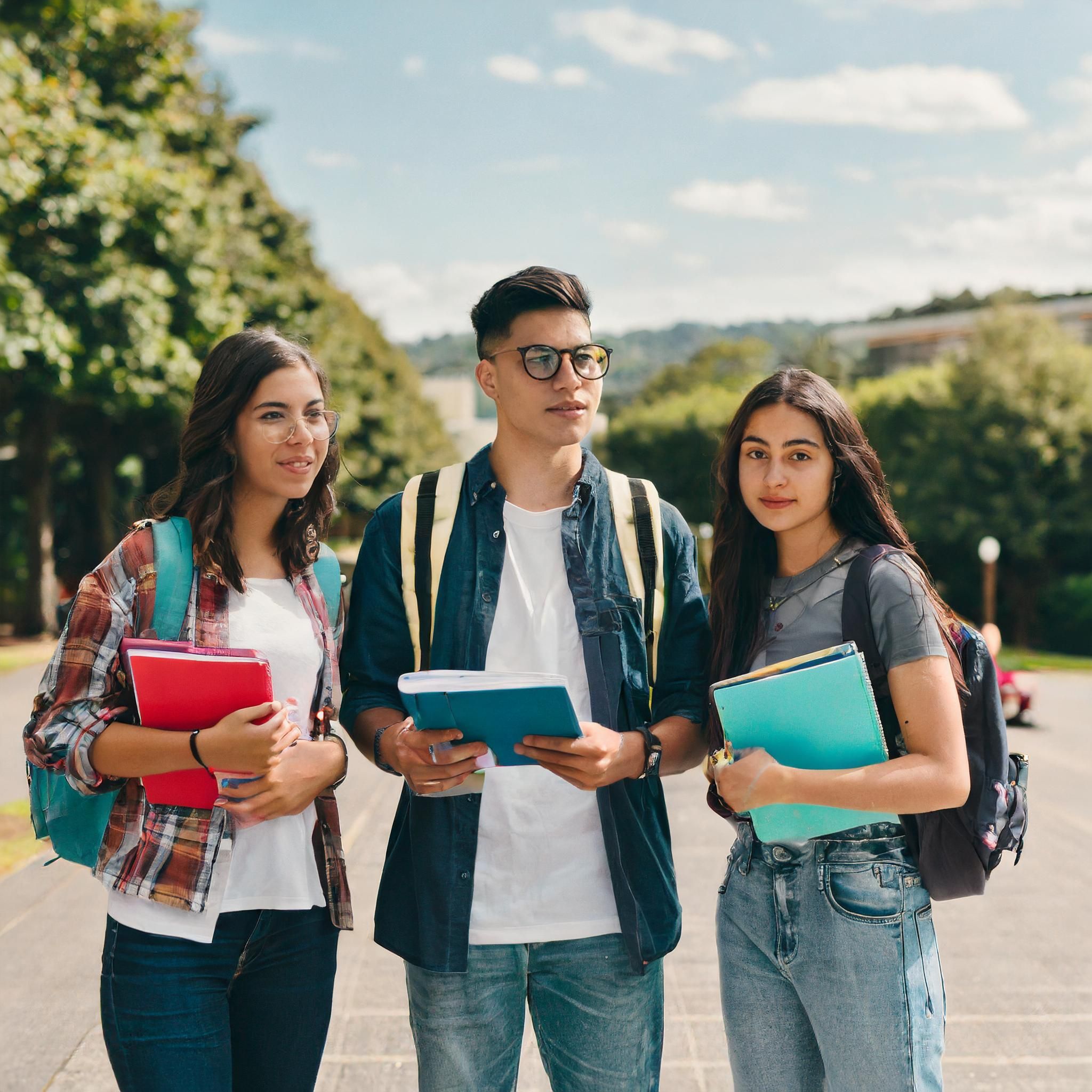}
\end{subfigure}
\hspace{1cm}
\begin{subfigure}[t]{0.27\linewidth}
\centering
    \subcaption{}
    \includegraphics[width=\linewidth]{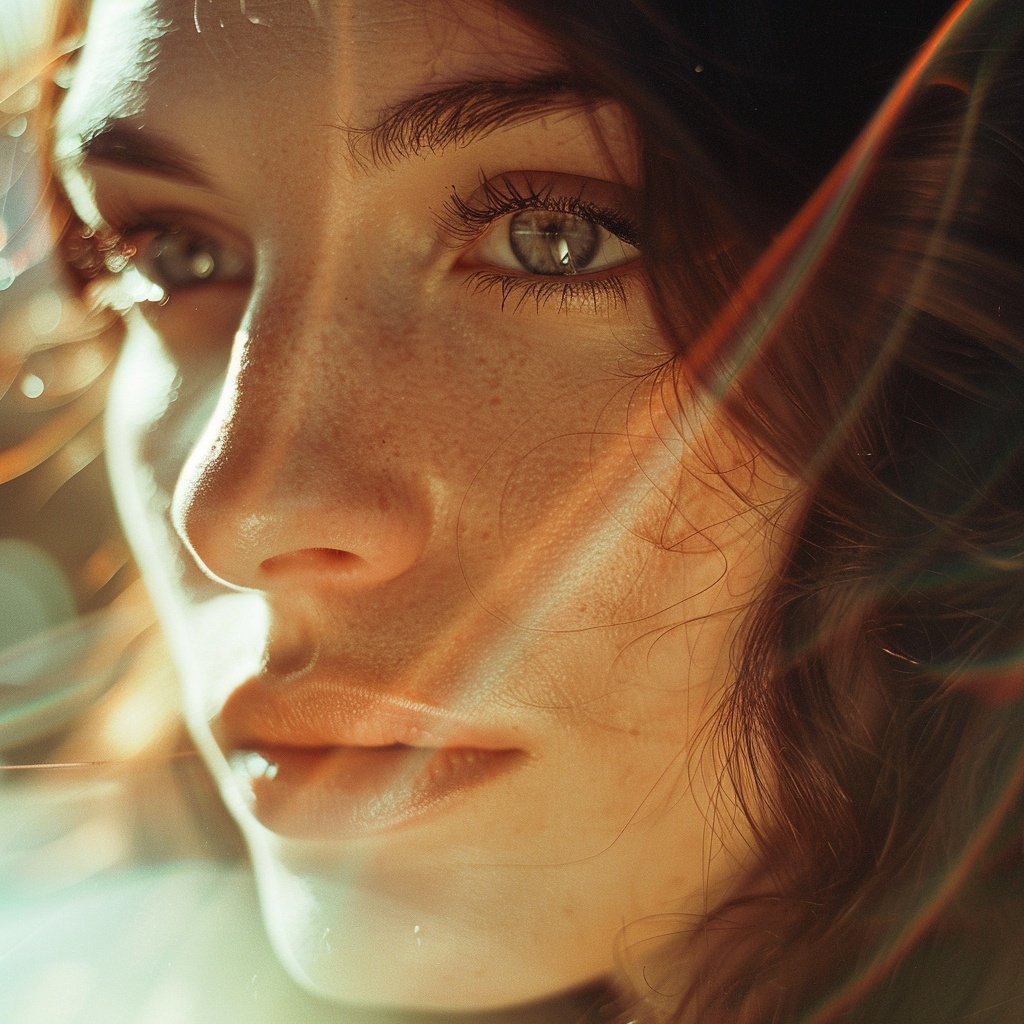}
\end{subfigure}
\hspace{1cm}
\begin{subfigure}[t]{0.27\linewidth}
\centering
    \subcaption{}
    \includegraphics[width=\linewidth]{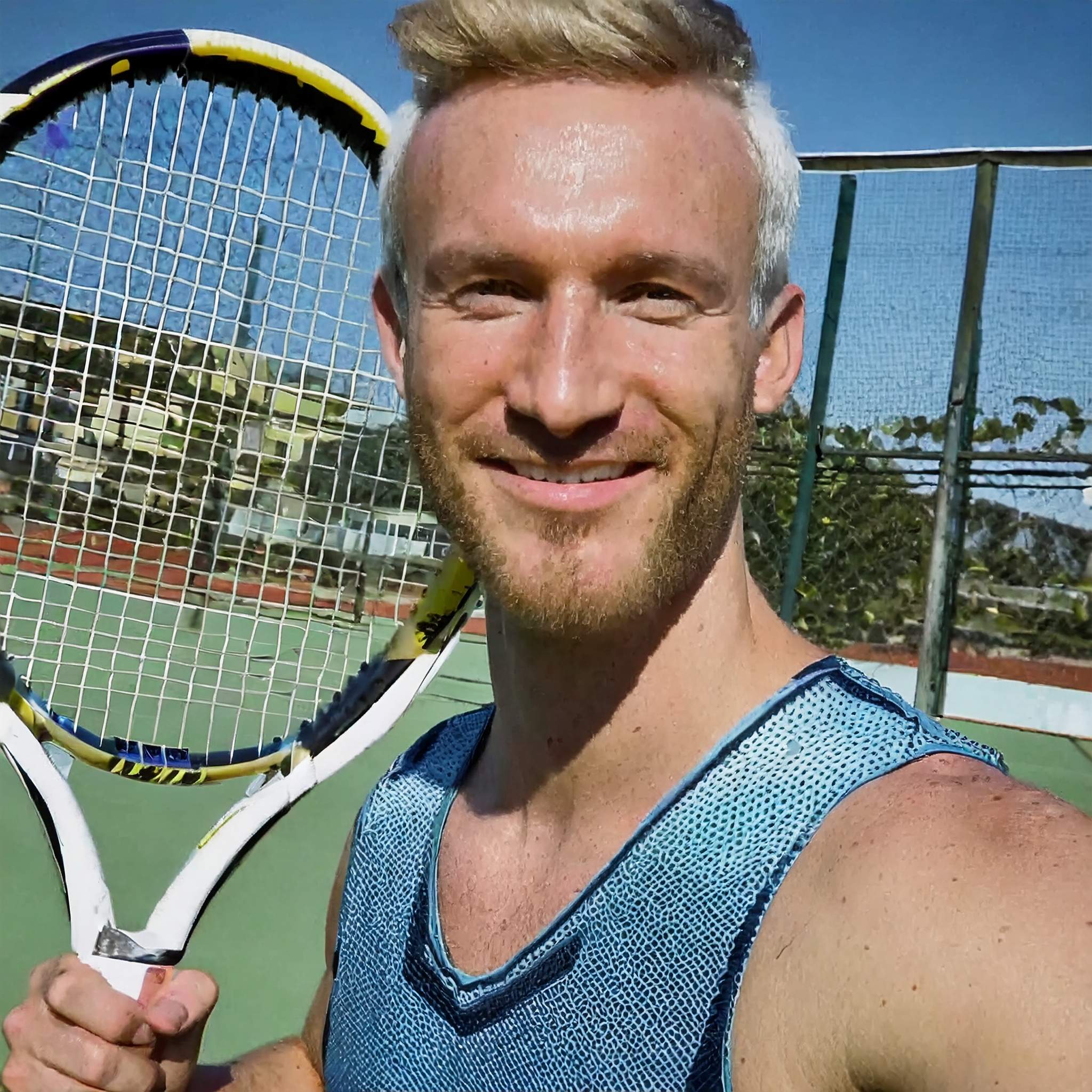}
\end{subfigure}

\caption{\textbf{Examples of images with varying artifact visibility.}  
\normalfont{\textbf{Top row (A--C):} Example images showcasing stylistic and functional artifacts with varying visibility.  
\textbf{A.} A subtle stylistic artifact in the soft and wispy textures of the woman's hair and a minor functional implausibility in the atypical design of her shirt collar (Accuracy: 47\%). \textbf{B.} An obvious stylistic artifact due to the overall cinematization of the image (Accuracy: 73\%). \textbf{C.} A combination of multiple artifacts, including anatomical implausibilities in the woman's hand, functional implausibilities in the table shape and wall panels, and a stylistic artifact in the soft texture of the woman's face (Accuracy: 38\%). \textbf{Bottom row (D--F):} Images with anatomical, stylistic, and functional artifacts of varying visibility. \textbf{D.} Anatomical implausibilities in the fingers of the three students (Accuracy: 84\%). \textbf{E.} A stylistic artifact in the cinematized look and plastic-like texture of the woman's skin (Accuracy: 62\%). \textbf{F.} No obvious anatomical or stylistic artifacts, but closer inspection reveals functional implausibilities: the tennis racket is asymmetrical, its strings are not taut, and the shirt has irregularly shaped designs with glitch-like inconsistencies (Accuracy: 54\%).}}

\label{fig:combined-varying-artifacts-visibility}

\Description{A composite figure showing six images with varying visibility of AI-generated artifacts.  
(A--C) The first row highlights stylistic and functional artifacts, including wispy hair, cinematized lighting, and a distorted table.  
(D--F) The second row focuses on anatomical, stylistic, and functional artifacts, including distorted fingers, plastic-like textures, and inconsistencies in objects like a tennis racket.}
\end{figure*}

In order to analyze accuracy by artifact type, we annotated images with diffusion model artifact categories from the taxonomy based on a three-step process. First, four co-authors independently annotated all 218 images with accuracy below 80\%, identifying artifacts and providing detailed explanations for their annotations. Second, each of these annotations was reviewed and edited by two additional co-authors. Third, a fifth co-author reviewed all annotations for consistency. Figures~\ref{fig:combined-varying-artifacts-visibility}A--C and \ref{fig:combined-varying-artifacts-visibility}D--F provide examples of how we annotated images, displaying the identified artifact categories, the reasoning behind their identification, and the associated detection accuracy for each image. During this process, we observed that the three main artifact types---anatomical implausibilities, stylistic artifacts, and functional artifacts---each appeared in nearly a third of the images we annotated. In contrast, violations of physics and sociocultural implausibilities were less common, appearing in only 20 and 12 images, respectively. In light of this distribution of artifacts, Figure~\ref{fig:combined-artifact-trends}A presents the distribution of accuracy scores across images containing at least the three listed artifact types.

Based on our annotations of artifacts in images, we find participants are less accurate on images with functional implausibilities than images with anatomical implausibilities or stylistic artifacts. The mean accuracy on images with at least one functional implausibility, one anatomical implausibility, and one stylistic artifact is 64.1\% (95\% CI: [63.8, 64.5]), 65\% (95\% CI: [64.6, 65.4]), and 64.9\% (95\% CI: [64.5, 65.3]), respectively. While the accuracy on images with functional implausibilities is lower than on images with other implausibilities and artifacts, the mean accuracy scores are similar. However, this similarity in means masks the differences in the distribution of accuracy scores, as shown in Figure~\ref{fig:combined-artifact-trends}A. We find that images with participant accuracy scores in the 40--60\% range (which represent images approaching indistinguishability between real and AI-generated) make up 32.8\% of images annotated with functional implausibilities compared to 21.4\% and 22.4\% of images annotated with anatomical implausibilities and stylistic artifacts, respectively.

We find that images that we annotated as containing multiple artifacts can still appear photorealistic enough to make detection difficult for most people. Artifacts vary in levels of visibility, as shown in Figure~\ref{fig:combined-varying-artifacts-visibility}A--C. While Figure~\ref{fig:combined-varying-artifacts-visibility}A and C contain stylistic artifacts, they are far more apparent in Figure~\ref{fig:combined-varying-artifacts-visibility}B, which is reflected in its higher detection accuracy. Despite Figure~\ref{fig:combined-varying-artifacts-visibility}A and C containing multiple artifact categories, they had low detection accuracy, suggesting that the presence of multiple artifacts does not necessarily make images easier to identify and that artifact visibility is also a contributing factor.

The visibility of artifacts is highly variable, and Figure~\ref{fig:combined-varying-artifacts-visibility}D--F present examples highlighting this variability. The anatomical implausibility in the fingers in image Figure~\ref{fig:combined-varying-artifacts-visibility}D is very noticeable, whereas the functional implausibilities in the tennis racket and shirt design of Figure~\ref{fig:combined-varying-artifacts-visibility}F are more subtle. The corresponding accuracy scores for these images--- 62\% for Figure~\ref{fig:combined-varying-artifacts-visibility}E and 54\% for Figure~\ref{fig:combined-varying-artifacts-visibility}F —reinforce the observation that anatomical artifacts tend to be more easily detected, while functional implausibilities often require closer attention and familiarity with depicted objects. The stylistic artifacts in the cinematization of Figure~\ref{fig:combined-varying-artifacts-visibility}E and plastic-like skin texture fall in between, further showing the spectrum of detectability across different artifact categories and visibility.

\subsection{Accuracy by Randomized Display Time}\label{sec:acc-time}
\begin{figure*}[h]
\centering
\captionsetup{justification=raggedright, singlelinecheck=false, skip=2pt, font=small}
\begin{subfigure}[t]{0.27\linewidth}
\centering
    \subcaption{}
    \includegraphics[width=\linewidth]{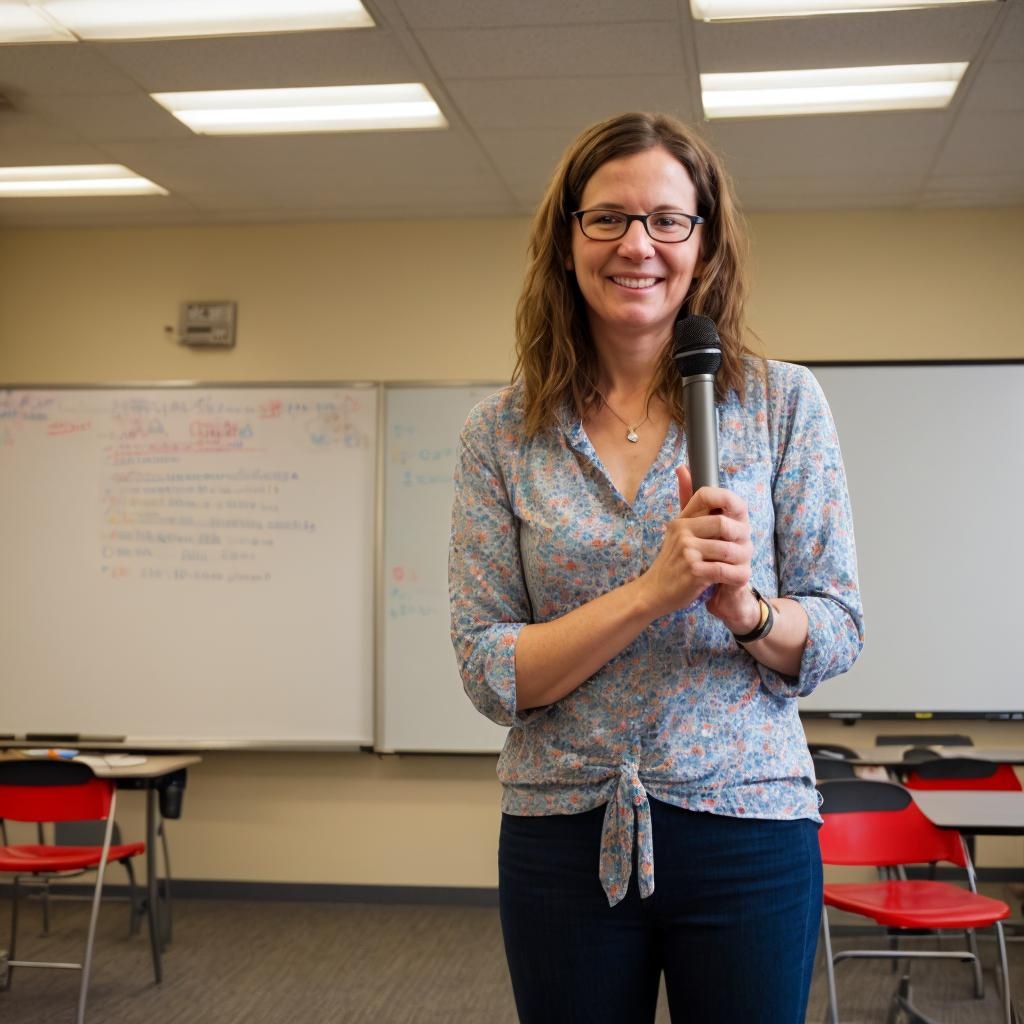}
\end{subfigure}
\hspace{1cm}
\begin{subfigure}[t]{0.27\linewidth}
\centering
  \subcaption{}
    \includegraphics[width=\linewidth]{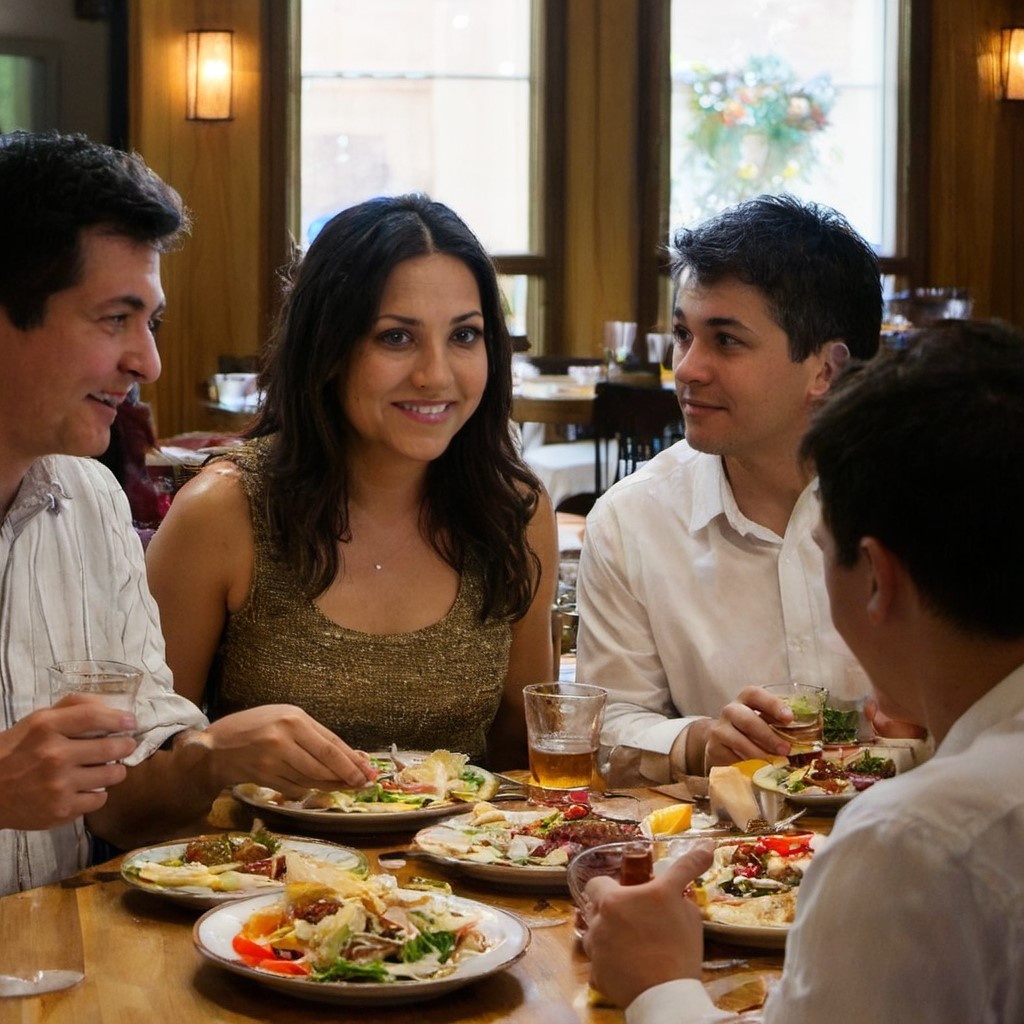}
\end{subfigure}
\hspace{1cm}
\begin{subfigure}[t]{0.27\linewidth}
\centering
  \subcaption{}
    \includegraphics[width=\linewidth]{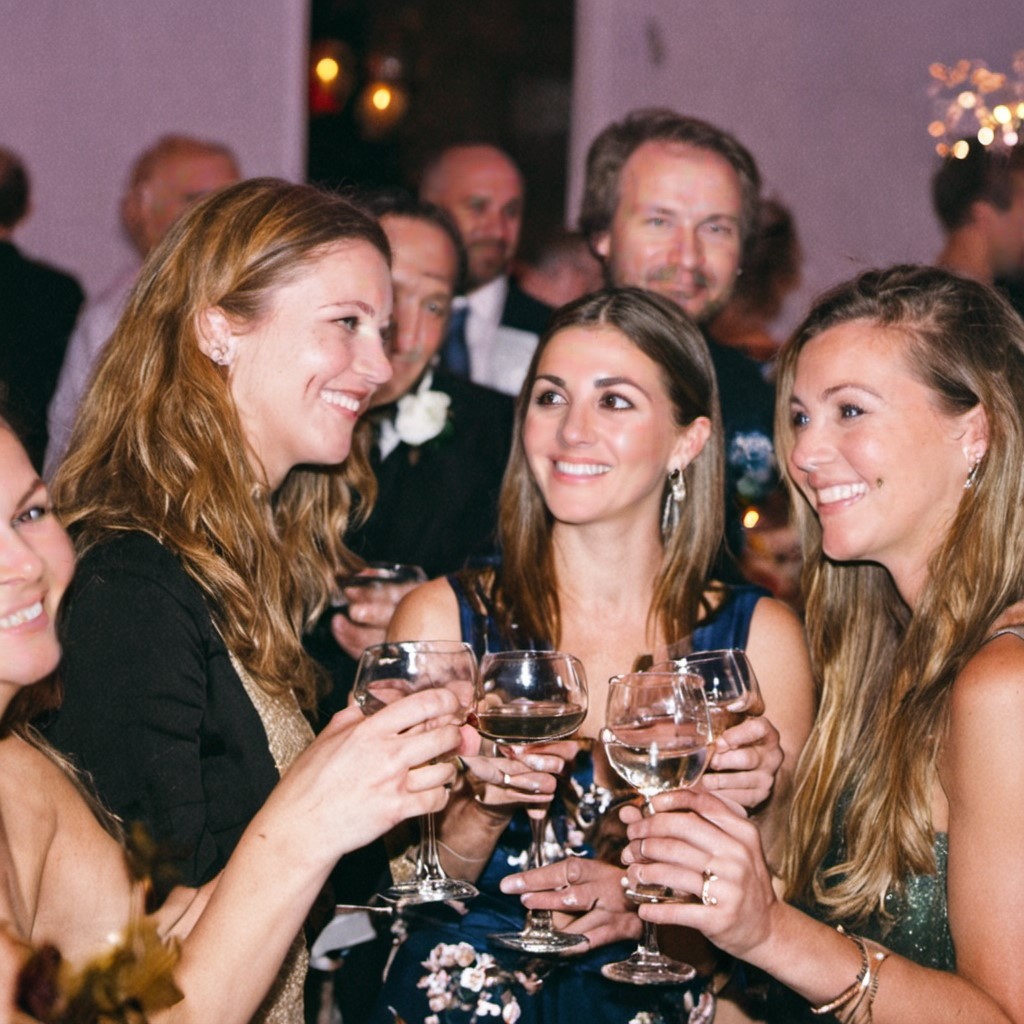}
\end{subfigure}
\caption{\textbf{Exemplar AI-generated images for which a closer look improves accuracy.} \normalfont{\textbf{A.} Accuracy: 38\% at 1 second display time to 65\% at 20 second display time. \textbf{B.} Accuracy: 44\% at 1 second display time to 82\% at 20 second display time. \textbf{C.} Accuracy: 27\% at 1 second display time to 70\% at 20 second display time.}}
\label{fig:displaytime}
\Description{Three images showing AI-generated images for which a closer look improves accuracy. (A) Image of woman generated by Stable Diffusion: Accuracy is 38\% at 1 second display time and it improves to 65\% at 20 second display time with  (B) Image of people generated by Stable Diffusion: Accuracy is 44\% at 1 second display time and it improves to 82\% at 20 second display time with (C) Image of people generated by Stable Diffusion: Accuracy is 27\% at 1 second display time and it improves to 70\% at 20 second display time with.}
\end{figure*}

By randomizing the display time of images in this experiment, our results support evaluating how viewing duration influences participants' accuracy. We find that longer viewing times improve performance. With just 1 second of display time, participants are  72\% accurate (95\% CI=[71.6, 72.5], 95\% CI=[71.3, 72.2]) on AI-generated and real images, respectively. With 5 seconds of display time, accuracy increases to 77\% (95\% CI=[77.0, 77.8], 95\% CI=[76.6, 77.4]) for both AI-generated and real images, respectively. While accuracy on real images appears to plateau by 5 seconds of display time, accuracy on AI-generated images increases up to 80\% (95\% CI=[79.6, 80.4]) at 10 seconds and 82\% (95\% CI=[81.2, 81.9]) at 20 seconds. Figure~\ref{fig:combined-artifact-trends}B presents the distribution of accuracy scores across display time conditions. Across the observations where display time was randomized, we find that the proportion of AI-generated images that are identified below random chance decreases from 43\% when participants only have 1 second to view the image to 30\%, 25\%, 17\%, and 17\% when participants have 5, 10, 20 seconds, and unlimited time to view the image.

In some images, AI artifacts can be noticed with a quick glance, but for others, careful attention to detail is necessary to spot the artifact. Figure~\ref{fig:displaytime} presents three images that require careful attention, as evidenced by the fact that most participants mark as real when they are limited to seeing the image for a second but
fake once they take into account the details of the scene.

Accuracy across all artifact types improved with increased display time. As shown in Figure~\ref{fig:combined-artifact-trends}C, participants showed higher accuracy when images were displayed for longer time (anatomical artifacts: 63\% at 5 seconds vs. 59\% at 1 second; stylistic artifacts: 63\% at 5 seconds vs. 60\% at 1 second; functional artifacts: 60\% at 5 seconds vs. 55\% at 1 second). For all artifacts, there is a significant improvement in detection accuracy when increasing display time from 5 seconds to unlimited. 

In Figure~\ref{fig:combined-artifact-trends}C, we observe that participants improved the most in identifying functional artifacts, with an 18\% improvement from 1 second to unlimited viewing time. In comparison, anatomical and stylistic artifacts showed smaller improvements of 11\% each over the same time interval. Unlike anatomical and stylistic implausibilities that can be identified at first glance, functional artifacts often require a closer look and familiarity with the elements in the image as they often appear in parts of the image that are not the main subject.

\subsection{Qualitative Analysis of Participant Comments}\label{sec:qualitative-analysis}

\begin{figure}[H]
\centering
\captionsetup{justification=raggedright, singlelinecheck=false, skip=2pt}
\begin{subfigure}[t]{0.9\linewidth}
    \includegraphics[width=\linewidth]{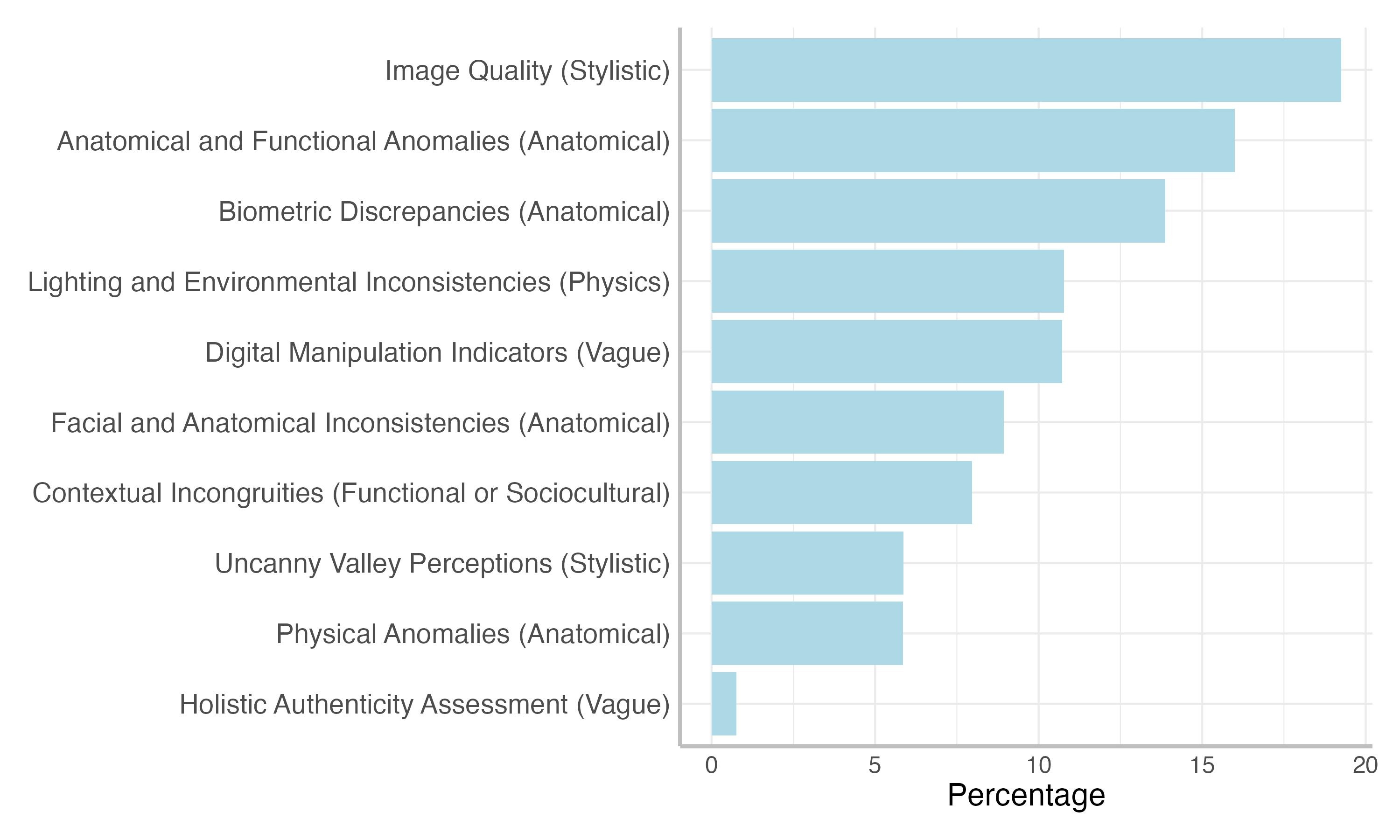}
\end{subfigure}
\caption{\mybold{Distribution of themes identified in participant comments.}}
\label{fig:themes}
\Description{A horizontal bar chart showing the distribution of themes identified in participant comments explaining their reasoning for AI image detection. The themes include Image Quality, Facial and Anatomical Inconsistencies, Anatomical and Functional Anomalies, Lighting and Environmental Inconsistencies, Digital Manipulation Indicators, Biometric Discrepancies, Uncanny Valley Perceptions, Contextual Incongruities, Physical Anomalies, and Holistic Authenticity Assessment.}
\end{figure}

\begin{figure}[htb]
\centering
\captionsetup{justification=raggedright, singlelinecheck=false, skip=2pt, font=small}
\begin{subfigure}[t]{0.22\textwidth}
    \subcaption{}\vtop{\vskip0pt\hbox{\includegraphics[width=\linewidth]{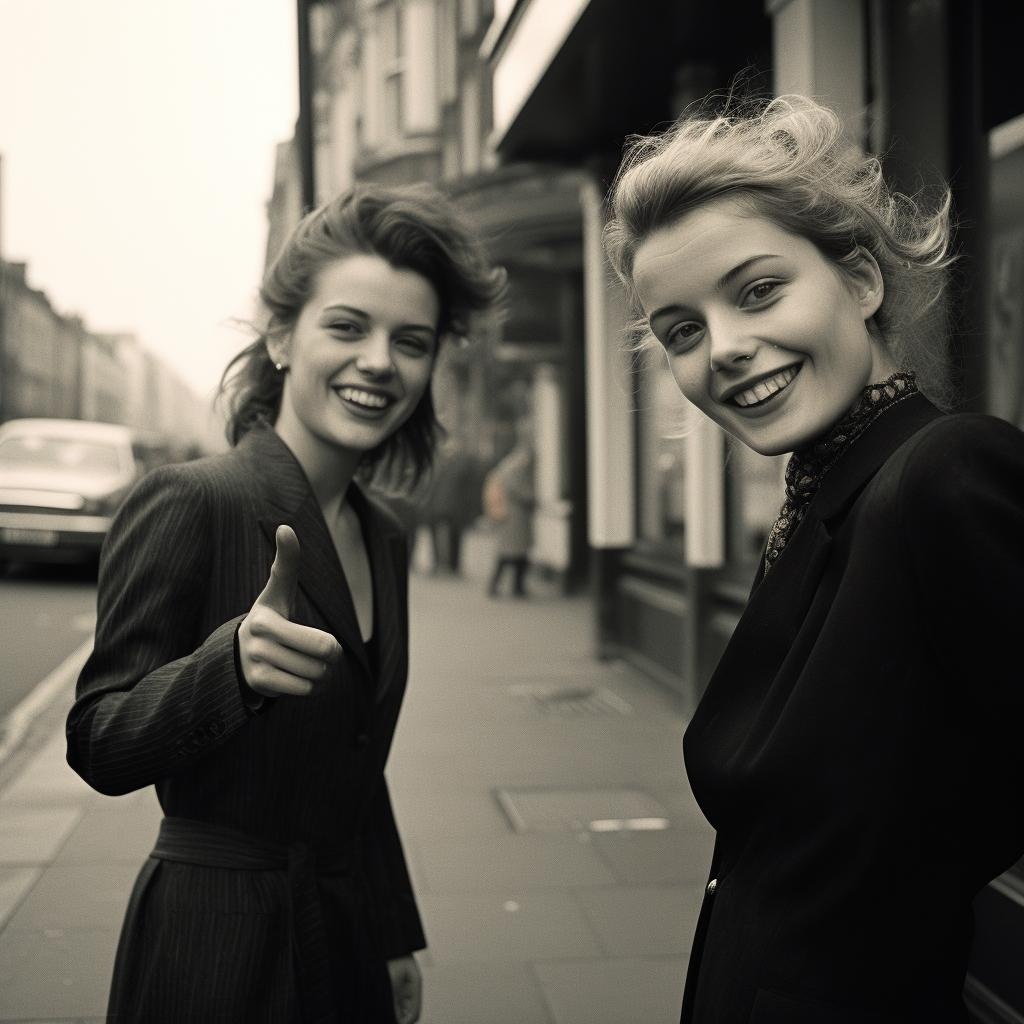}}}
\end{subfigure}
\hfill
\begin{subfigure}[t]{0.22\textwidth}
    \subcaption{}\vtop{\vskip0pt\hbox{\includegraphics[width=\linewidth]{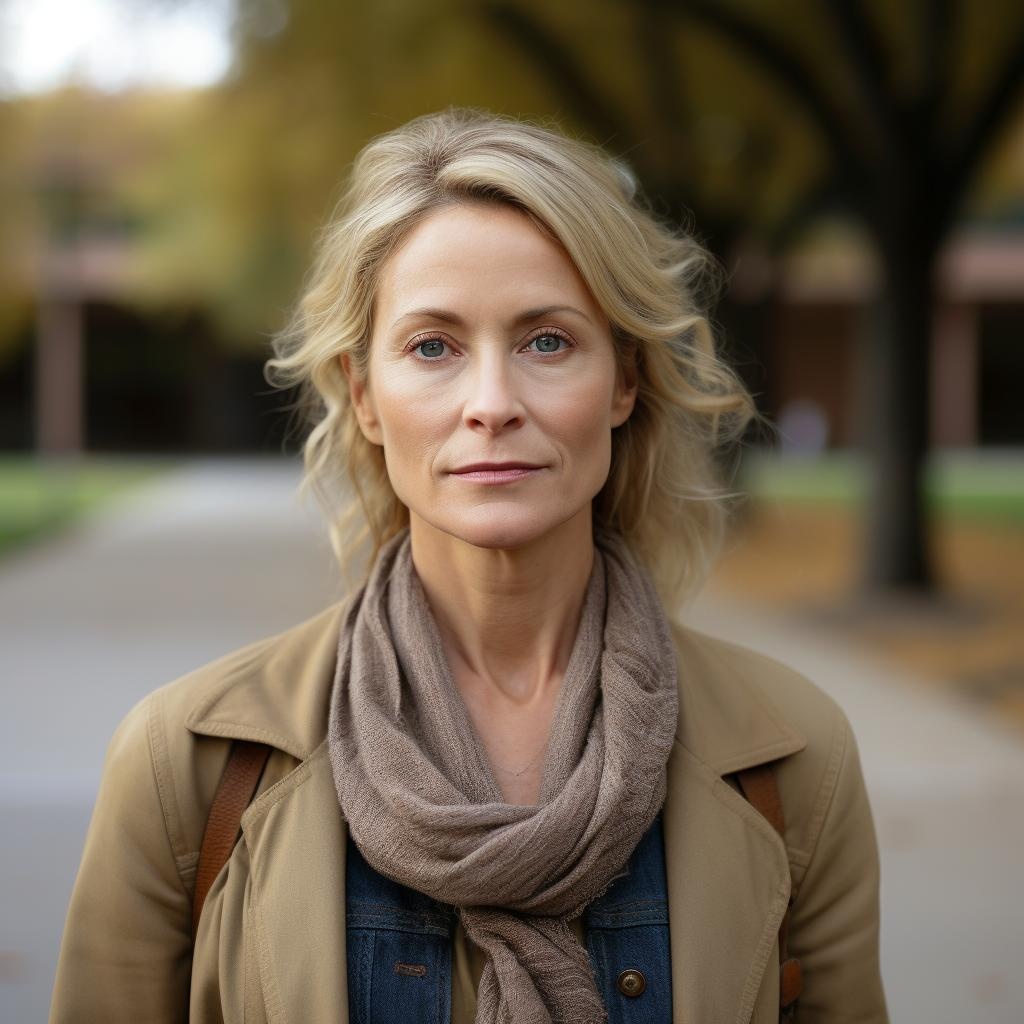}}}
\end{subfigure}
\hfill
\begin{subfigure}[t]{0.22\textwidth}
    \subcaption{}\vtop{\vskip0pt\hbox{\includegraphics[width=\linewidth]{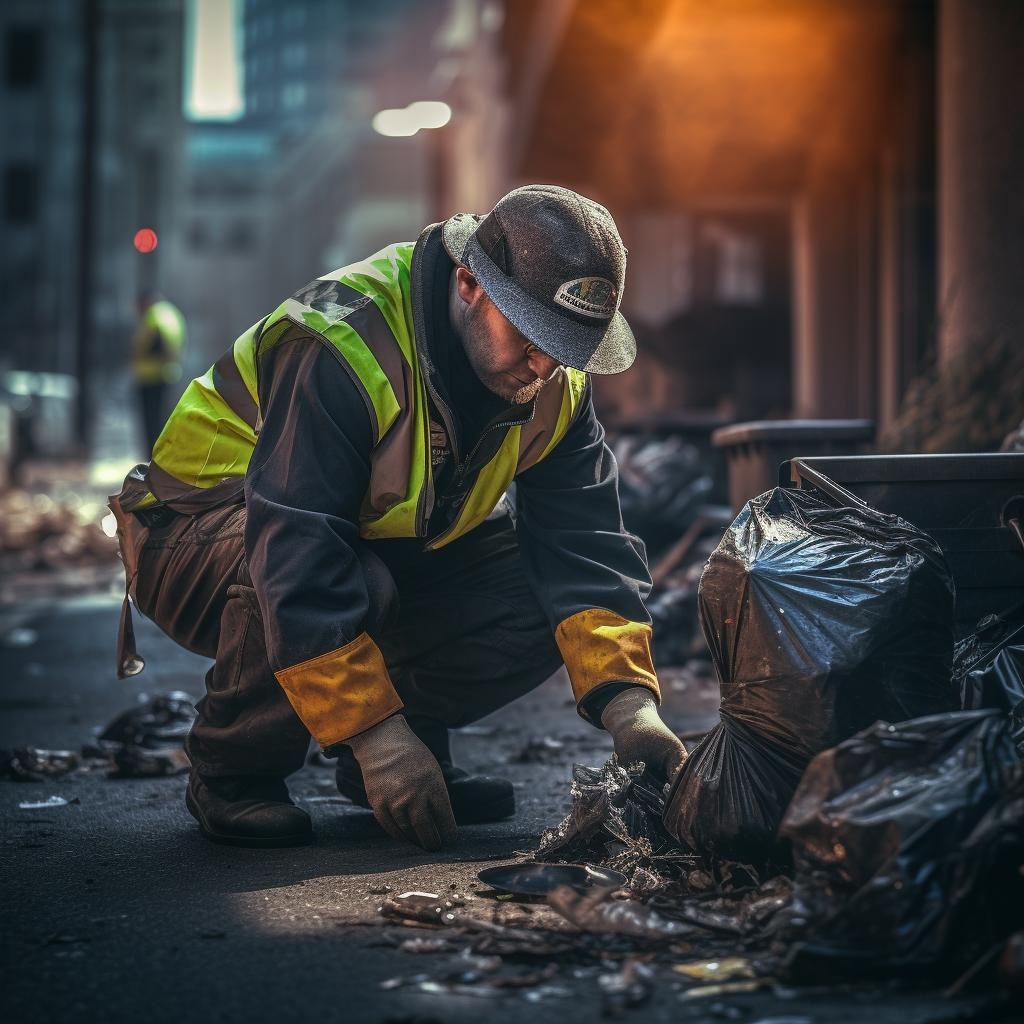}}}
\end{subfigure}
\hfill
\begin{subfigure}[t]{0.22\textwidth}
    \subcaption{}\vtop{\vskip0pt\hbox{\includegraphics[width=\linewidth]{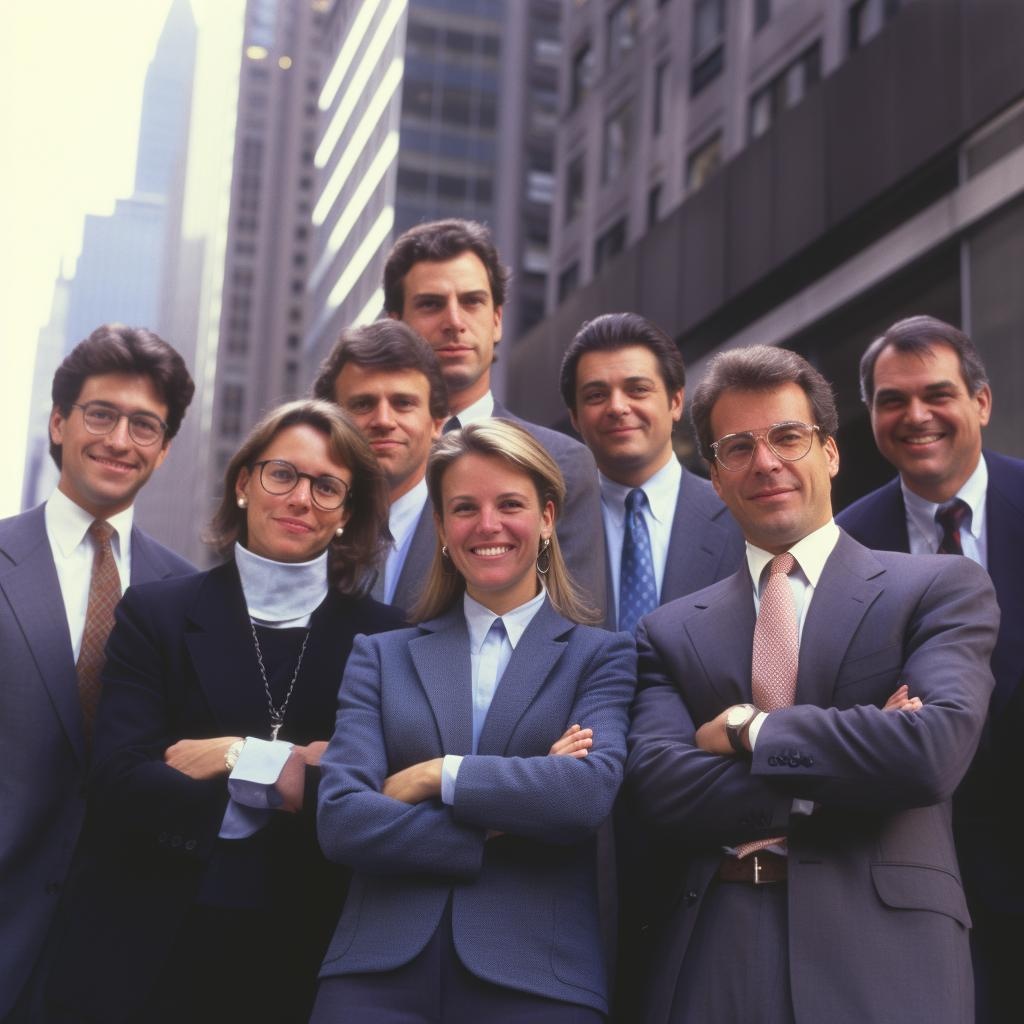}}}
\end{subfigure}
\caption{\mybold{Examples of participant comments mapped to themes.} \normalfont{\textbf{A.} ``Cosmetic style out of character with vintage setting": Contextual Incongruities. \textbf{B.} ``Skin too smooth, depth of field shallow.": Image Quality, Lighting Inconsistencies. \textbf{C.} ``If this is not AI then it is a staged photograph like a movie set because of the lighting and he is an actor.": Lighting inconsistencies, Contextual Incongruities.
\textbf{D.} ``Group looks pasted onto background.": Digital Manipulation Indicators.}}
\label{fig:comments-all}
\Description{Four images with participant comments mapped to themes.(A) AI-generated candid image with a comment on cosmetic style being out of character with a vintage setting.(B) AI image with smooth skin, with a comment on skin being too smooth and shallow depth of field.(C) AI-generated full body shot of a man, with a comment suggesting it resembles a staged photograph due to lighting.
(D) AI image of a group, with a comment on the group looking pasted onto the background.}
\end{figure}

We collected 34,675 comments from participants who filled out the optional text input box asking participants: ``If you think this is AI-generated, please explain why.'' In order to identify themes from these 34,675 comments, we prompted GPT-3.5 Turbo to identify 10 main themes across these comments. GPT-3.5 Turbo responded with the following ten themes, which we manually reviewed and refined to mitigate the ambiguities and generalization typical of large language models \cite{stephan2024rlvflearningverbalfeedback}: (1) Image quality focusing on the overall appearance, smoothness, and sometimes unrealistic perfection of image elements; (2) Facial and anatomical inconsistencies where participants pointed to irregularities in eyes, mouths, noses, skin texture, expressions, and general human anatomy; (3) Anatomical and functional anomalies such as deformities, misplaced body parts, and irregularities in objects or environments; (4) Lighting and environmental inconsistencies including unnatural lighting, inconsistent shadows, and reflections; (5) Digital manipulation indicators suggesting suspicions of AI-generation or digital alteration; (6) Biometric discrepancies particularly unnatural or imperfect body parts like hands and fingers; (7) Uncanny valley perceptions where images almost looked human but had subtle unnatural features that caused discomfort; (8) Contextual incongruities such as unrealistic scenarios and mismatched social elements; (9) Physical anomalies highlighting illogical physical interactions within the images; and (10) holistic authenticity assessment making overall judgments based on a combination of multiple cues and inconsistencies. Based on these ten main themes, we prompted GPT-3.5 to label each comment with one of the ten themes. Figure~\ref{fig:comments-all} illustrates examples of participant comments for four images and how they were categorized into themes. Figure~\ref{fig:themes} displays the distribution of themes across the comments and the related concept from our taxonomy in parentheses.

Based on GPT-3.5 Turbo, we find that 61\% of participants' comments mentioned relying on anatomical implausibilities. The next most common concept referred to is stylistic artifacts, which is mentioned in 30\% of comments. Participants mentioned functional implausibilities in 21\% of comments, violations of physics in 15\% of comments, and sociocultural implausibilities in only 4\% of comments. 

Based on the authors' annotations of artifacts, we find functional implausibilities to be the most prevalent, appearing in 58.7\% of images, followed by anatomical implausibilities in 51.4\% and stylistic artifacts in 39.0\% of images. We identify violations of physics and sociocultural implausibilities in only 9.17\% and 5.50\% of images, respectively. 
While functional artifacts were the most prevalent in human researcher annotated images, they were less frequently mentioned in participant comments annotated by GPT--3.5. Conversely, anatomical artifacts were emphasized more in participant comments than in their prevalence in annotated images. 

\subsection{Accuracy by Models} \label{sec:acc-model}

\begin{figure}[H]
    \centering
    \includegraphics[width=0.8\linewidth]{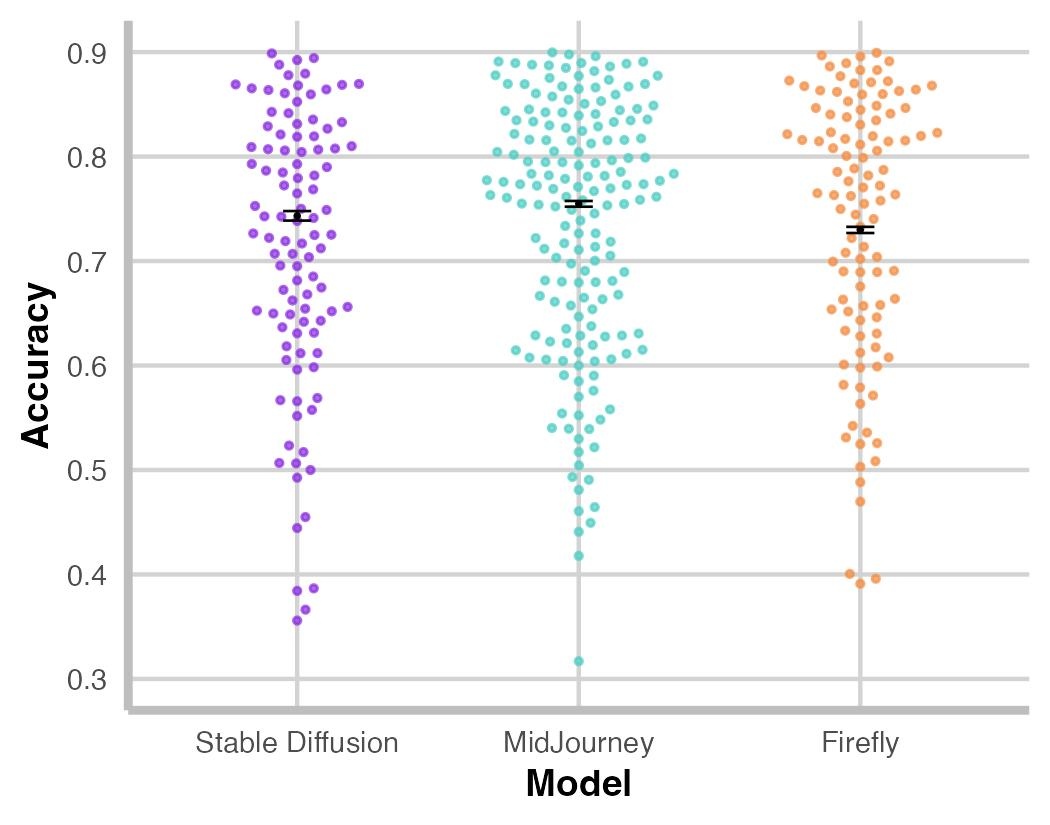} 
    \caption{\textbf{Accuracy across generative AI models} \normalfont{Each point represents an image. The black dots and error bars show the mean accuracy and 95\% bootstrapped confidence intervals for each model}}
    \label{fig:models}
    \Description{Bee swarm chart showing accuracy across different generative AI models. The chart compares the accuracy rates for identifying AI-generated content among various models along with bootstrapped 95\% confidence intervals}
\end{figure}

In the process of generating the images for this experiment's stimuli set, we noticed that Midjourney, Firefly, and Stable Diffusion have different capabilities and limitations. For example, we noticed that Midjourney often produced images with persistent stylistic artifacts that were challenging to eliminate. Firefly, on the other hand, frequently exhibited a tendency toward synthetic emotional expressions, with subjects often appearing unnaturally and overly cheerful, necessitating multiple iterations to produce more realistic results. Stable Diffusion struggled significantly with generating group images, often introducing artifacts such as anatomical inconsistencies. In light of the limitations to generate non-portrait images with Stable Diffusion, 75\% of the Stable Diffusion-generated stimuli in this experiment were portraits. On the other hand, 30\% of Midjourney and Firefly-generated images in this experiment depict portraits. In order to compare the three models fairly, we focus our comparison on portrait images. Figure~\ref{fig:models} presents accuracy shown on portraits by each of the three models and reveals that participants' mean accuracy on Midjourney, Stable Diffusion, and Firefly were  76\% (95\% CI: [75.2, 75.8]), 74\% (95\% CI: [73.9, 74.8]), and 73\% (95\% CI: [72.7, 73.3]), respectively.

\subsection{Accuracy on Human Curated Images vs. Uncurated Images} \label{sec:human-curation}
\begin{figure*}[h]
\centering
\captionsetup{justification=raggedright, singlelinecheck=false, skip=2pt, font=small}
\begin{subfigure}[t]{0.24\linewidth}
    \subcaption{}\vtop{\vskip0pt\hbox{\includegraphics[width=\linewidth]{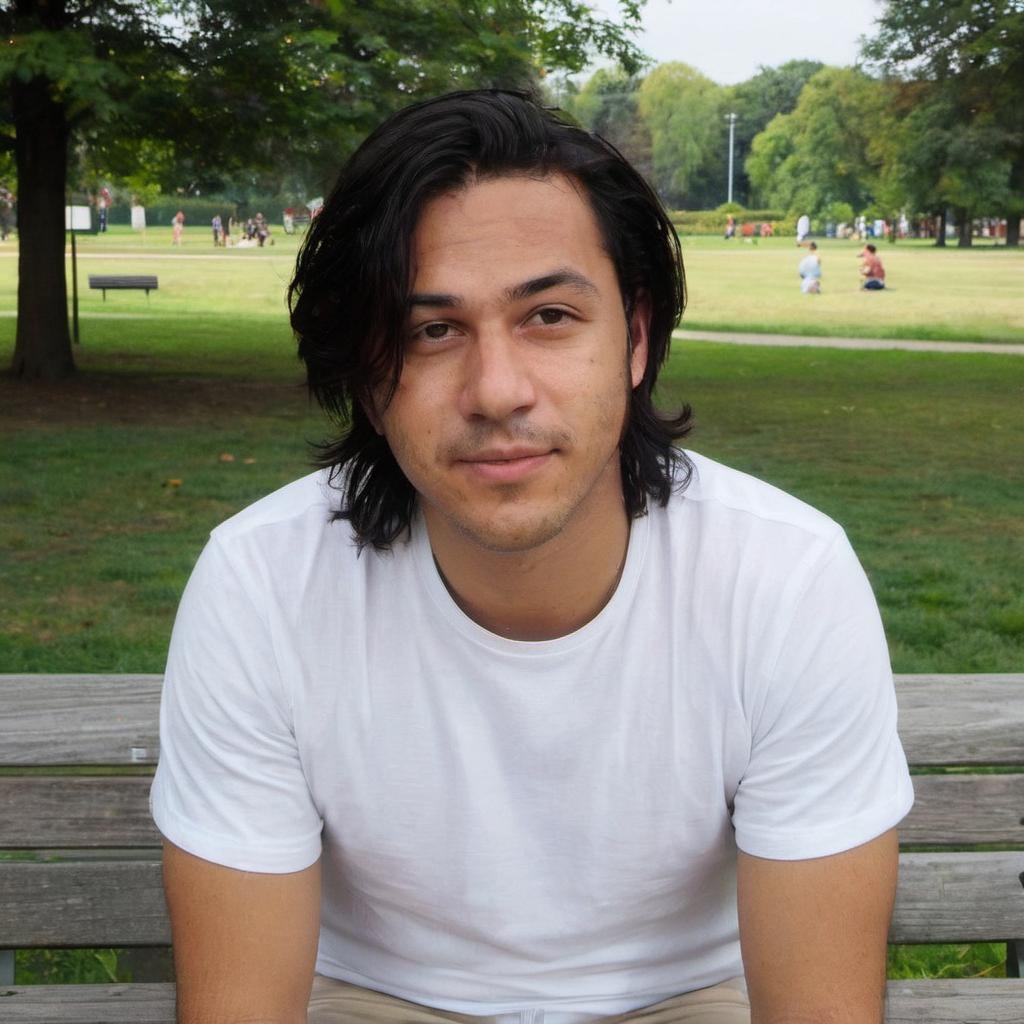}}}
\end{subfigure}
\hfill
\begin{subfigure}[t]{0.24\linewidth}
    \subcaption{}\vtop{\vskip0pt\hbox{\includegraphics[width=\linewidth]{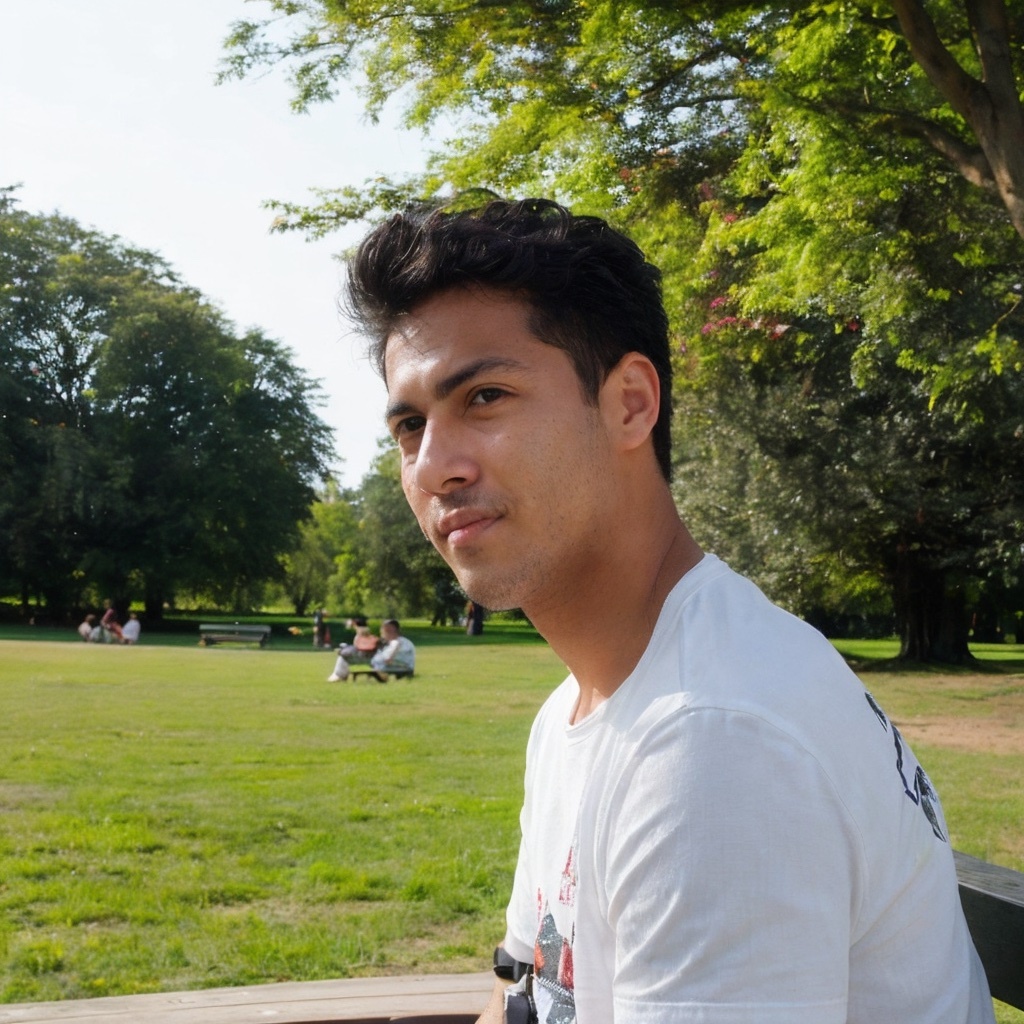}}}
\end{subfigure}
\hfill
\begin{subfigure}[t]{0.24\linewidth}
    \subcaption{}\vtop{\vskip0pt\hbox{\includegraphics[width=\linewidth]{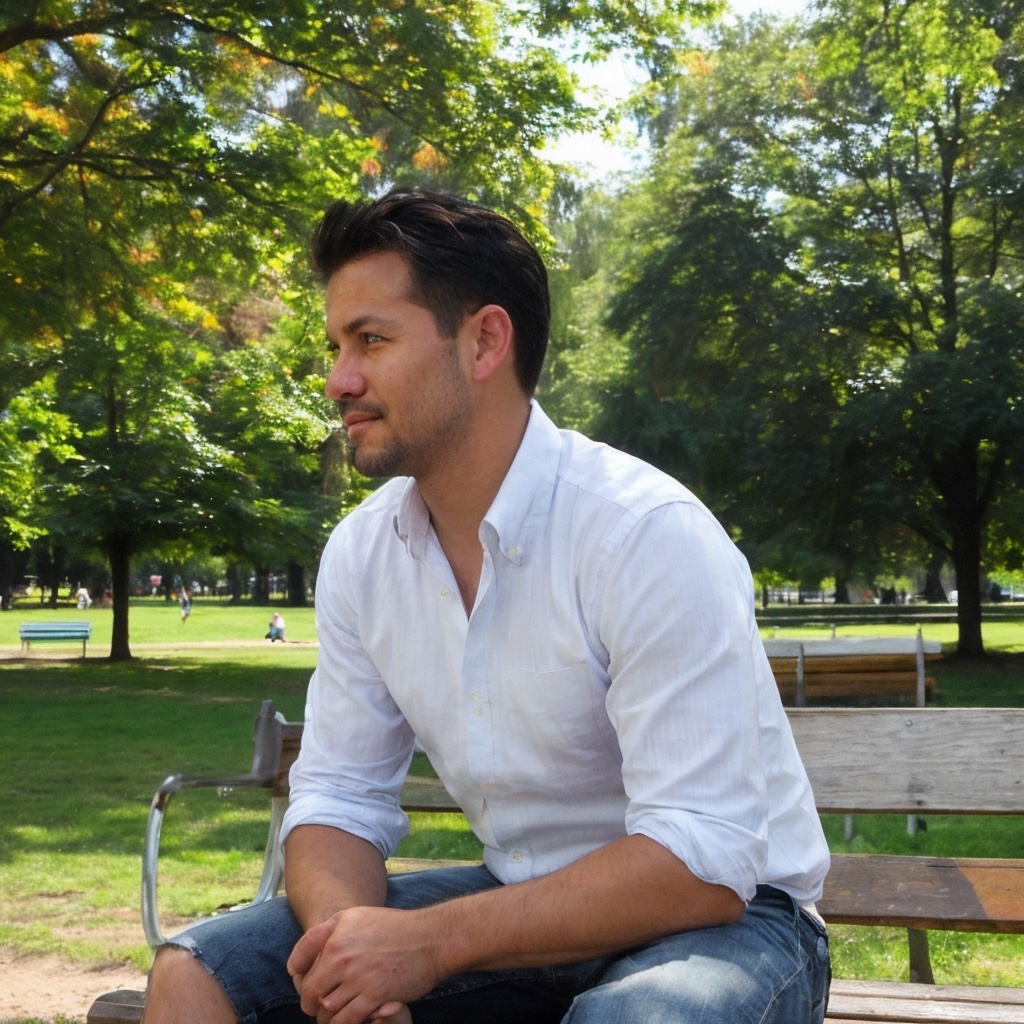}}}
\end{subfigure}
\hfill
\begin{subfigure}[t]{0.24\linewidth}
    \subcaption{}\vtop{\vskip0pt\hbox{\includegraphics[width=\linewidth]{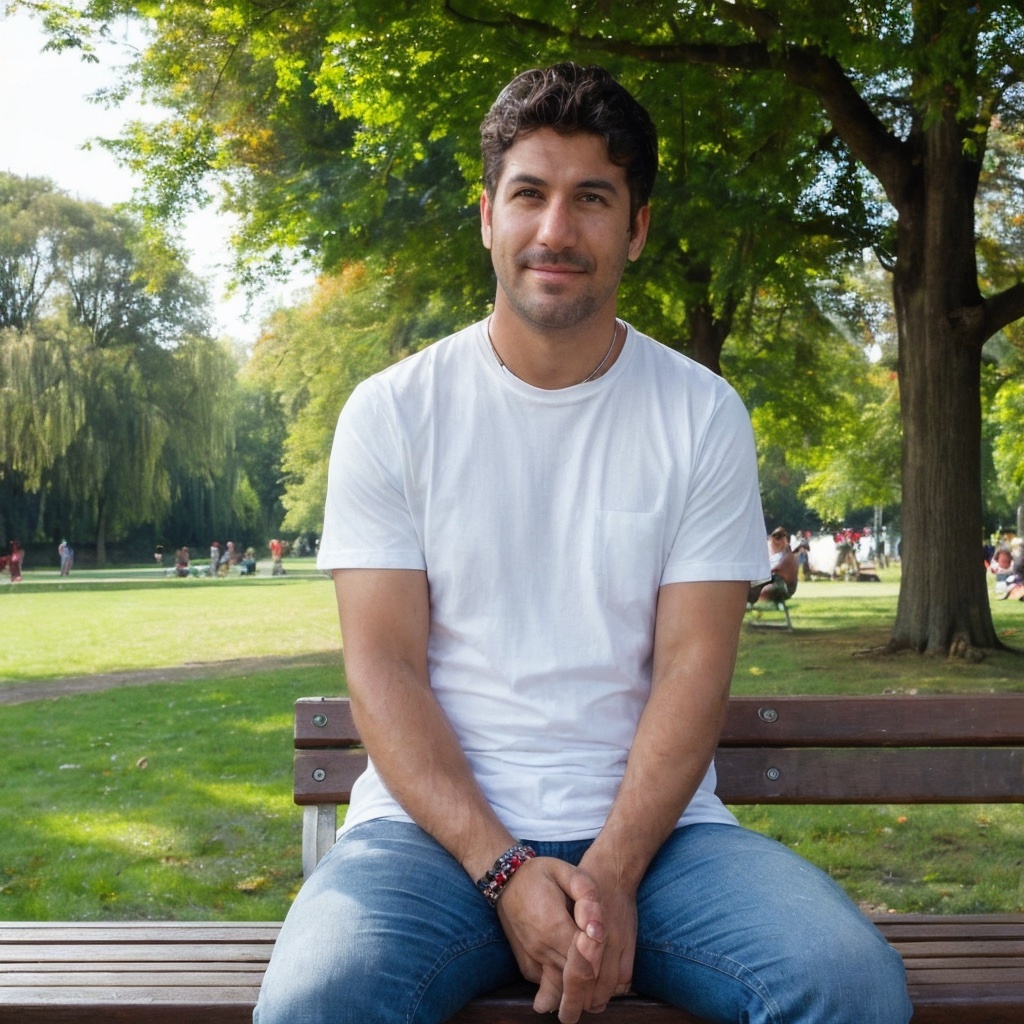}}}
\end{subfigure}
\caption{\mybold{Re-generated images from the same prompt.} \normalfont{ \textbf{A.} Stage 1 image generated by Stable Diffusion and curated by our team (37\% accuracy) \textbf{B.} Most photorealistic of 12 prompt-matched image generations by Stable Diffusion (42\% accuracy) \textbf{C.} Median photorealistic of 12 prompt-matched image generations by Stable Diffusion (59\% accuracy) \textbf{D.} Least photorealistic of 12 prompt-matched image generations by Stable Diffusion(83\% accuracy)}}
\label{fig:regeneration}
\Description{Four images showing re-generated outputs from the same prompt.\textbf{A.} Stage 1 image generated by Stable Diffusion and curated by our team (37\% accuracy) \textbf{B.} Most photorealistic of 12 prompt-matched image generations by Stable Diffusion (42\% accuracy) \textbf{C.} Median photorealistic of 12 prompt-matched image generations by Stable Diffusion (59\% accuracy) \textbf{D.} Least photorealistic of 12 prompt-matched image generations by Stable Diffusion(83\% accuracy)}
\end{figure*}

\begin{figure*}[h]
\centering
\captionsetup{justification=raggedright, singlelinecheck=false, skip=2pt}
\begin{subfigure}[t]{0.48\textwidth}  
\subcaption{}
\vtop{\vskip0pt\hbox{\includegraphics[width=\linewidth]{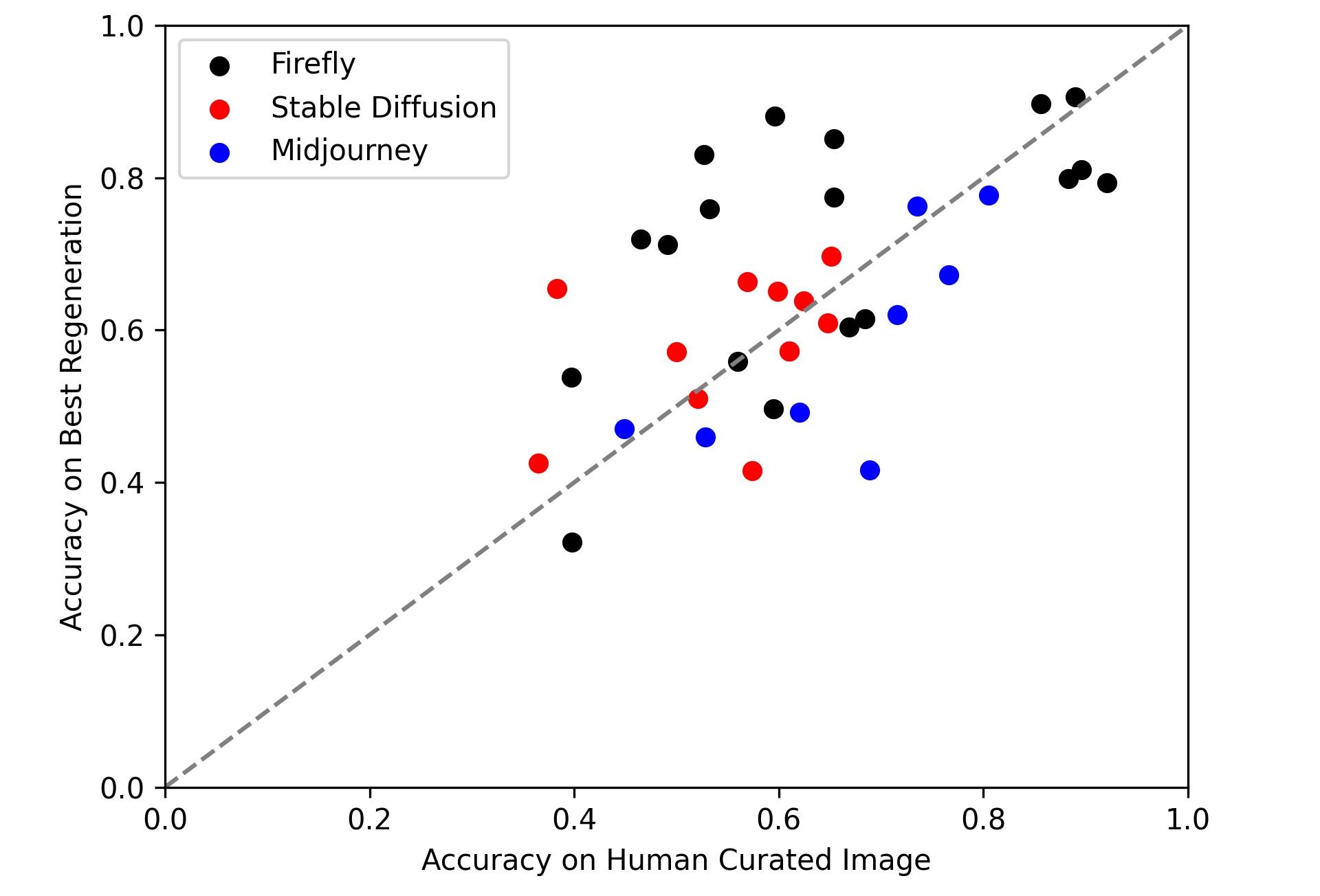}}}
\end{subfigure}
\hfill
\begin{subfigure}[t]{0.48\textwidth}
\subcaption{}
\vtop{\vskip0pt\hbox{\includegraphics[width=\linewidth]{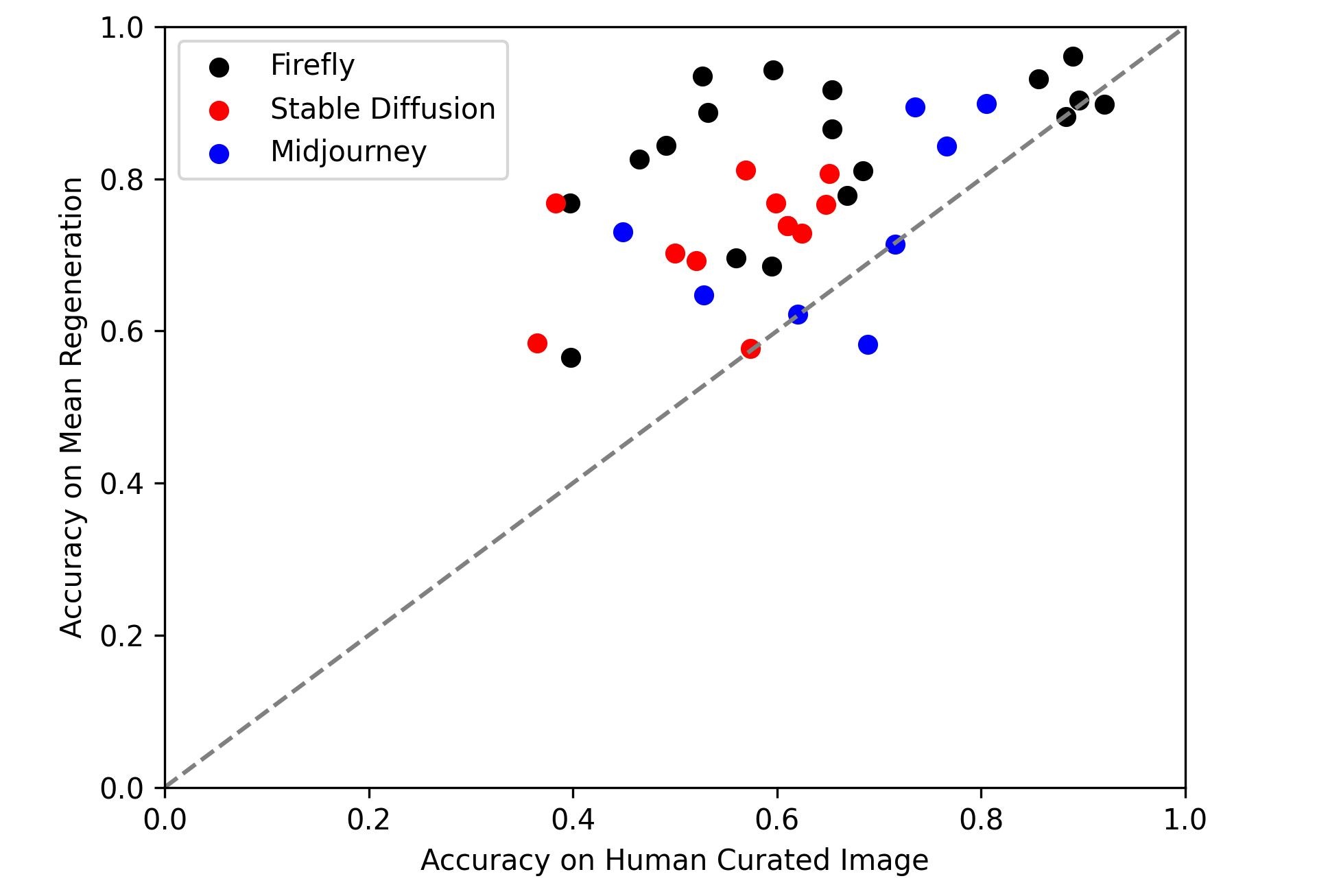}}}
\end{subfigure}
\caption{\mybold{Comparing accuracy scores on curated images and uncurated prompt-matched images.} \normalfont{\textbf{A.} Scatterplot showing human detection accuracy of the original curated image compared to human detection accuracy of its most photorealistic regeneration out of 11 to 24 prompt-matched images labeled as re-generations. \textbf{B.} Scatterplot showing human detection accuracy of the original curated image compared to human detection accuracy of its mean photorealistic regeneration out of 11 to 24 re-generations.}}
\label{fig:curation-value}
\Description{Two scatterplots showing curation value analysis.(A) Minimum curation value added.(B) Mean curation value added. Each chart illustrates the impact of curation on the overall value added to the dataset.}
\end{figure*}

Generating photorealistic AI-generated images involves three key ingredients: the diffusion model, the prompt, and human curation. In this section, we examine how human curation of diffusion model-generated images influences the aggregate accuracy scores of human participants. In order to show this influence, we compare diffusion model images from the main experiment, which were curated by our research team, with multiple diffusion model images generated from the same prompt as the curated images. This comparison reveals the increase in photorealism (as measured by the decrease in participants' accuracy) on the curated images relative to the prompt-matched images.

In this second phase of the experiment, we randomly sampled 39 AI-generated images from the main stimuli set, where the sample was stratified on 10 percentage point wide bins on human detection accuracy. For each of these 39 images, we generated at least 11 prompt-matched images using Midjourney, Firefly, and the same pipeline in Stable Diffusion. Figure~\ref{fig:regeneration} displays a Stable Diffusion-generated image from our original stimuli set and three of the twelve generations using the same prompt. We generated 482 total additional images, with at least 11 per prompt. These 482 images were included alongside the 149 real images on the experiment website.

In Figure~\ref{fig:curation-value}, we present scatterplots comparing human detection accuracy on the initial curated images and the best prompt-matched images in panel A, and mean prompt-matched images in panel B. We find the human-curated images have lower human detection accuracy than the best regenerated image in 18 of 39 instances and the mean re-generated image in 35 of 39 instances. In total, the human-curated images were perceived to be more photorealistic than 408 of the 482 (84\%) uncurated prompt-matched images. Specifically, we find the marginal value added by human curation for images that were initially detected in the range of 30\% to 50\% is 31 percentage points, 50 to 60\% is 23 percentage points, 60-70\% is 11 percentage points, 70-80\% is 8 percentage points, and 80+\% is 4 percentage points. Across the stimuli selected from Midjourney, Firefly, and Stable Diffusion, the marginal value of human curation is 7.8, 19.0, and 16.9 percentage points, respectively.

The two panels in Figure~\ref{fig:curation-value} illustrate the positive correlation between accuracy on the human-curated image and accuracy on the regeneration. This reveals how the prompt influences photorealism. The Pearson Correlation Coefficient between accuracy on curated images and their best, mean, and worst re-generations are .58, .53, and .32, respectively. This positive correlation suggests the choice of a prompt plays a significant role in the photorealism of an image. Figure~\ref{fig:goodandbadprompt} displays two original curated images where A is generated by a prompt in which re-generations achieved low human detection accuracy (a `good' prompt), and B is generated by a prompt in which re-generations achieved a high human detection accuracy (a `bad' prompt). Prompts that consistently generate easily detectable images often have elements that are difficult to generate and result in artifacts. The prompt ``Persian woman astronaut in astronaut clothes, family photo with husband and two toddlers, high resolution, realistic" for Figure~\ref{fig:goodandbadprompt}B  generates a posed group image that tends to be easy to detect. On the other hand, the prompt ``American woman faculty portrait, not a close-up, blond" for Figure~\ref{fig:goodandbadprompt}A generates a portrait image that tends to be perceived as more photorealistic.
\section{Discussion}\label{sec:disc}

While diffusion models can generate highly realistic images, most of the images they produce still contain visible artifacts. In particular, we find that only 17\% of diffusion model-generated images are misclassified as real at rates consistent with random guessing. Notably, this misclassification rate increases to 43\% when the viewing duration is restricted to 1 second. By curating a dataset of 599 images and conducting a large scale digital experiment, we can begin to answer fundamental questions about what drives the appearance of photorealism in diffusion model-generated images. 

First, we find that images with greater scene complexity tend to introduce more opportunities for artifacts to appear, making it easier for participants to detect AI-generated images. Our results reveal that participants were less accurate at identifying AI-generated portraits compared to more complex scenes, such as those involving multiple people in candid settings.  Based on qualitative analysis of the images, we identify three main reasons for this difference. First, portraits often feature a single person against a blurred background, which can obscure details and provide fewer cues compared to full-body or group images. Second, portraits typically involve fewer and simpler poses, focusing only on the face and torso, leaving fewer opportunities for errors or inconsistencies to be apparent. Third, the prevalence of edited and retouched portraits in real-world photography complicates the distinction between real and AI-generated portraits, addressing the question of how subject type and context (e.g., unknown people vs. public figures) influence the perceived authenticity of an image. In contrast, more complex images, like full-body or group shots, involve a greater number of elements, increasing the likelihood of noticeable errors or inconsistencies. Similar to our results on AI-generated images, we find that real images with lower scene complexity are also harder to identify as real. 

Second, we identify five high-level categories of artifacts and implausibilities and find that the easiest images to identify as diffusion model generated are the ones with anatomical implausibilities, such as unrealistic body proportions and stylistic artifacts like overly glossy or waxy features.

Third, by randomizing display time, we identify the relationship between how long an individual looks at an image and their accuracy at distinguishing between real and AI-generated images. Specifically, we find that participants' accuracy at identifying an AI-generated image upon a quick glance of 1 second is 72\% and increases by 5 percentage points with just an additional 4 seconds of viewing time and 10 percentage points when unconstrained by time. Given the nature of rapid scrolling on social media and how much time people have to see advertisements as they pass by billboards on a highway, these results reveal the importance of attentive viewing of images before making judgments about an image's veracity.

Fourth, we find that human curation had a notable negative impact on participants' accuracy compared to uncurated images generated by the same prompts as the human-curated AI-generated images. In particular, the images curated by our research team were harder to identify as AI-generated than 84\% of the uncurated images generated using the same prompts as the curated images. This finding reveals the limitation of state-of-the-art diffusion models in producing images of consistent quality. It also suggests that human curation is a bottleneck to generating fake images at scale. The process of generating high-quality AI images is inherently iterative---users refine prompts and select outputs until they achieve their desired result. This fundamental aspect of AI image generation is evident across all applications, from advertising and marketing to education and beyond. While concerns exist about fake images being used to mislead or impersonate, many use cases exist for business and educational applications~\cite{vartiainen2023using, hartmann2023power, gvirtz2023text}. The critical role of human curation in this iterative process further emphasizes how the photorealism of images produced by diffusion models depends not only on the capabilities of the diffusion model but also on the quality of human curation, choice of prompts, and context of the scene. Given the importance of these factors beyond the generative AI model, these results reveal the importance of considering these factors in research examining human perception of AI-generated images. Without considering these elements, it is possible to produce biased findings showing AI-generated images are more or less realistic than they really appear in real-world settings. 

The taxonomy offers a practical framework on which to build AI literacy tools for the general public. We synthesized information from diverse sources such as social media posts, scientific literature, and our online behavioral study with 50,444 participants to systematically categorize artifacts in AI-generated images. Through this process, we identify five key categories: anatomical implausibilities, which involve unlikely artifacts in individual body parts or inconsistent proportions, particularly in images with multiple people;  stylistic artifacts, referring to overly glossy, waxy, or picturesque qualities of specific elements of an image; functional implausibilities, arising from a lack of understanding of real-world mechanics and leading to objects or details that appear impossible or nonsensical; violations of physics, which include inconsistencies in shadows, reflections, and perspective that defy physical logic; and sociocultural implausibilities, focusing on scenarios that violate social norms, cultural context, or historical accuracy. Our taxonomy builds upon the Borji 2023 taxonomy \cite{borji2023qualitative} and focuses on images that appear more realistic at first glance, which is useful for comparing and contrasting real photographs with diffusion model generated images for revealing the nuances of the artifacts and implausibilities~\cite{kamali2024distinguish}. Moreover, this taxonomy offers a shared language by which practitioners and researchers can communicate about artifacts commonly seen in AI-generated images and exposes the persistent challenges that can help people identify AI-generated images. 

\subsection{Future Work and Limitations}

In addition to aiding in identifying AI-generated content, the taxonomy offers insights into the open problems for producing realistic AI-generated images. Future work may explore integrating such taxonomies into model evaluation frameworks to provide iterative feedback during the development of generative models. As models advance to address the weaknesses presented in this taxonomy, new and more subtle artifacts may emerge, requiring future updates to this taxonomy. This dynamic interplay between detection and generation capabilities demonstrates why we need to maintain robust human detection abilities even as models evolve. We acknowledge the potential dual use of these insights to create more deceptive synthetic media, and we believe that transparent documentation of artifacts does more good than harm by offering detection strategies and an opportunity to develop general awareness in the public.

Large-scale digital experiments with participants who participate based on their own interests come with certain limitations. First, we did not collect demographic data from participants. Participants were not recruited for this experiment; instead, participants found the experiment organically and participated. Given the organic nature of the participation, we prioritized maximizing engagement, which involves questions unrelated to distinguishing AI-generated and real images like demographic questions. While this approach enabled substantial data collection, it limits analysis by excluding factors like age, gender, and cultural background that may influence detection. 

Second, we provided feedback on the correct answer after each participant made an observation, which has the potential to introduce learning effects. Future research could address these open questions by collecting demographic data to design more inclusive AI literacy tools and evaluating how performance changes with and without feedback. 

This research focused on images generated by state-of-the-art generative models available in 2024, and the findings are inherently tied to the state of diffusion models and generative AI technologies as of 2024. In the future, models are likely to change, and the somewhat visible errors that emerge will also likely change. Past state-of-the-art GAN models such as StyleGAN2~\cite{karras2020analyzingimprovingimagequality} and BigGAN~\cite{brock2018biggan}, often produced more noticeable artifacts in facial features, color balance, and overall photorealism, making their outputs more easily distinguishable. Nonetheless, the current taxonomy on diffusion models points out elements like anatomical implausibilities and stylistic artifacts that can be mapped to the facial feature and color balance cues. These recurring issues offer evidence of the taxonomy’s robustness to differences across model generations, but future studies should explore how the taxonomy may need to adapt to these changes, which may involve adding or removing categories or may involve further identifying nuances within these categories. As an example of how this taxonomy may be applied to AI-generated video, Figure~\ref{fig:sora} presents an example of an anatomical implausibility that we never saw in diffusion model-generated images because it involves a temporal inconsistency. Future research on the realism of AI-generated audio and video may also consider following the three-step process involved in building this taxonomy for images generated by diffusion models. Based on first surveying AI literacy resources, academic literature, and social media, second generating media with state-of-the-art models, and third collecting empirical data on the human ability to distinguish AI-generated media from authentically recorded media, researchers can build empirical insights towards characterizing realism and categorizing the artifacts in AI-generated media.  
 
The empirical insights on the photorealism of AI-generated images and the resulting taxonomy designed to help people better navigate real and synthetic images online lead to a practical research question: How can AI literacy interventions improve people's ability to distinguish real photographs and AI-generated images? Future research may address this question via randomized experiments comparing a control group with no intervention to a treatment group that receives training based on the taxonomy presented in this paper. Likewise, future research may explore this with just-in-time interventions to direct people's attention to the cues identified in the taxonomy.
\section{Conclusion}\label{sec:con}

Our work contributes empirical insights on the photorealism of AI-generated images and a taxonomy of artifacts commonly found in AI-generated images, organized into five categories: anatomical implausibilities, stylistic artifacts, functional implausibilities, violations of physics, and sociocultural implausibilities. We find that the photorealism of AI-generated images depends on the scene complexity of the image, the kind of artifacts and implausibilities, if any, detectable in an image, the duration of visual attention to an image, and the quality of human effort to select appropriate prompts and curate images. A question such as ``How photorealistic are state-of-the-art diffusion models'' may sound simple, but the answer is more complex and depends on many details, including what images are generated and selected, how photorealism is measured, what real images are included in the experiment, and how much time, skill, and effort a human participant has and willing to offer. This paper offers an initial exploration into how we can address this question and develops a practical taxonomy that offers scaffolding for building AI--literacy interventions to help people navigate the capabilities and limitations of diffusion models and whether an image is AI-generated or not. 

\begin{acks}

This material is based upon work supported by Robert Pozen, and in part with funding from the Department of Defense (DoD). Any opinions, findings, conclusions, or recommendations expressed in this material are those of the authors and do not necessarily reflect the views of the DoD or any agency or entity of the United States Government. We thank Will Thompson from Kellogg Research Support for performing a replication check.
\end{acks}



\bibliographystyle{ACM-Reference-Format}
\bibliography{bibliography}

\appendix
\appendix
\beginsupplement
\clearpage
\onecolumn
\section{Further Methodological Details} \label{sec:appendix-methodol}
\FloatBarrier
\begin{figure*}[ht]
\captionsetup{justification=raggedright, singlelinecheck=false, skip=2pt}
\centering
\begin{subfigure}[t]{0.3\linewidth}
   
    \subcaption{}
    \includegraphics[width=\linewidth]{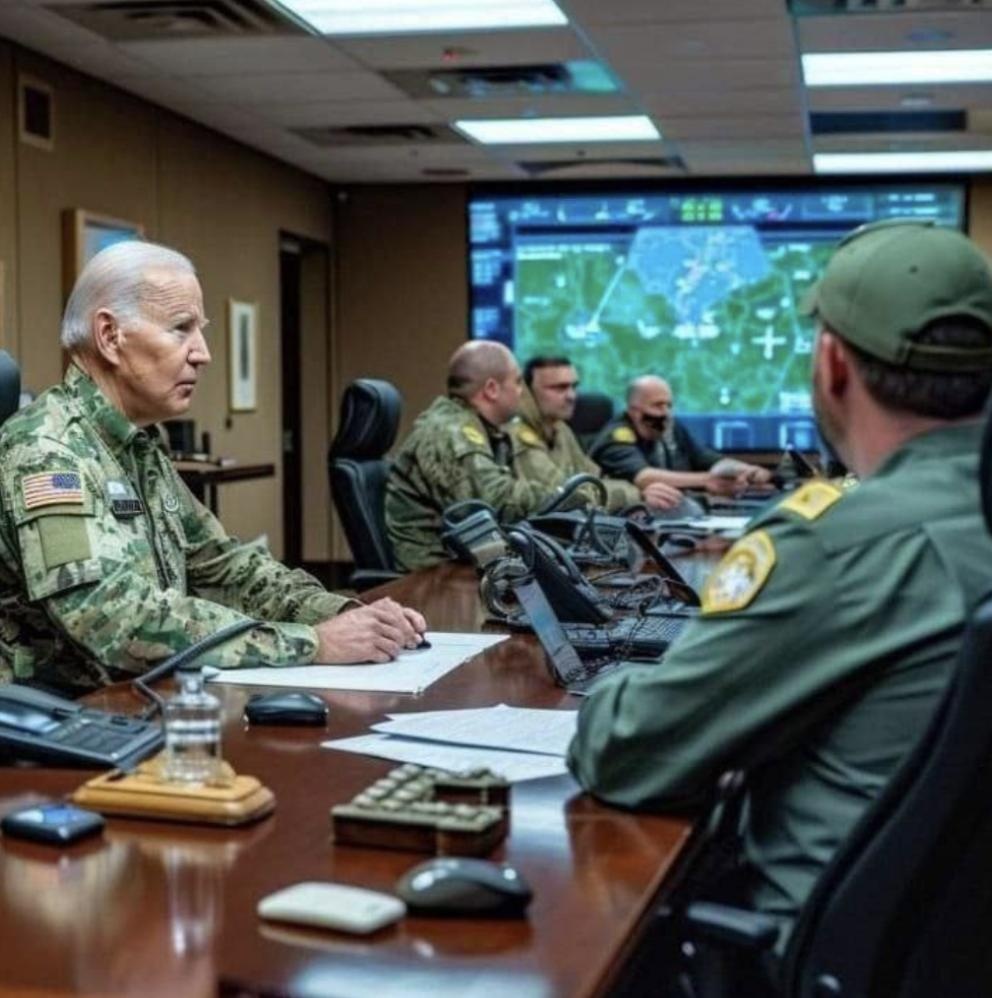}
\end{subfigure}
\hspace{1cm}
\begin{subfigure}[t]{0.31\linewidth}

   \subcaption{}
   \includegraphics[width=\linewidth]{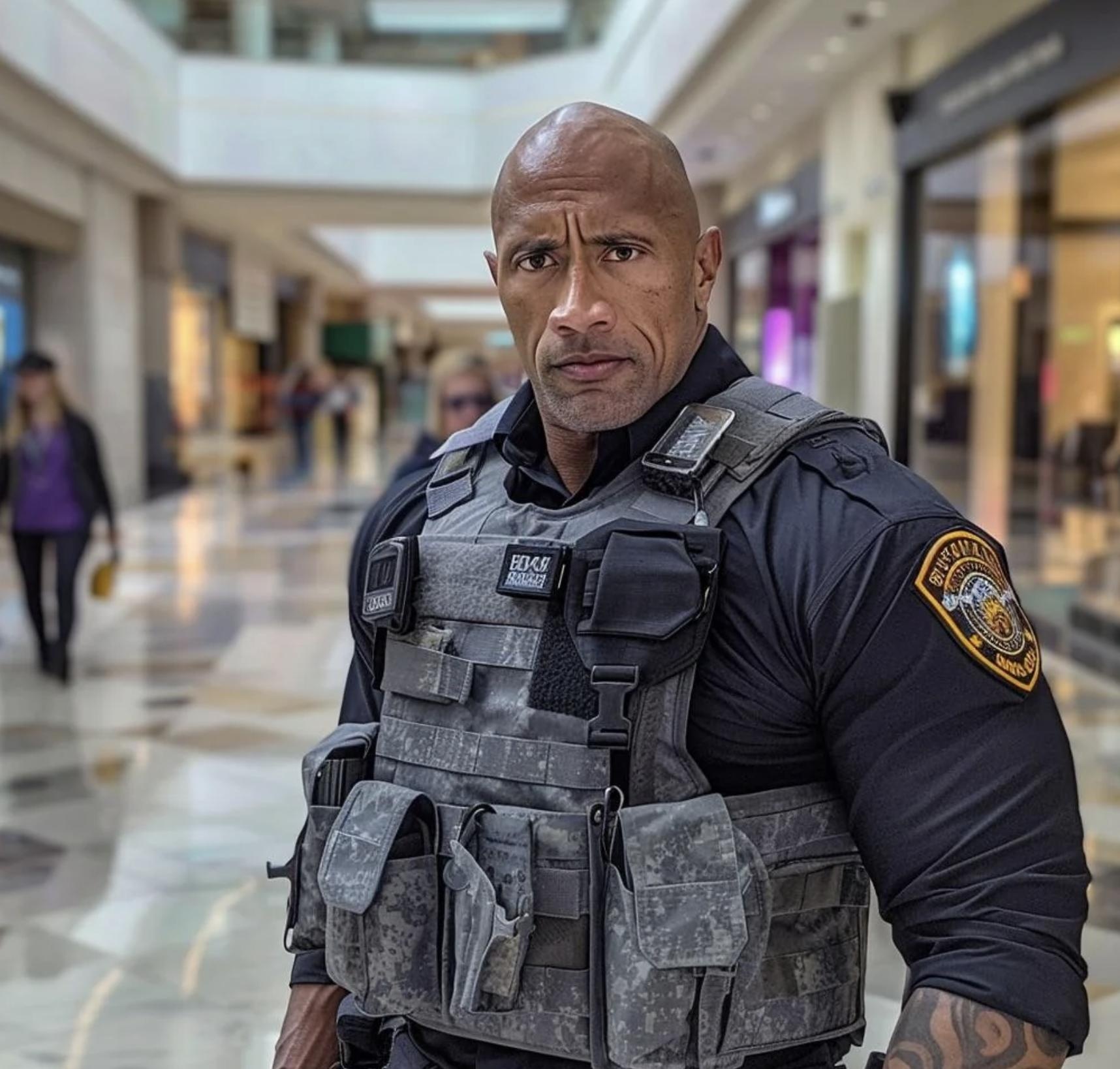}
\end{subfigure}
\caption{\mybold{AI-Generated Images from New York Times Quiz} \normalfont{\textbf{A}. NYT's explanation for evidence pointing to this image as AI-generated is: ``Though the resemblance to President Biden is striking, he would not be wearing military fatigues as a civilian.''~\cite{nytimes2024deepfake} \textbf{B}. NYT's explanation for evidence pointing to this image as AI-generated is ``One giveaway in this image is the badge, which includes garbled text.''~\cite{nytimes2024deepfake}}}
\label{fig:nytimes}
\Description{AI-generated image of Joe Biden in a conference room and an AI-generated image of the Rock in military uniform in a mall.}
\end{figure*}
\begin{figure*}[ht]
\centering
\captionsetup{justification=raggedright, singlelinecheck=false, skip=2pt}
\begin{subfigure}[t]{0.65\textwidth}
    \caption{}
    \vtop{\vskip0pt\hbox{\includegraphics[width=\linewidth]{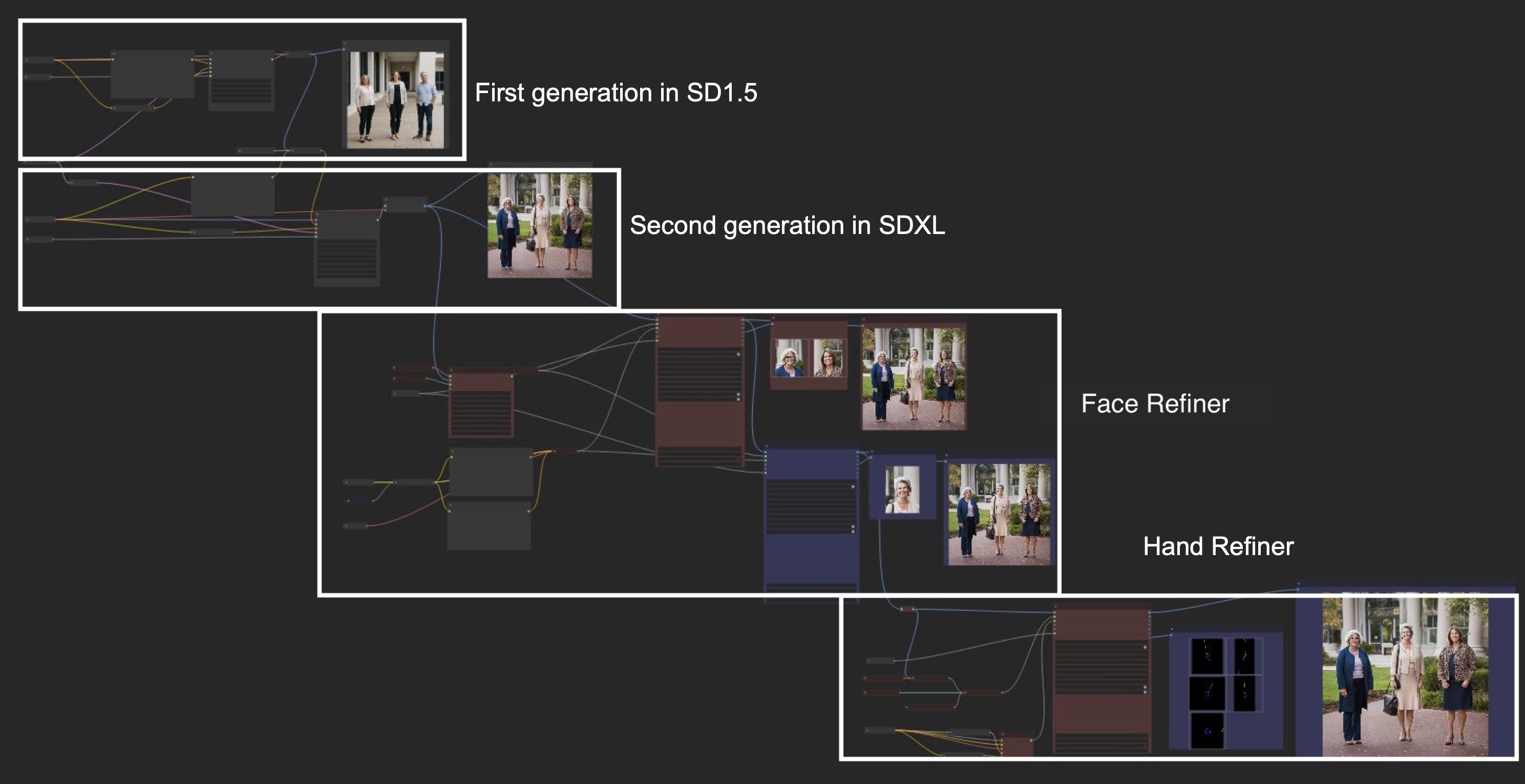}}}
\end{subfigure}
\hspace{0.01\textwidth} 
\begin{minipage}[t]{0.23\textwidth}
    \begin{subfigure}[t]{\textwidth}
        \caption{}
        \vtop{\vskip0pt\hbox{\includegraphics[width=\linewidth]{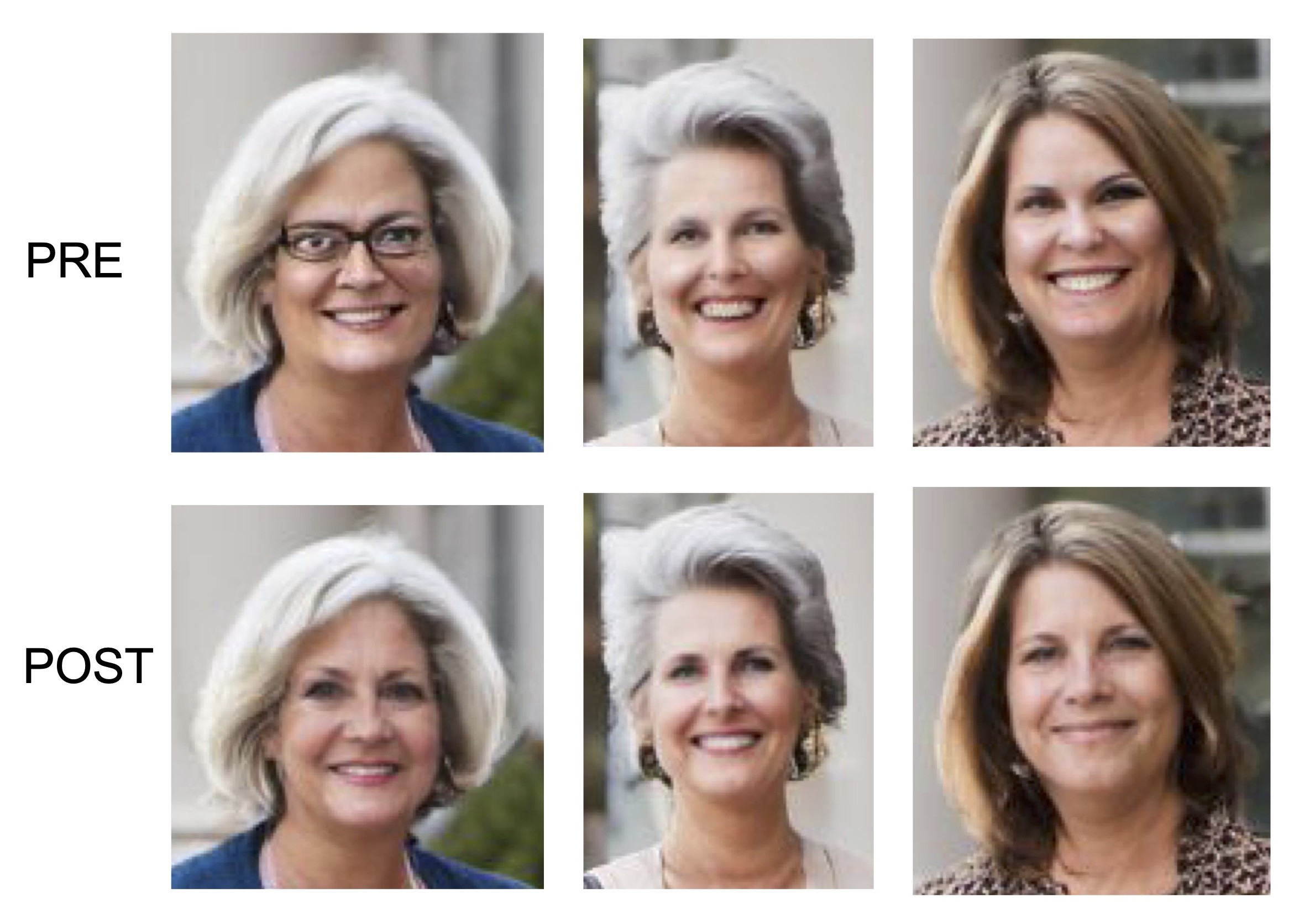}}}
    \end{subfigure}
    \vspace{0.5cm}
    \begin{subfigure}[t]{\textwidth}
        \caption{}\hfill\vtop{\vskip0pt\hbox{\includegraphics[width=0.8\linewidth]{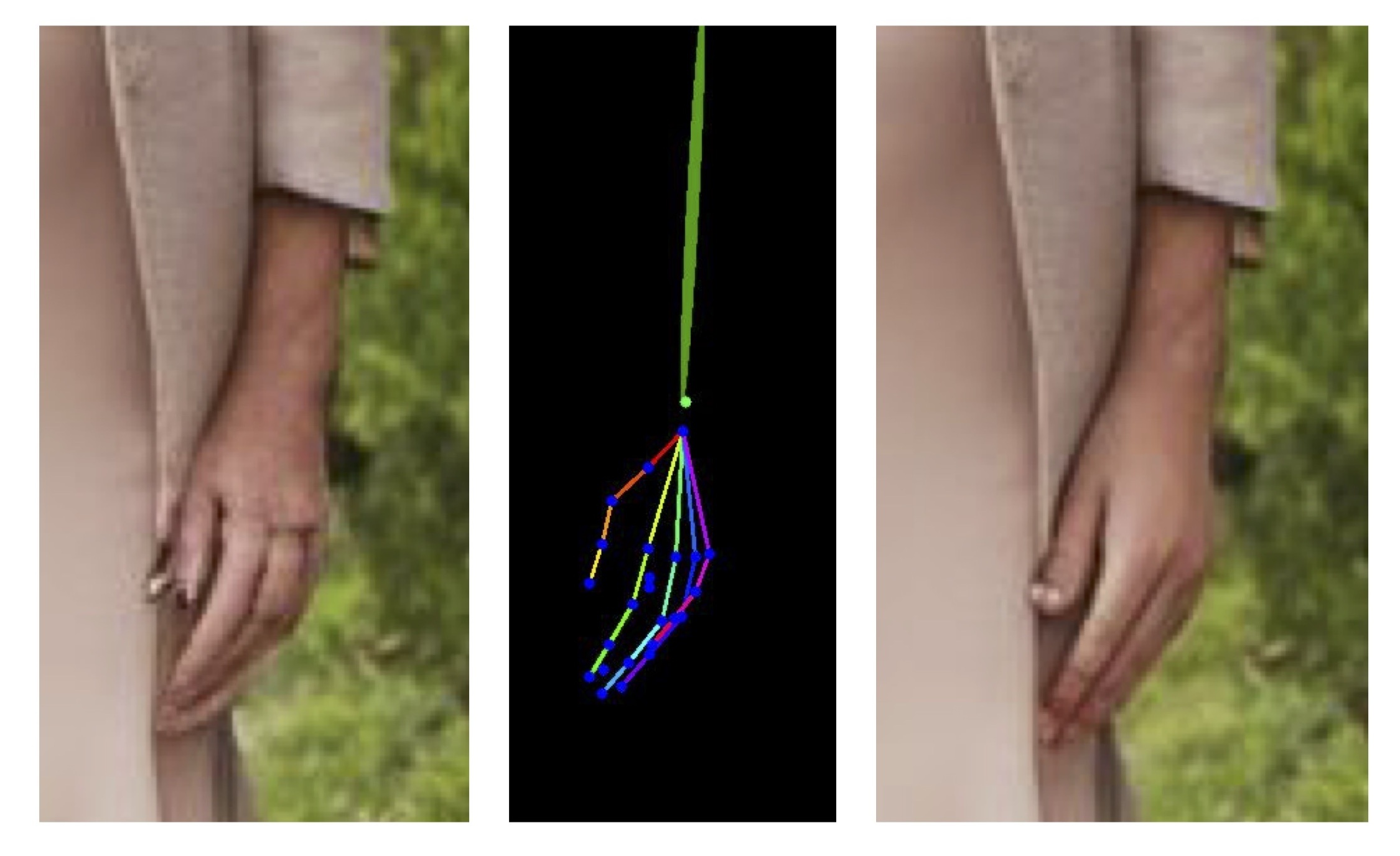}}}
    \end{subfigure}
\end{minipage}
 \caption{\textbf{Image generation process in Stable Diffusion} A. \normalfont{Four stage image generation pipeline where the image is first generated in SD1.5. The output image is then encoded as latent and upscaled to be re-generated in SDXL with ControlNets applied for pose consistency. This is passed to the face refiner \cite{comfyuiimpactpack} which detects dominant and background faces in the image via YOLOv8 \cite{yolov8} and re-generates them using an SDXL pipeline. Finally, the resulting image is passed to the hand refiner \cite{comfyuiimpactpack} which detects hands in the image via YOLOv8  and predicts the hand pose used to guide the re-generation of the hands. \textbf{B}. Faces in the image before and after the face refining process \textbf{C}. Hand refining process. The left image shows the initial generation of the hand. The center image shows a predicted skeleton for the hand that is used for a ControlNet that guides the re-generation of the hand shown in the image on the right.}}
\label{fig:refiningpipe}
\Description{4 stage image generation pipeline}
\end{figure*}
\FloatBarrier
\twocolumn
According to a New York Times (NYT) quiz, qualities that typically signify AI generation include missing fingers, misaligned eyes, repeated elements, and garbled or nonsensical details~\cite{nytimes2024deepfake}.  Examples are shown in \ref{fig:nytimes}. The NYT quiz also discusses qualities that may cause a real image to look AI-generated, such as repeated cropping and compression that often happens over social media.

A screenshot of the pipeline, along with images before and after refinement, is shown in Figure~\ref{fig:refiningpipe}.
\FloatBarrier

\begin{figure}[H]
\centering
\resizebox{0.85\textwidth}{!}{
\begin{minipage}{\textwidth}
\begin{subfigure}{0.22\linewidth}
    \includegraphics[width=\linewidth]{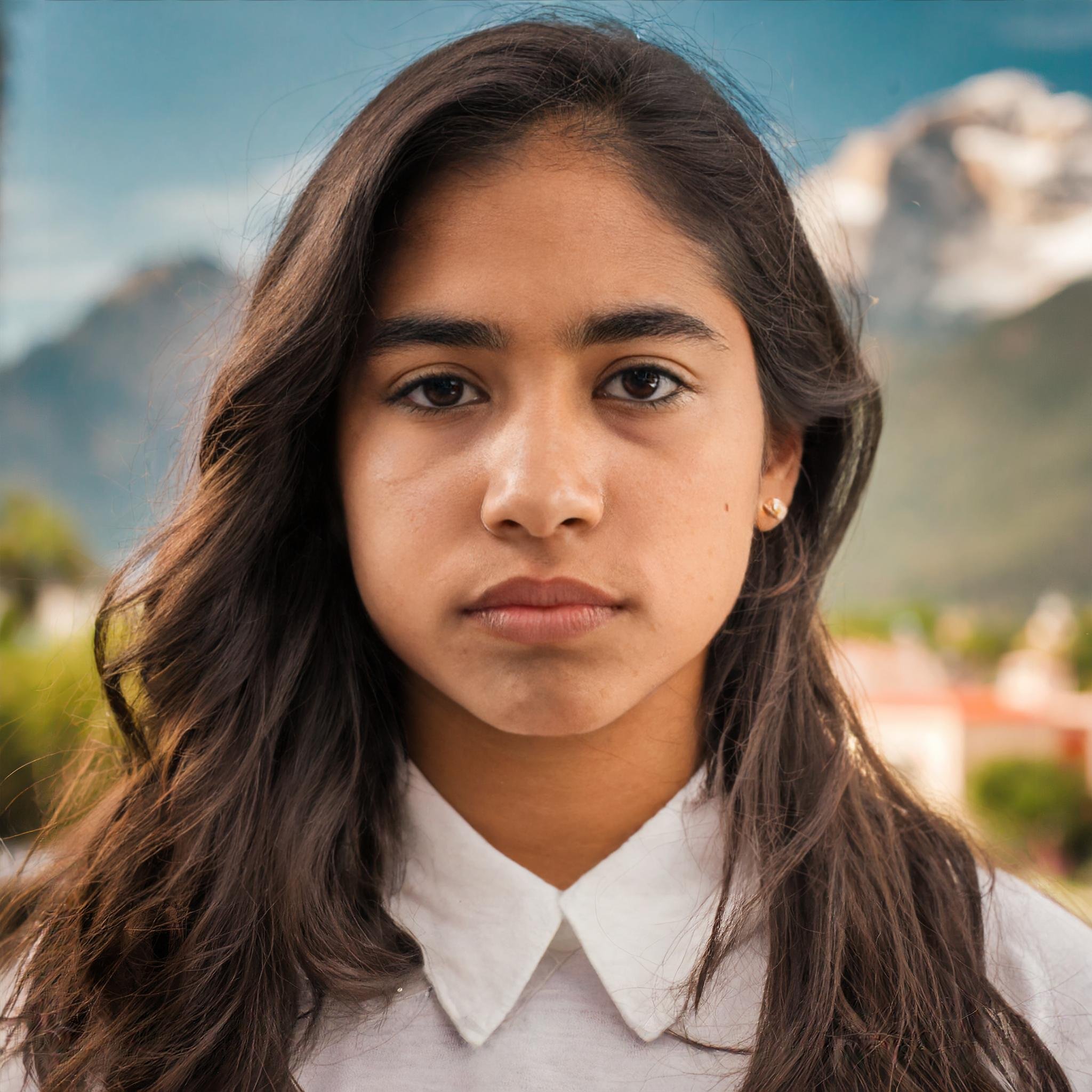}
    \caption{Acc: 25\%}
\end{subfigure}
\hfill
\begin{subfigure}{0.22\linewidth}
    \includegraphics[width=\linewidth]{sections/images/sd_portrait3_003.jpg}
    \caption{Acc: 37\%}
\end{subfigure}
\hfill
\begin{subfigure}{0.22\linewidth}
    \includegraphics[width=\linewidth]{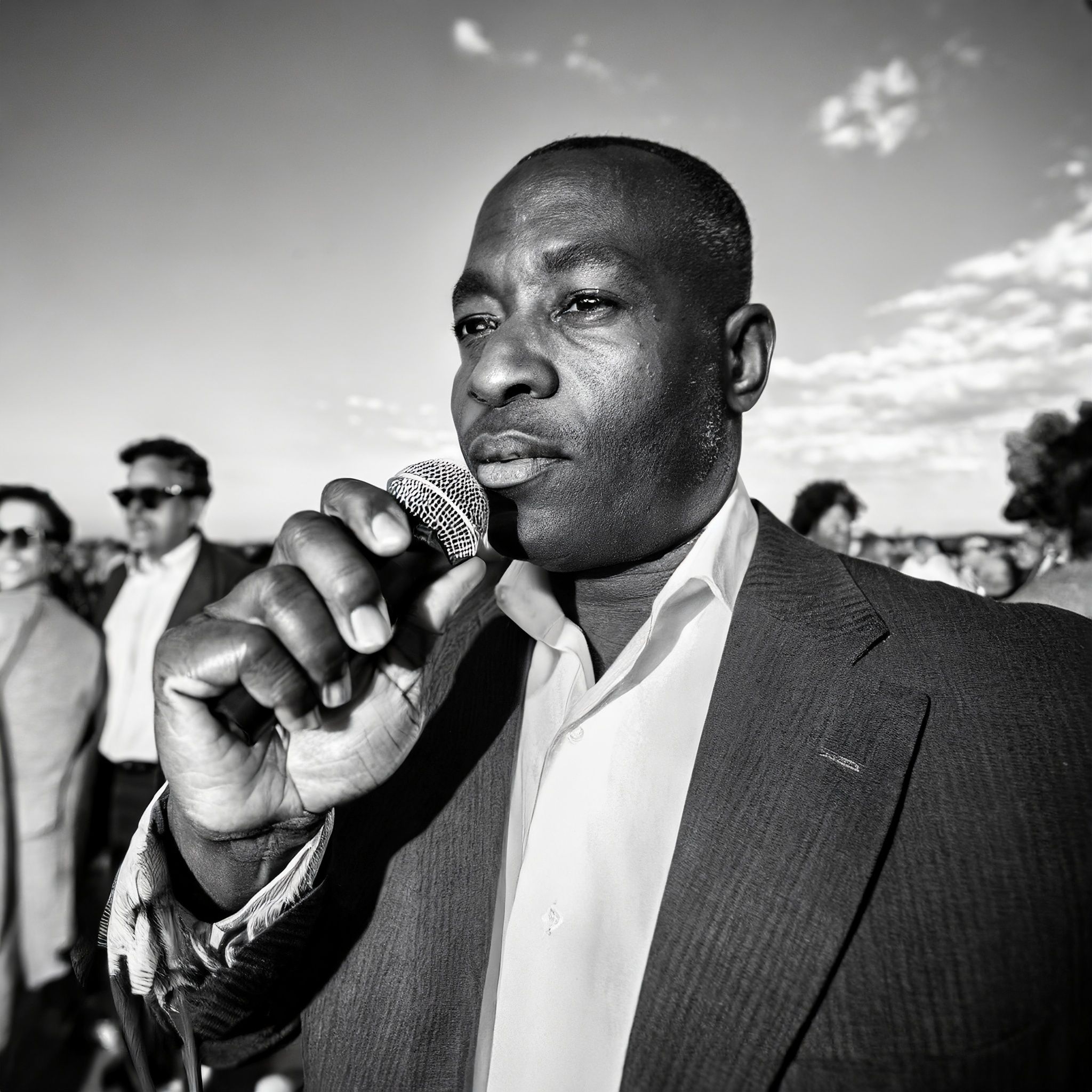}
    \caption{Acc: 66\%}
\end{subfigure}
\hfill
\begin{subfigure}{0.22\linewidth}
    \includegraphics[width=\linewidth]{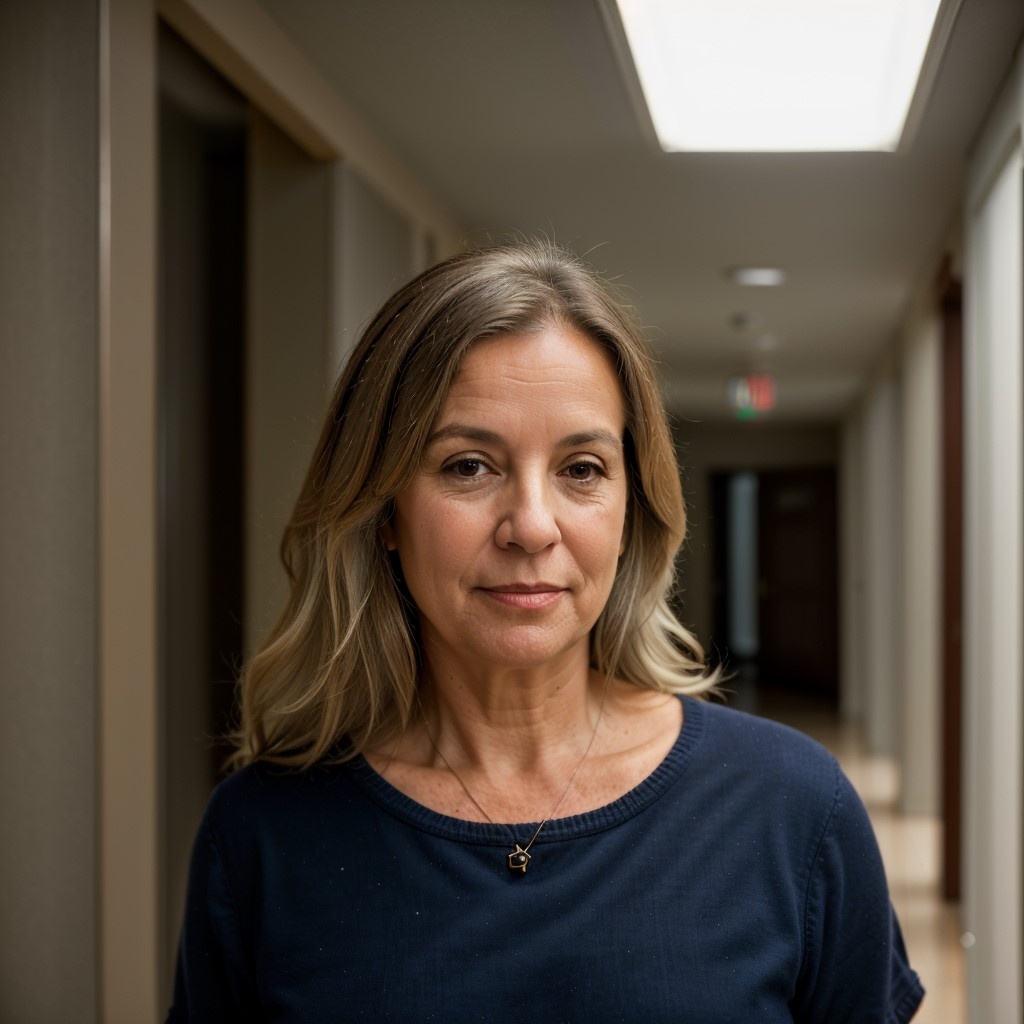}
    \caption{Acc: 80\%}
\end{subfigure}

\vspace{0.3cm}

\begin{subfigure}{0.22\linewidth}
    \includegraphics[width=\linewidth]{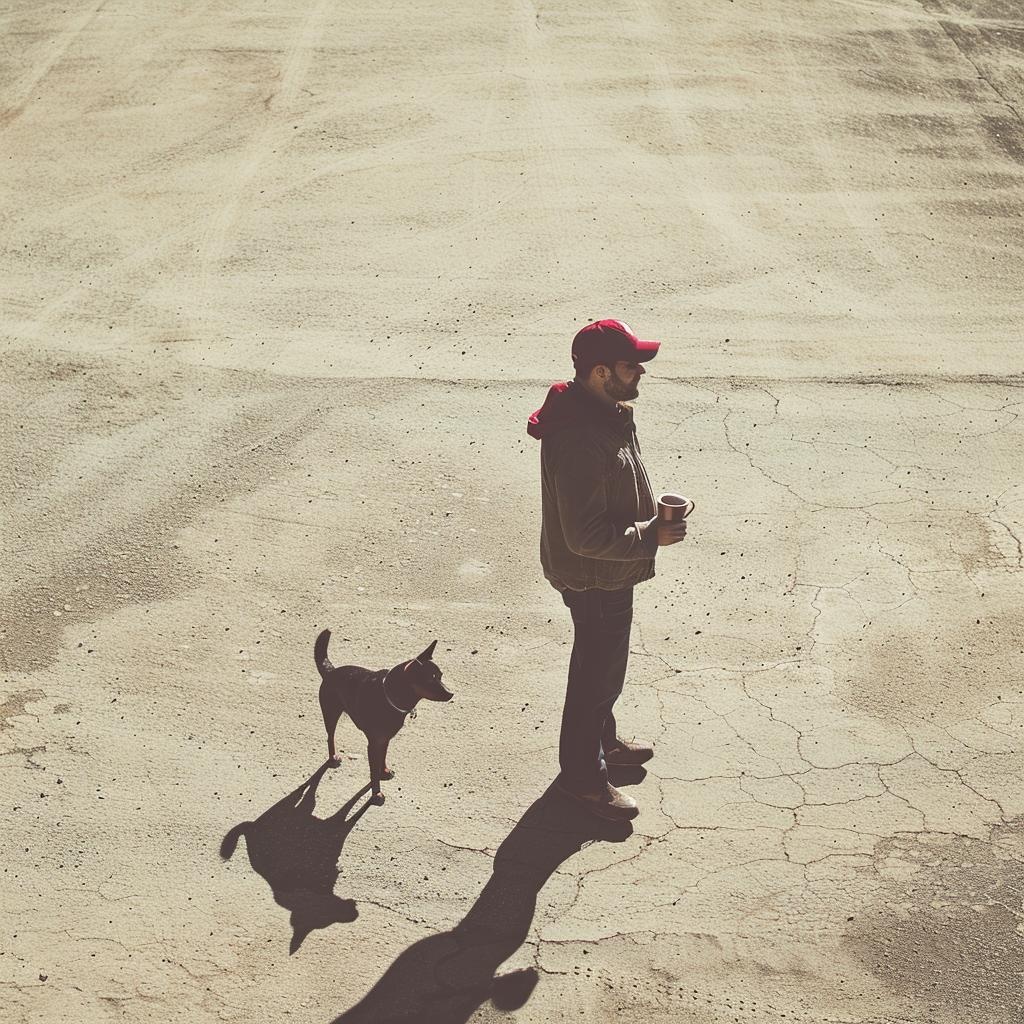}
    \caption{Acc: 37\%}
\end{subfigure}
\hfill
\begin{subfigure}{0.22\linewidth}
    \includegraphics[width=\linewidth]{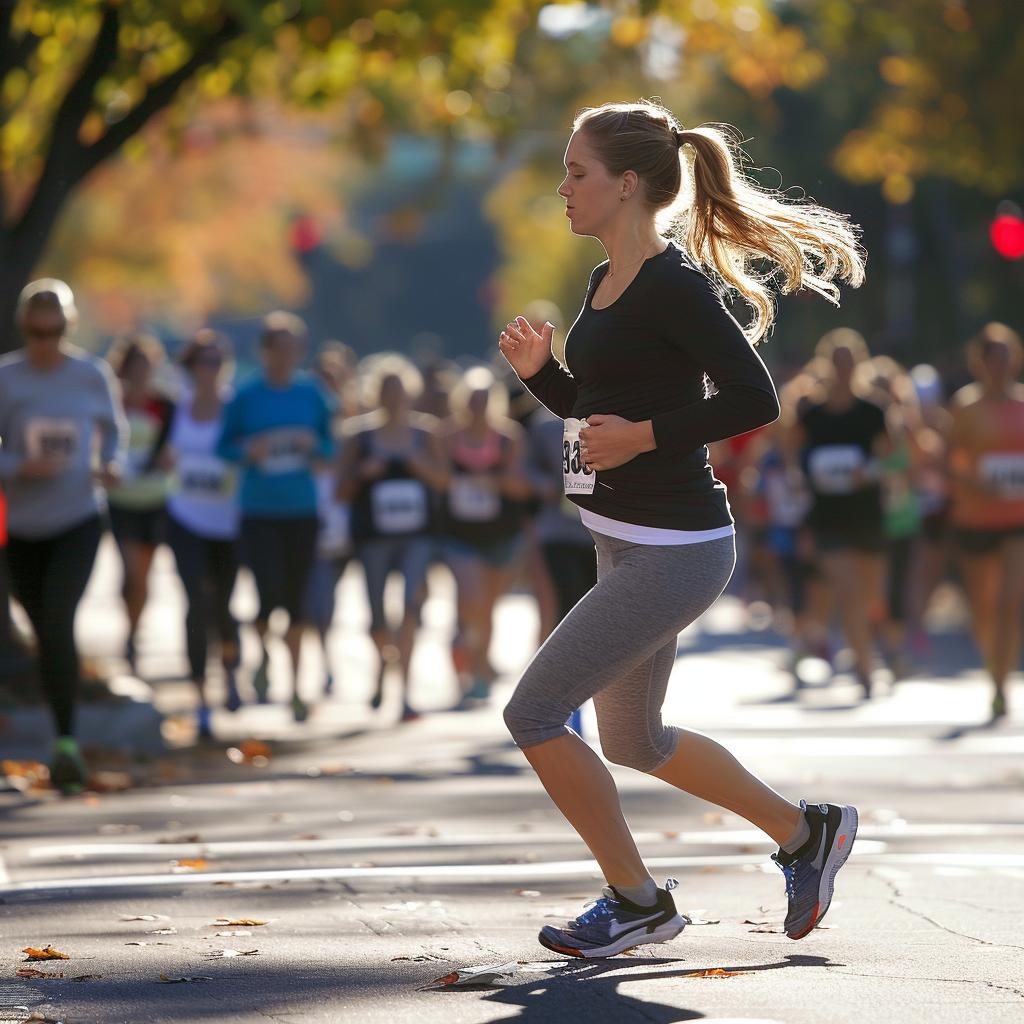}
    \caption{Acc: 57\%}
\end{subfigure}
\hfill
\begin{subfigure}{0.22\linewidth}
    \includegraphics[width=\linewidth]{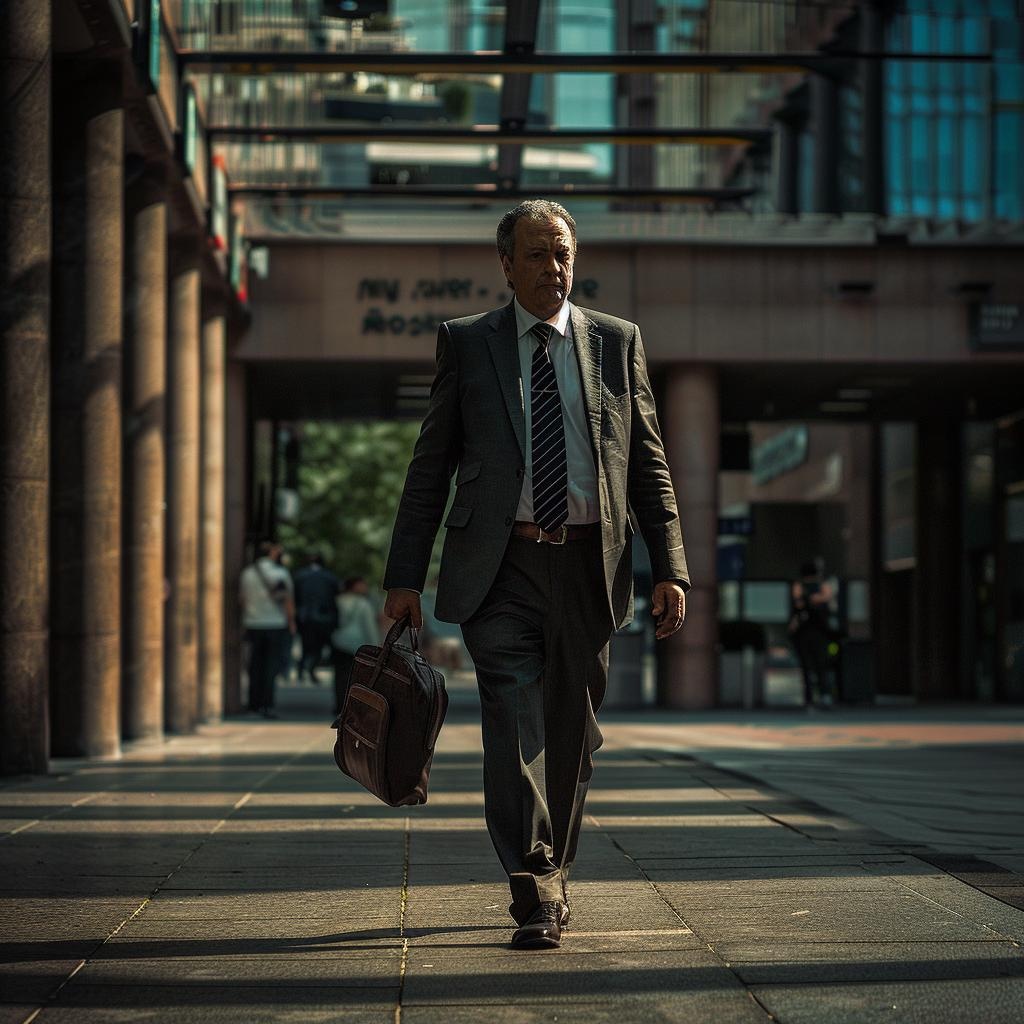}
    \caption{Acc: 66\%}
\end{subfigure}
\hfill
\begin{subfigure}{0.22\linewidth}
    \includegraphics[width=\linewidth]{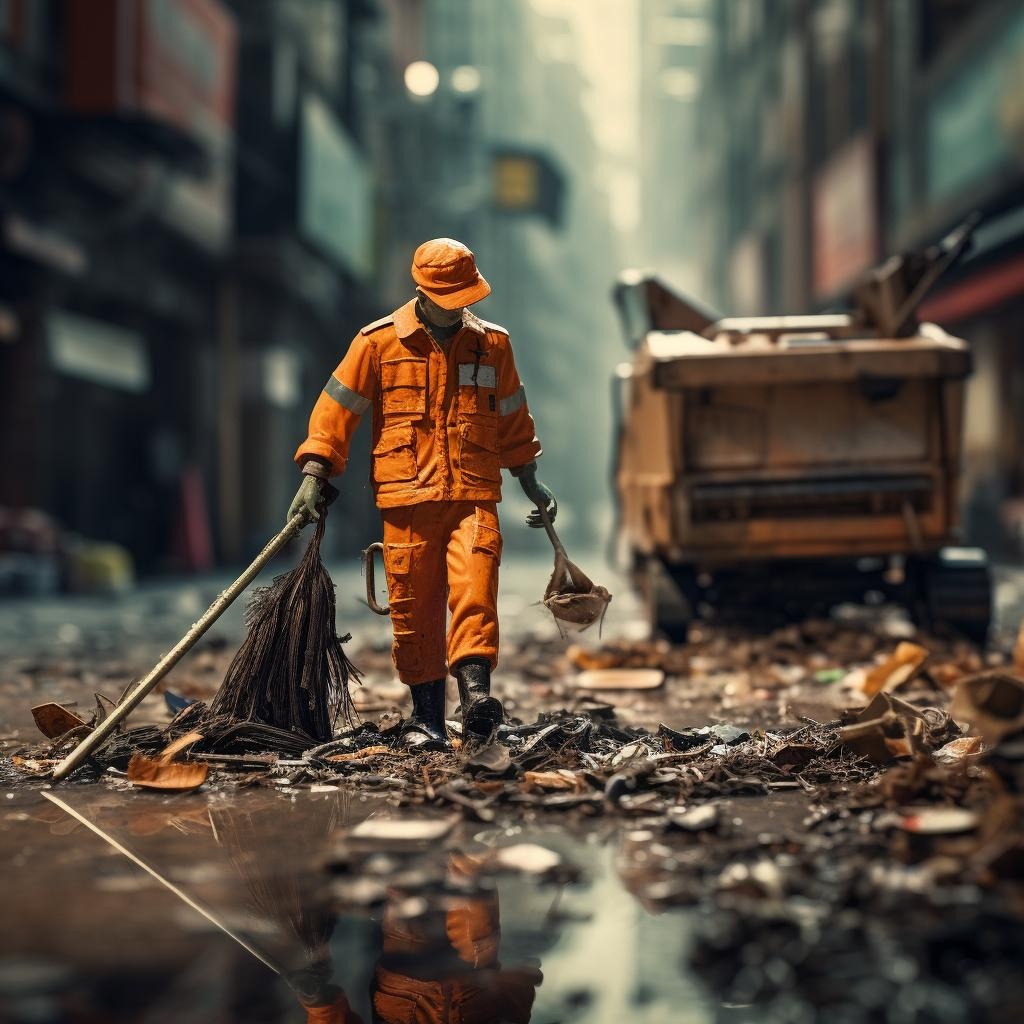}
    \caption{Acc: 83\%}
\end{subfigure}

\vspace{0.3cm}

\begin{subfigure}{0.22\linewidth}
    \includegraphics[width=\linewidth]{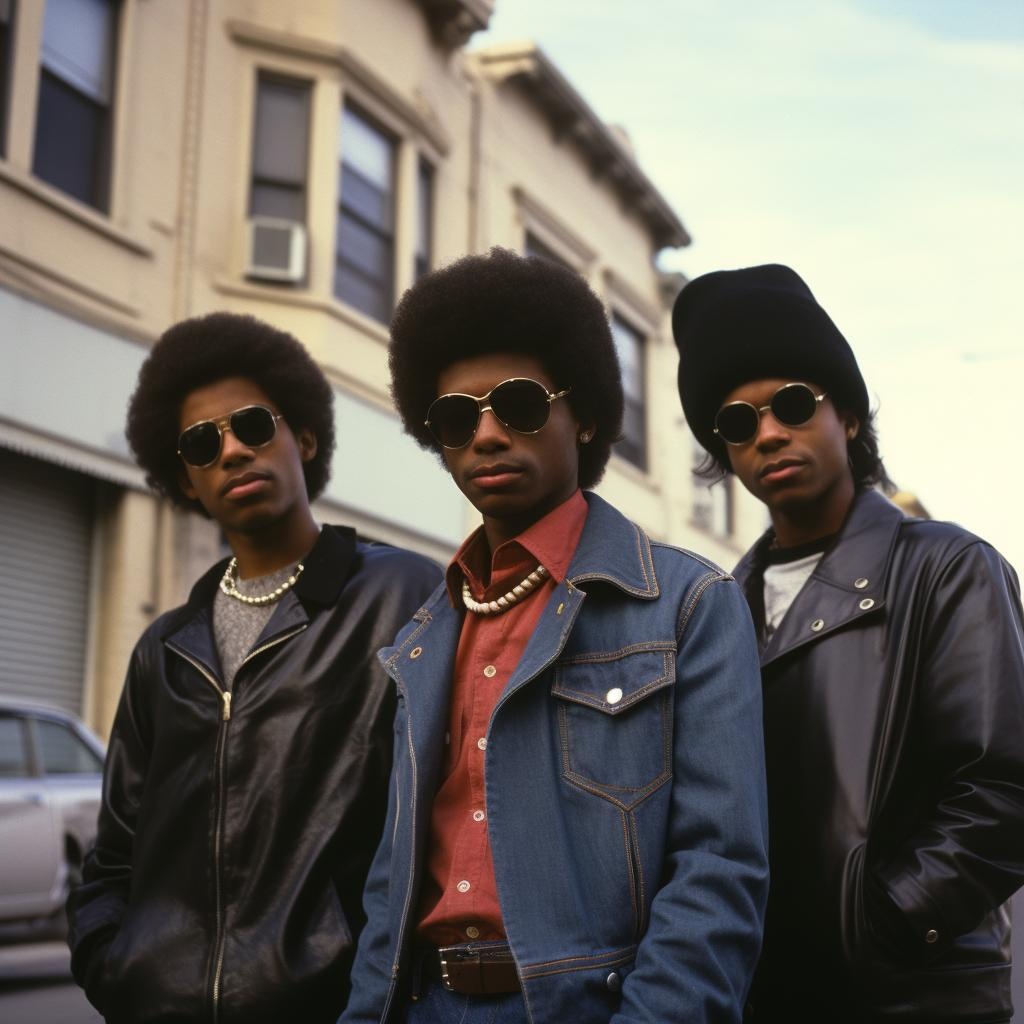}
    \caption{Acc: 37\%}
\end{subfigure}
\hfill
\begin{subfigure}{0.22\linewidth}
    \includegraphics[width=\linewidth]{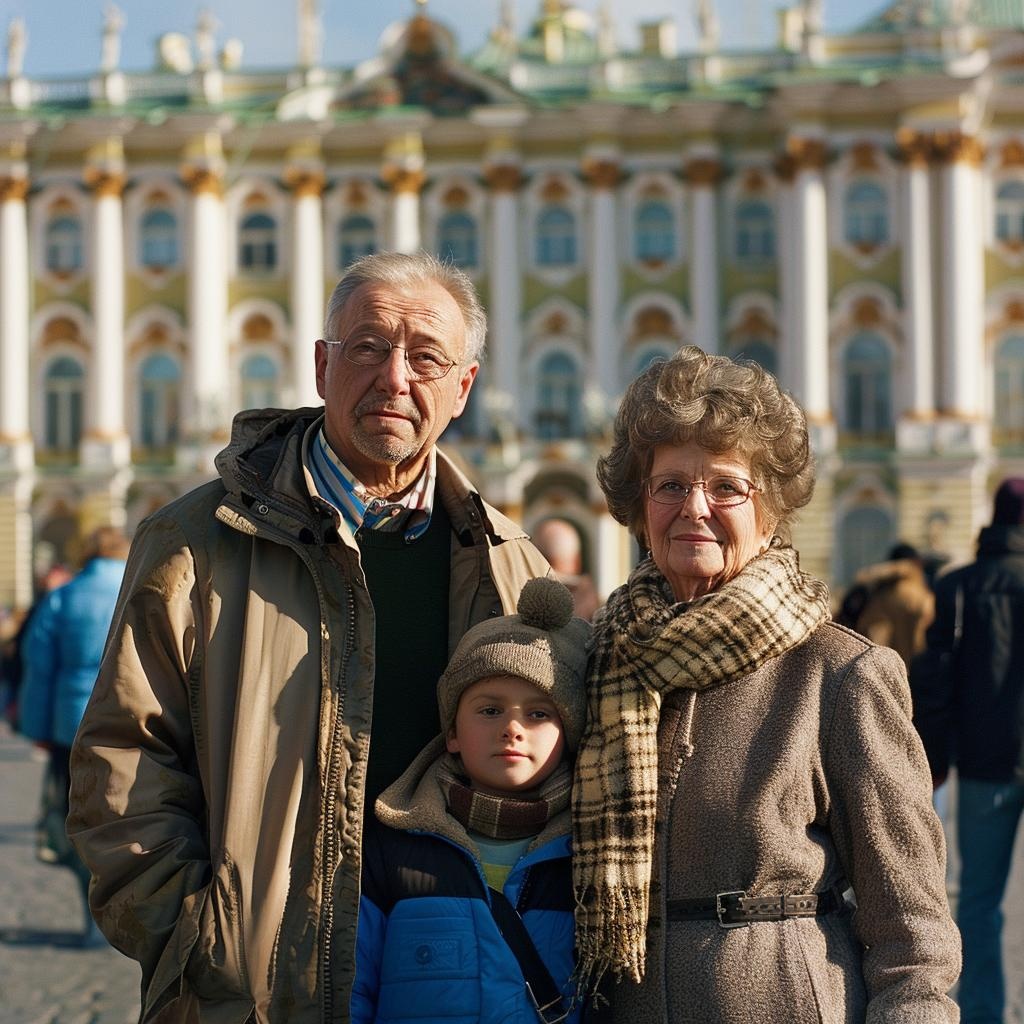}
    \caption{Acc: 57\%}
\end{subfigure}
\hfill
\begin{subfigure}{0.22\linewidth}
    \includegraphics[width=\linewidth]{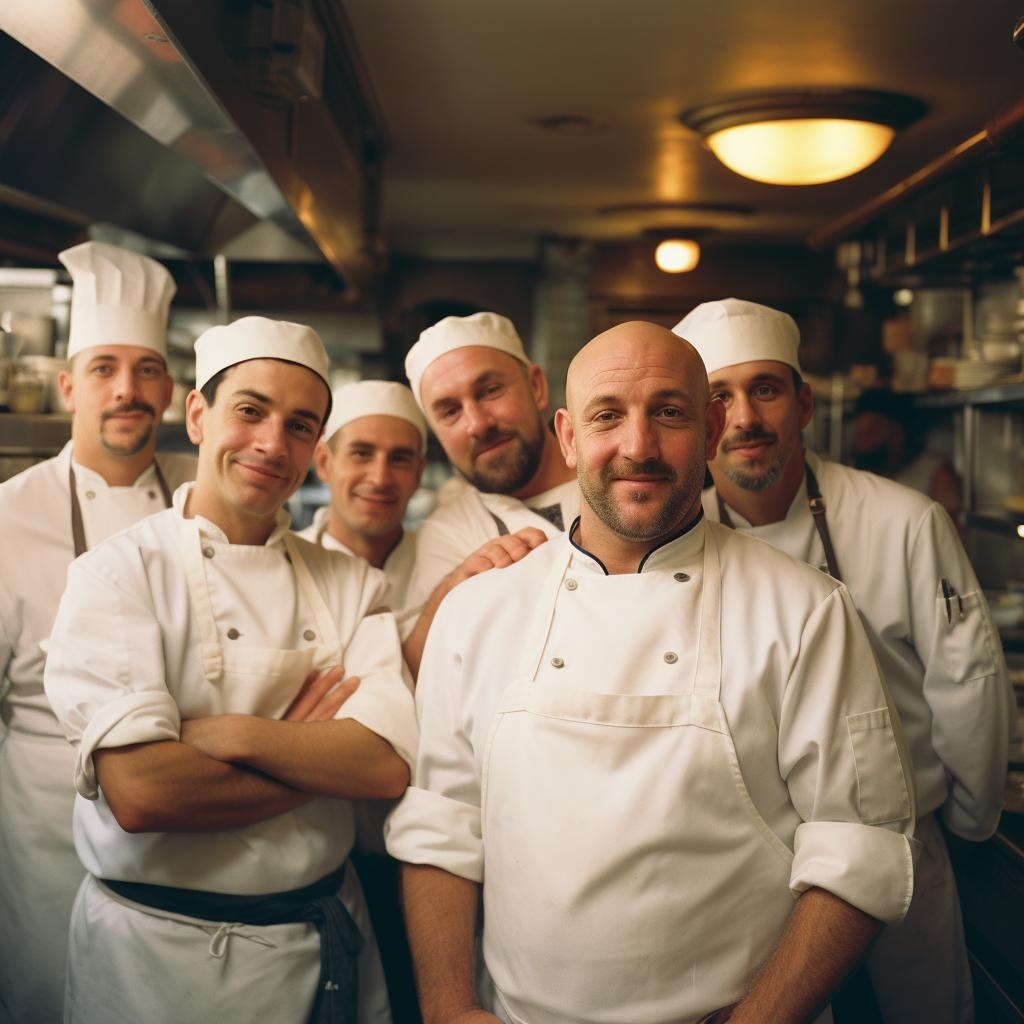}
    \caption{Acc: 66\%}
\end{subfigure}
\hfill
\begin{subfigure}{0.22\linewidth}
    \includegraphics[width=\linewidth]{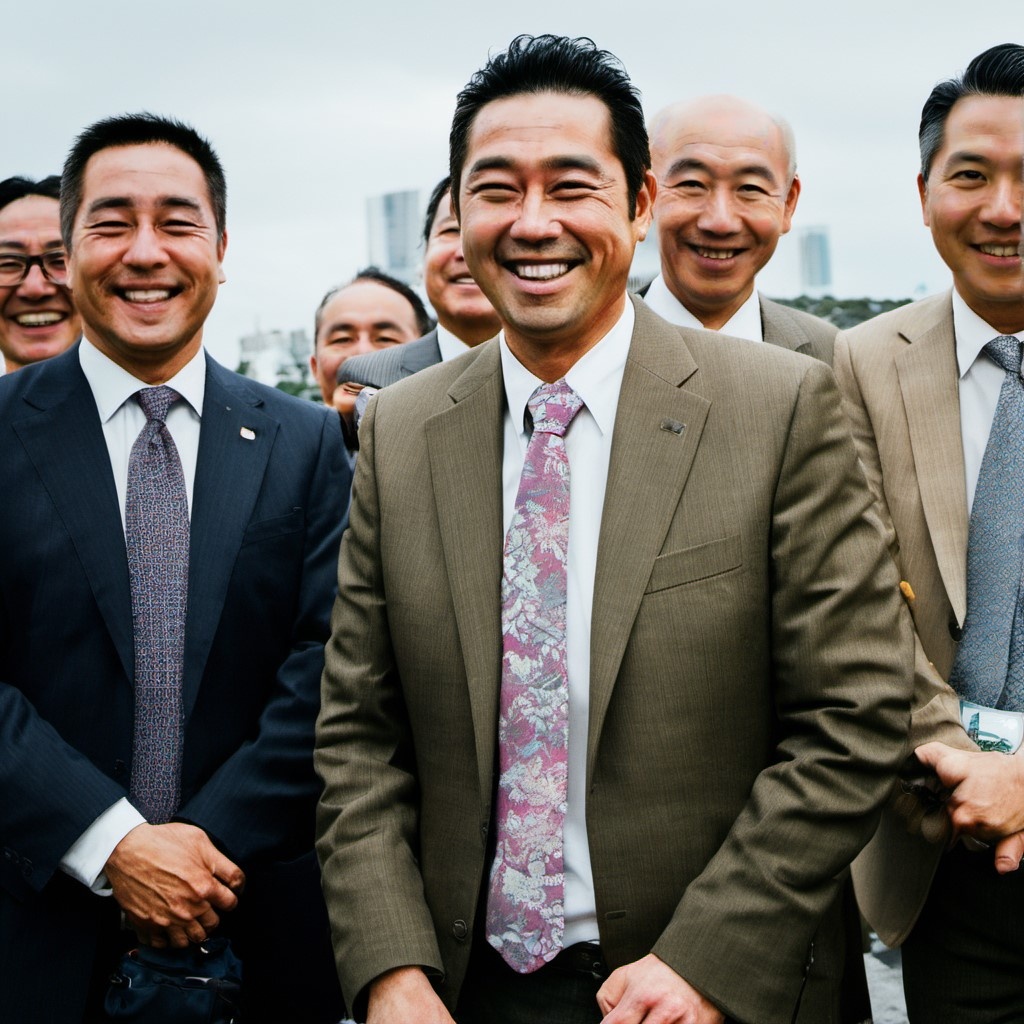}
    \caption{Acc: 83\%}
\end{subfigure}

\vspace{0.3cm}

\begin{subfigure}{0.22\linewidth}
    \includegraphics[width=\linewidth]{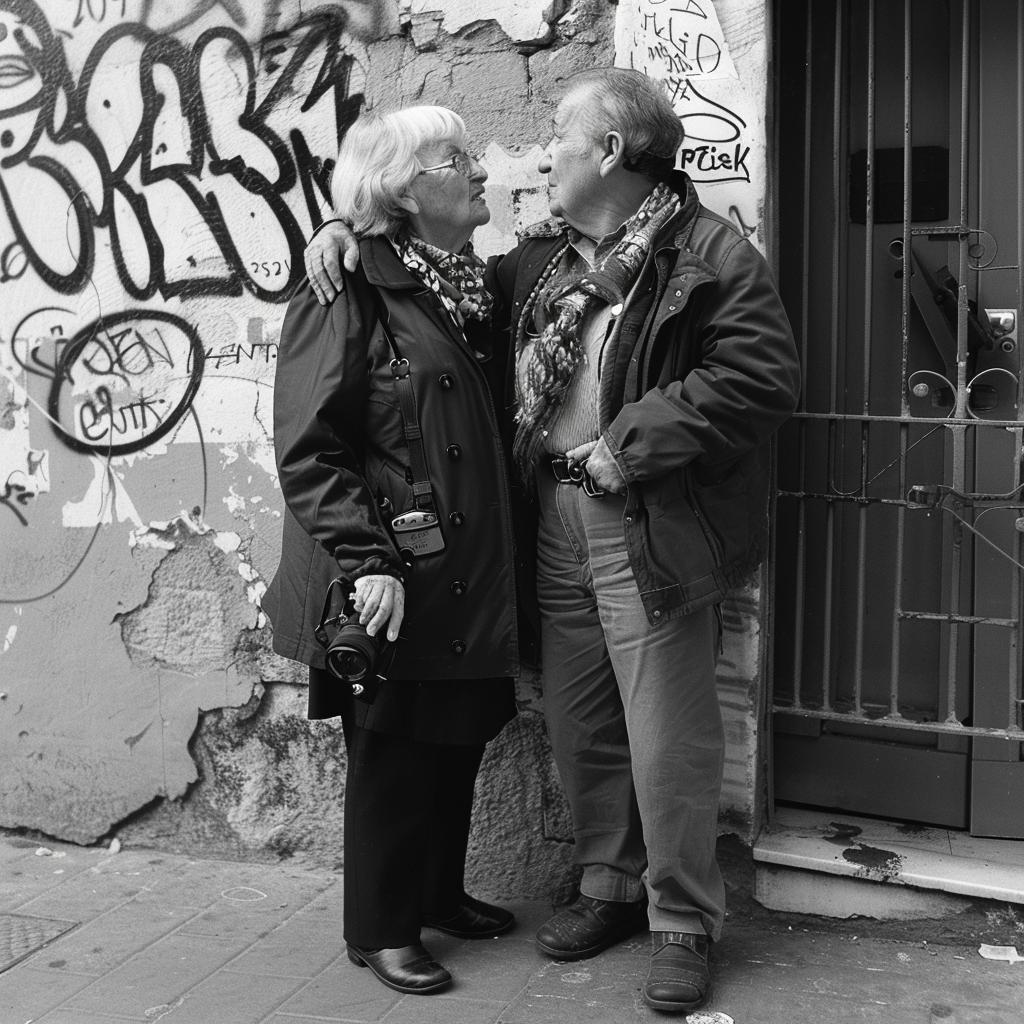}
    \caption{Acc: 31\%}
\end{subfigure}
\hfill
\begin{subfigure}{0.22\linewidth}
    \includegraphics[width=\linewidth]{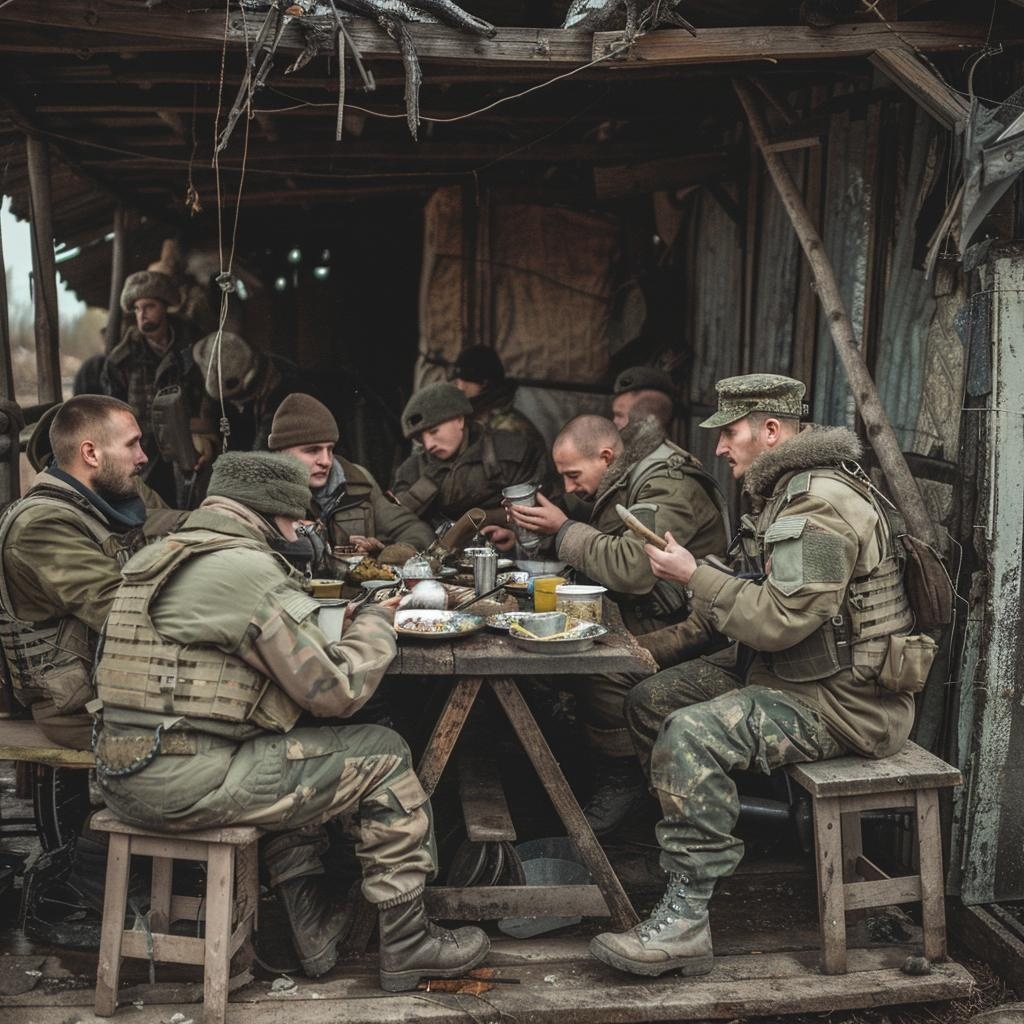}
    \caption{Acc: 66\%}
\end{subfigure}
\hfill
\begin{subfigure}{0.22\linewidth}
    \includegraphics[width=\linewidth]{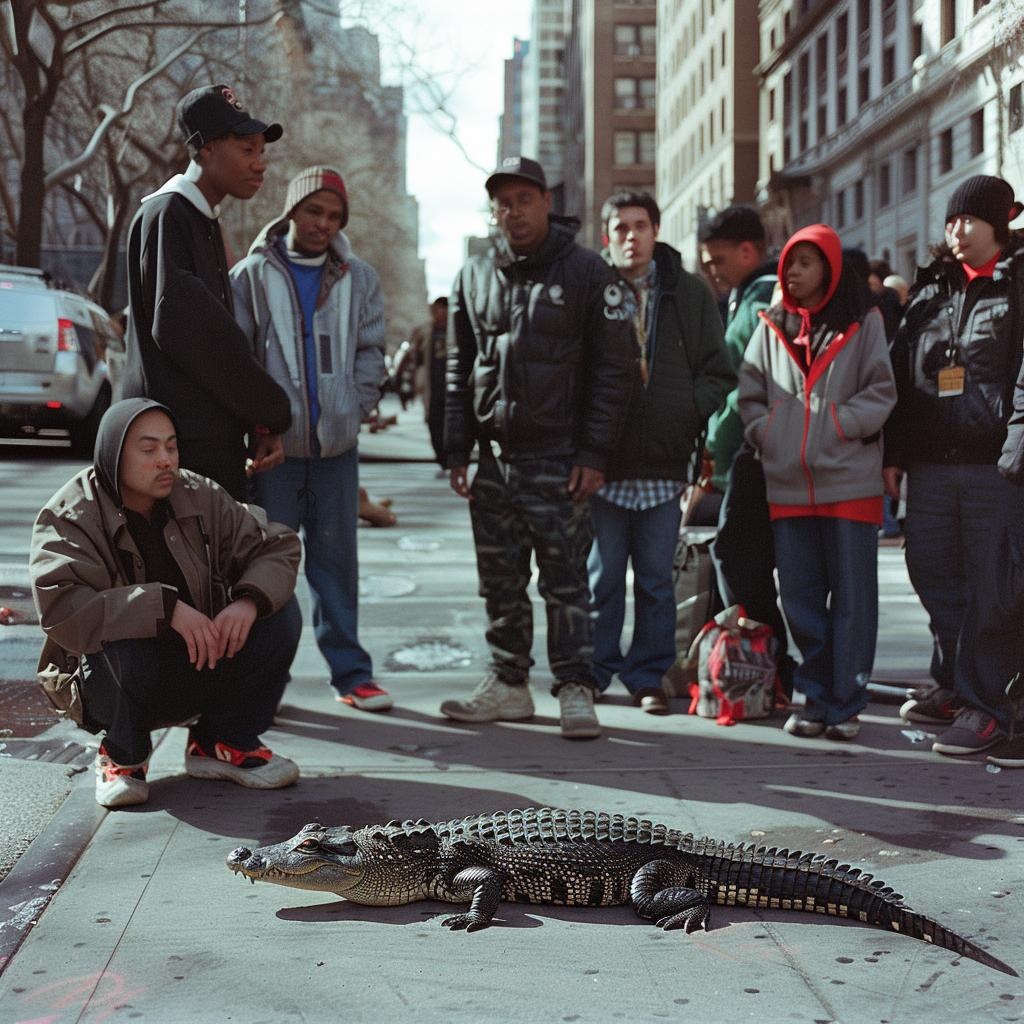}
    \caption{Acc: 75\%}
\end{subfigure}
\hfill
\begin{subfigure}{0.22\linewidth}
    \includegraphics[width=\linewidth]{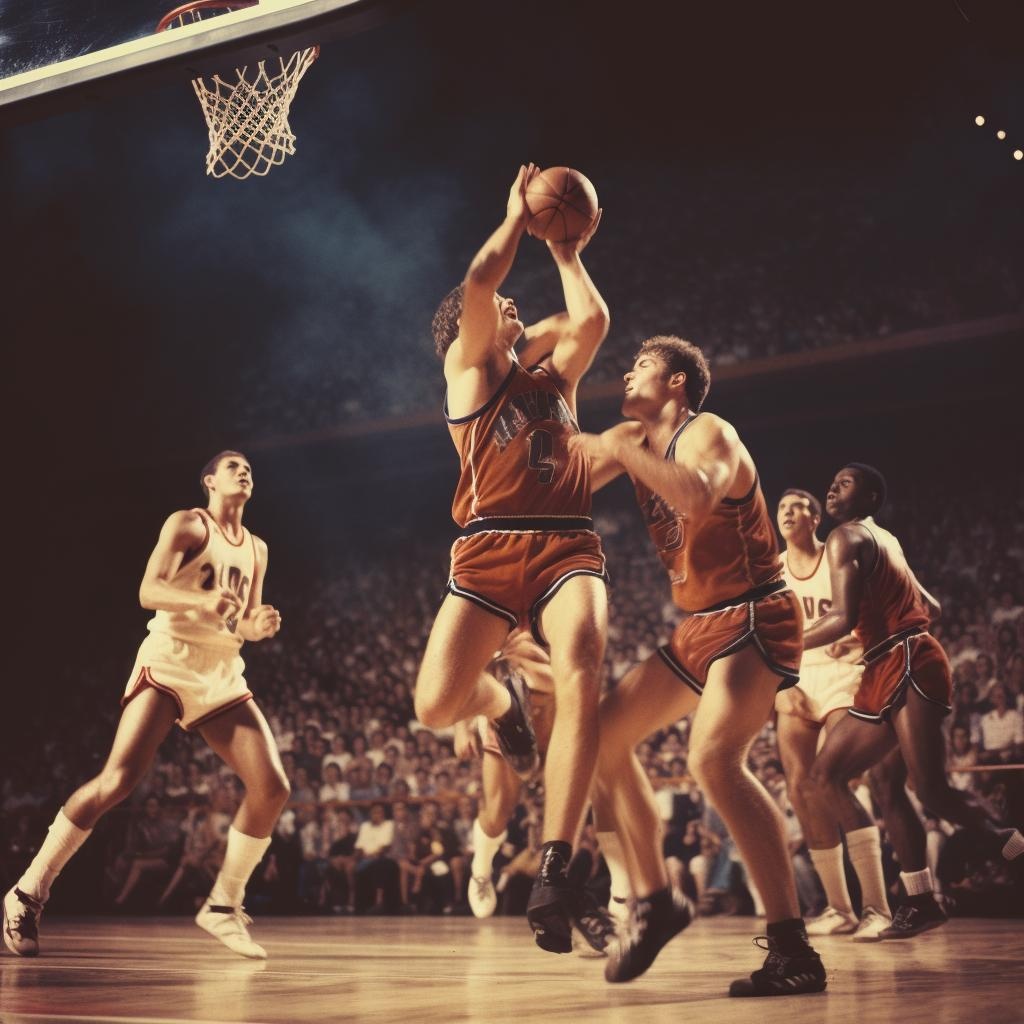}
    \caption{Acc: 87\%}
\end{subfigure}

\end{minipage}
}
\vspace{-2mm}
\caption{\textbf{More examples of the four pose complexities and their average accuracies.} \normalfont{The first row shows Portraits, the second row Full Body images, the third row Posed Groups, and the last row Candid Groups.}}
\label{fig:pose-comprehensive}
\Description{Examples of AI-generated images in different pose complexities: Portraits, Full Body, Posed Groups, and Candid Groups, with participant accuracy percentages.}
\end{figure}

\FloatBarrier
Figure~\ref{fig:pose-comprehensive} displays more examples of the four pose complexities and their average accuracies.
\onecolumn
\FloatBarrier
\twocolumn
\clearpage
\subsection{Robustness Check: Dataset Comparison}

To ensure the validity of our conclusions, we conducted a robustness check comparing the results from our full dataset against a subset excluding data collected before May 10th, 2024. This comparison addresses potential biases introduced by the initial experimental design, which did not implement stratified randomization as mentioned in Section \ref{exp-design}.

Table~\ref{tab:accuracy-comparison-dataset} presents the accuracy metrics for both the full dataset and the dataset excluding pre-May 10th data. The table includes overall accuracy, as well as specific accuracy for AI-generated and real images, along with their respective 95\% confidence intervals.

\begin{table}[H]
\centering
\caption{Comparison of accuracy: Full Dataset vs. Dataset excluding data before May 10th}
\label{tab:accuracy-comparison-dataset}
\resizebox{\linewidth}{!}{
\begin{tabular}{lcccccc}
\hline
Dataset & \multicolumn{2}{c}{Overall} & \multicolumn{2}{c}{AI-generated} & \multicolumn{2}{c}{Real} \\
 & Accuracy & 95\% CI & Accuracy & 95\% CI & Accuracy & 95\% CI \\
\hline
Full Dataset & 0.75 & [0.74, 0.76] & 0.76 & [0.74, 0.77] & 0.73 & [0.71, 0.75] \\
Dataset excluding data before May 10th & 0.75 & [0.74, 0.76] & 0.76 & [0.75, 0.77] & 0.7201 & [0.70, 0.74] \\
\hline
\end{tabular}}
\Description{A robustness check by comparing accuracy in full dataset vs. dataset excluding data before May 10th}
\end{table}

\begin{figure}[H]
\centering
\includegraphics[width=\linewidth]{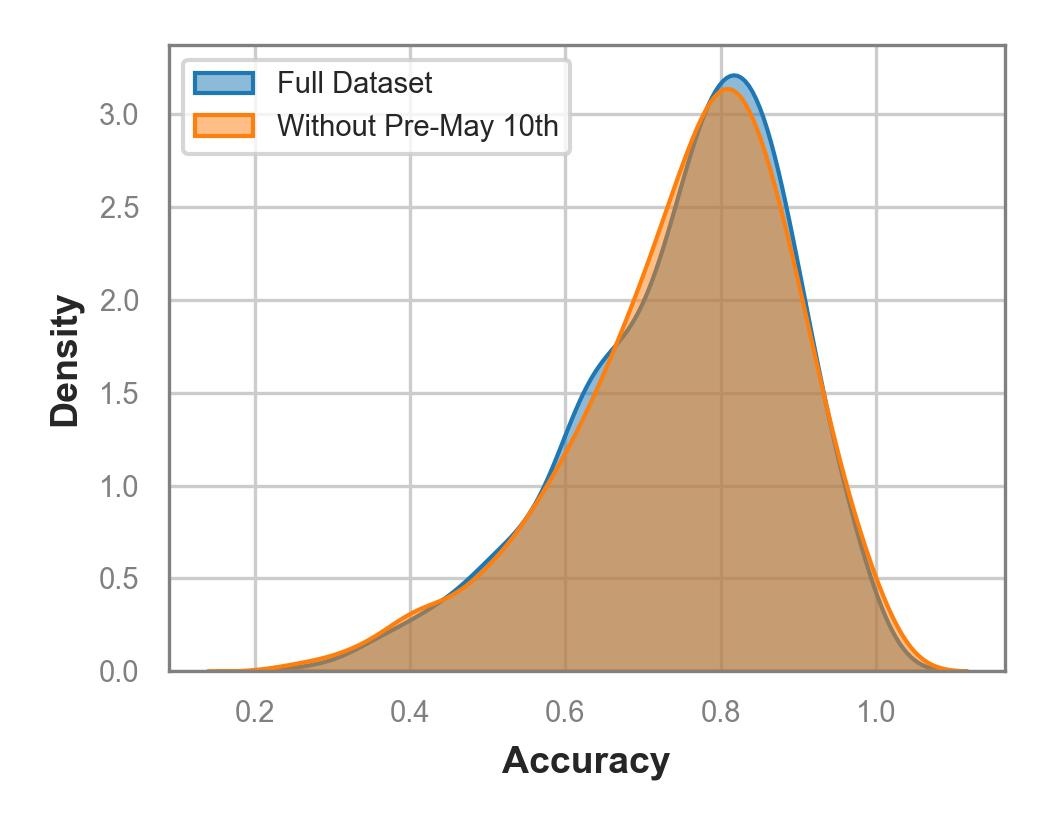}
\vspace{-10mm}
\caption{Comparison of accuracies between the full dataset and the dataset excluding data before 10th data.}
\label{fig:accuracy-comparison-dstaset}
\Description{Figure compares the distribution of accuracy of images from the full dataset vs. from the dataset excluding data before May 10th}
\end{figure}

Figure~\ref{fig:accuracy-comparison-dstaset} visualizes the distribution of image accuracies for both datasets. This comparison allows for direct observation of any potential shifts in accuracy distributions between the full dataset and the subset, excluding early data.
 This robustness check supports the validity of using the full dataset in our main analysis.

\clearpage
\onecolumn
\section{Curated and Uncurated AI-generated Images}

\begin{figure*}[!htb]
\centering
\resizebox{1.0\textwidth}{!}{ 
\begin{minipage}{\textwidth} 
\captionsetup{justification=raggedright, singlelinecheck=false, skip=2pt}


\begin{subfigure}[t]{0.23\linewidth}  
\subcaption{}
\vtop{\vskip0pt\hbox{\includegraphics[width=\linewidth]{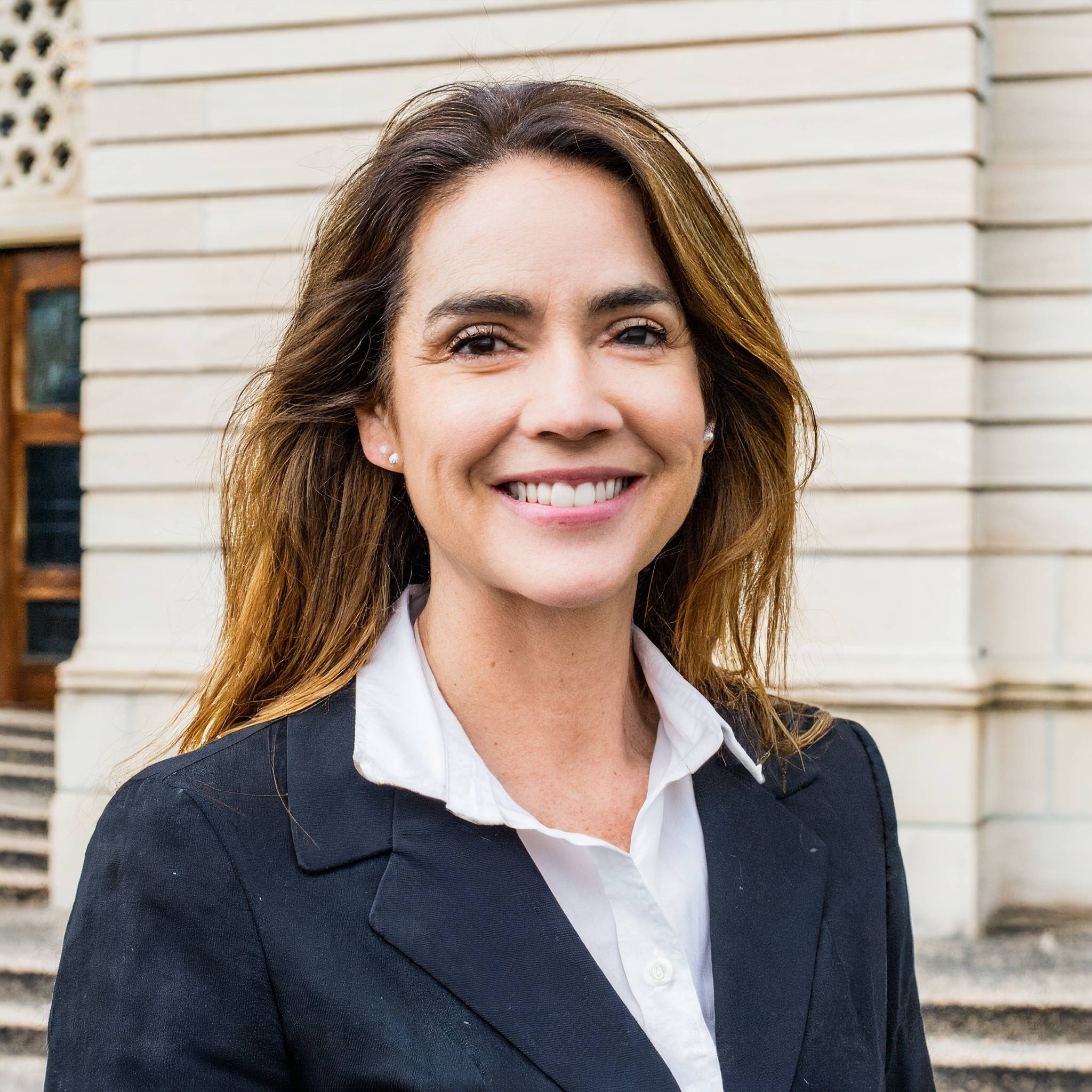}}}
\end{subfigure}
\hfill
\begin{subfigure}[t]{0.23\linewidth}  
\subcaption{}
\vtop{\vskip0pt\hbox{\includegraphics[width=\linewidth]{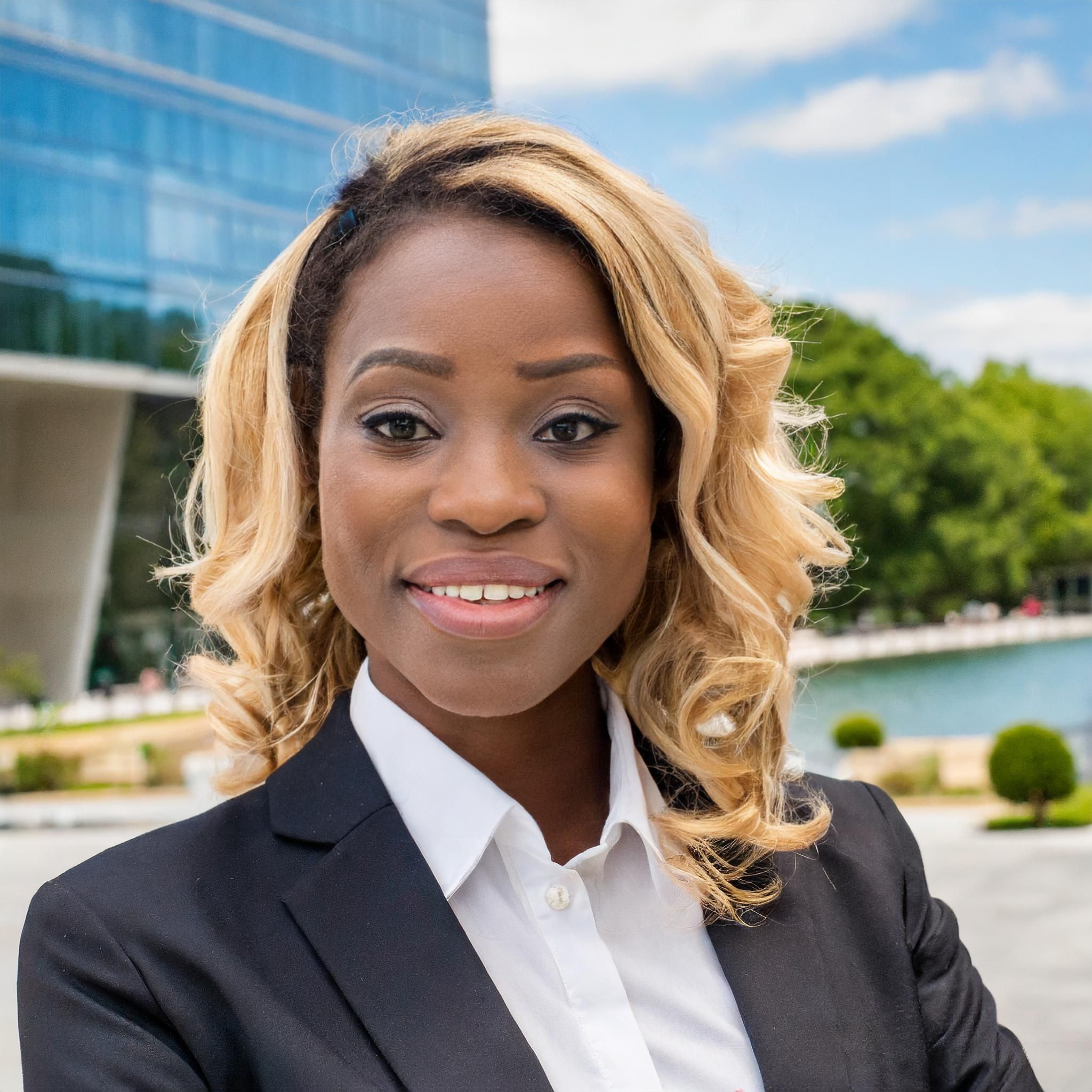}}}
\end{subfigure}
\hfill
\begin{subfigure}[t]{0.23\linewidth}  
\subcaption{}
\vtop{\vskip0pt\hbox{\includegraphics[width=\linewidth]{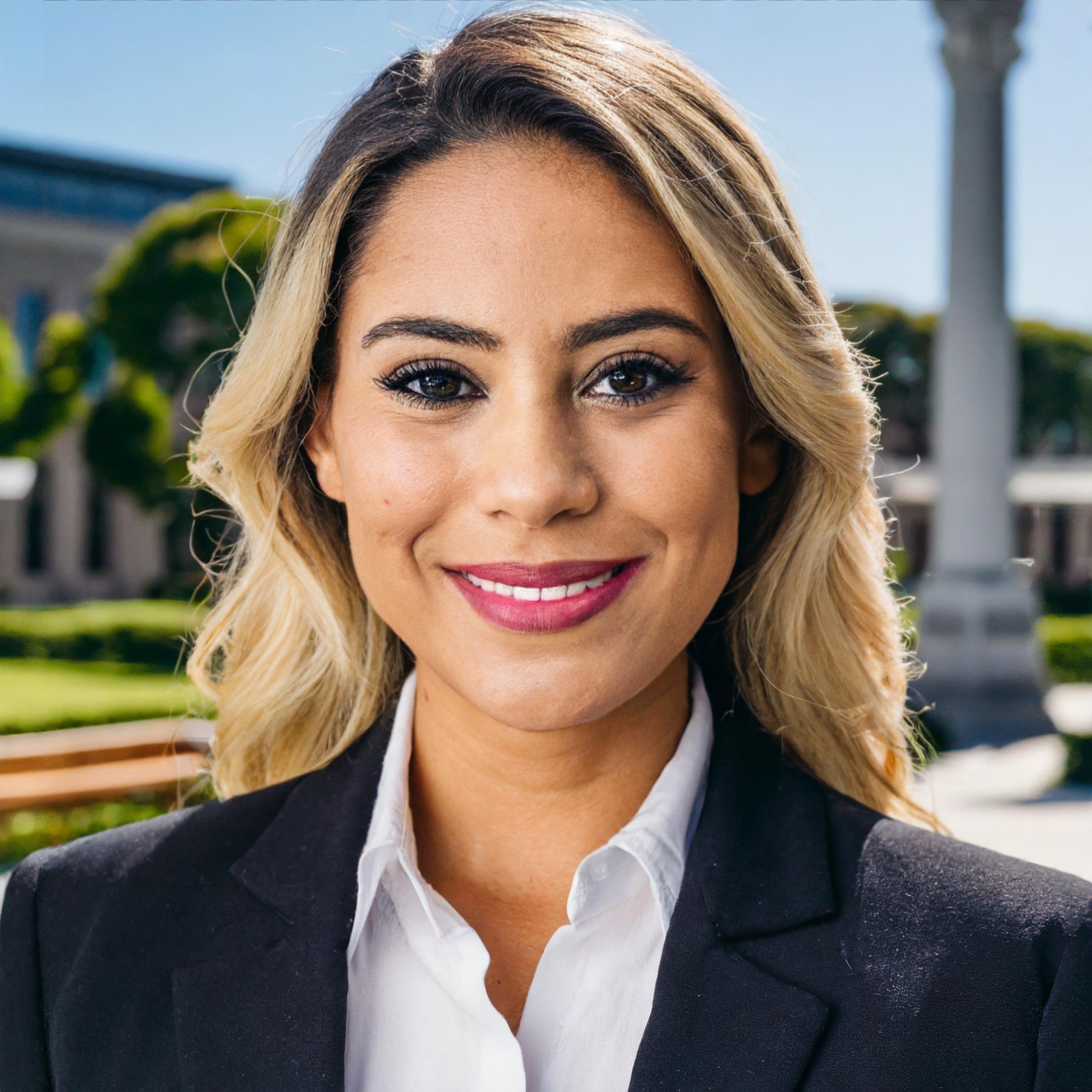}}}
\end{subfigure}
\hfill
\begin{subfigure}[t]{0.23\linewidth}  
\subcaption{}
\vtop{\vskip0pt\hbox{\includegraphics[width=\linewidth]{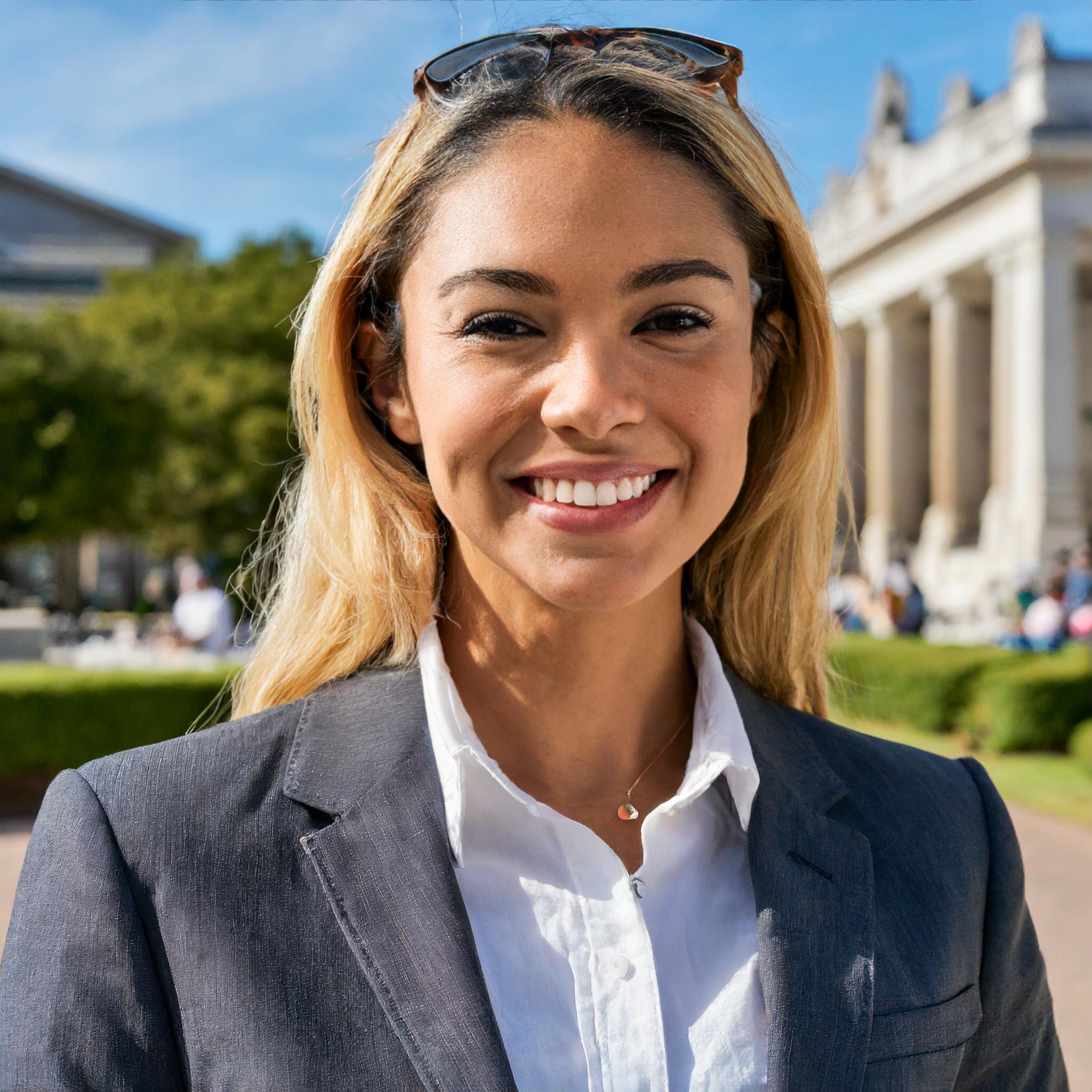}}}
\end{subfigure}

\vspace{8pt} 

\begin{subfigure}[t]{0.23\linewidth}  
\subcaption{}
\vtop{\vskip0pt\hbox{\includegraphics[width=\linewidth]{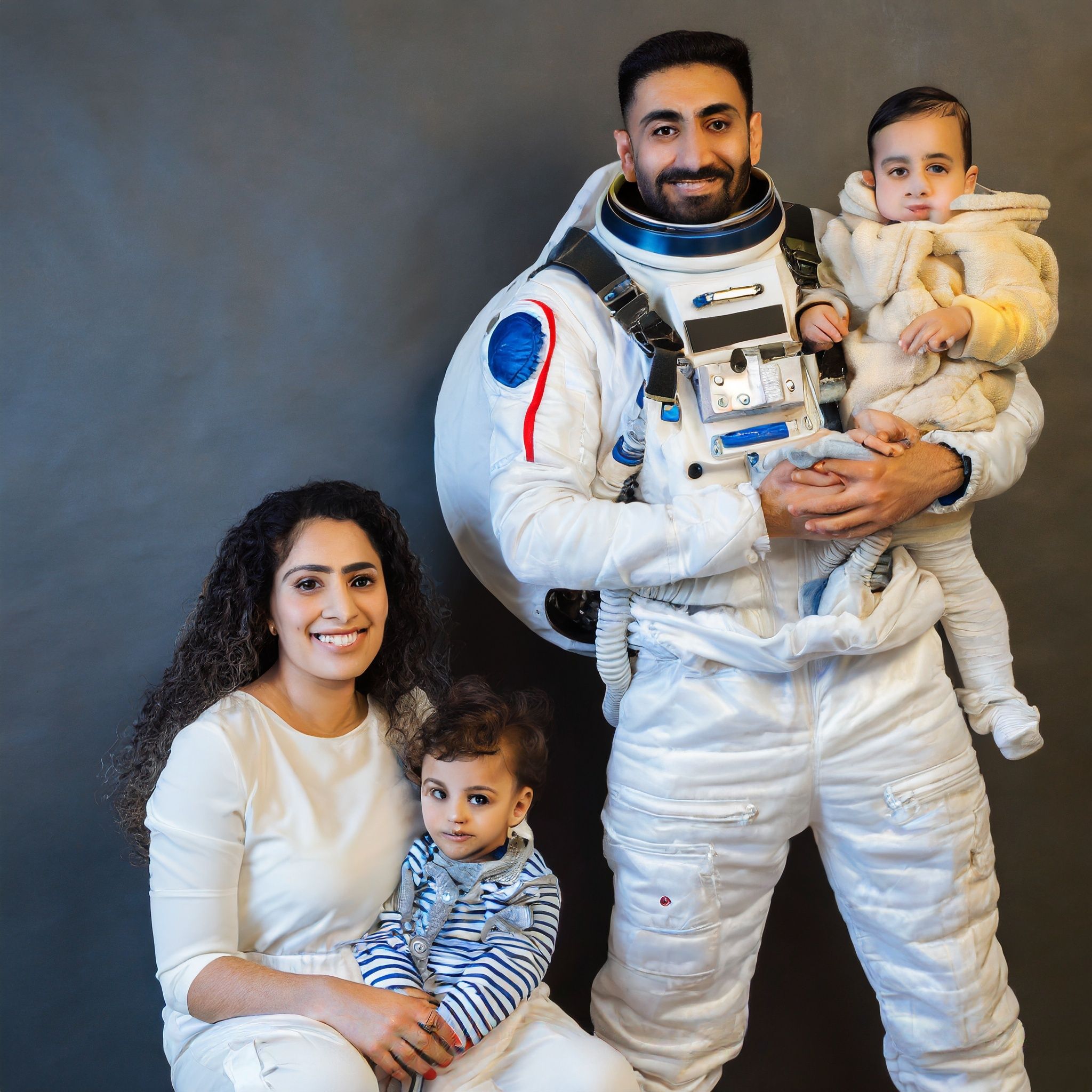}}}
\end{subfigure}
\hfill
\begin{subfigure}[t]{0.23\linewidth}  
\subcaption{}
\vtop{\vskip0pt\hbox{\includegraphics[width=\linewidth]{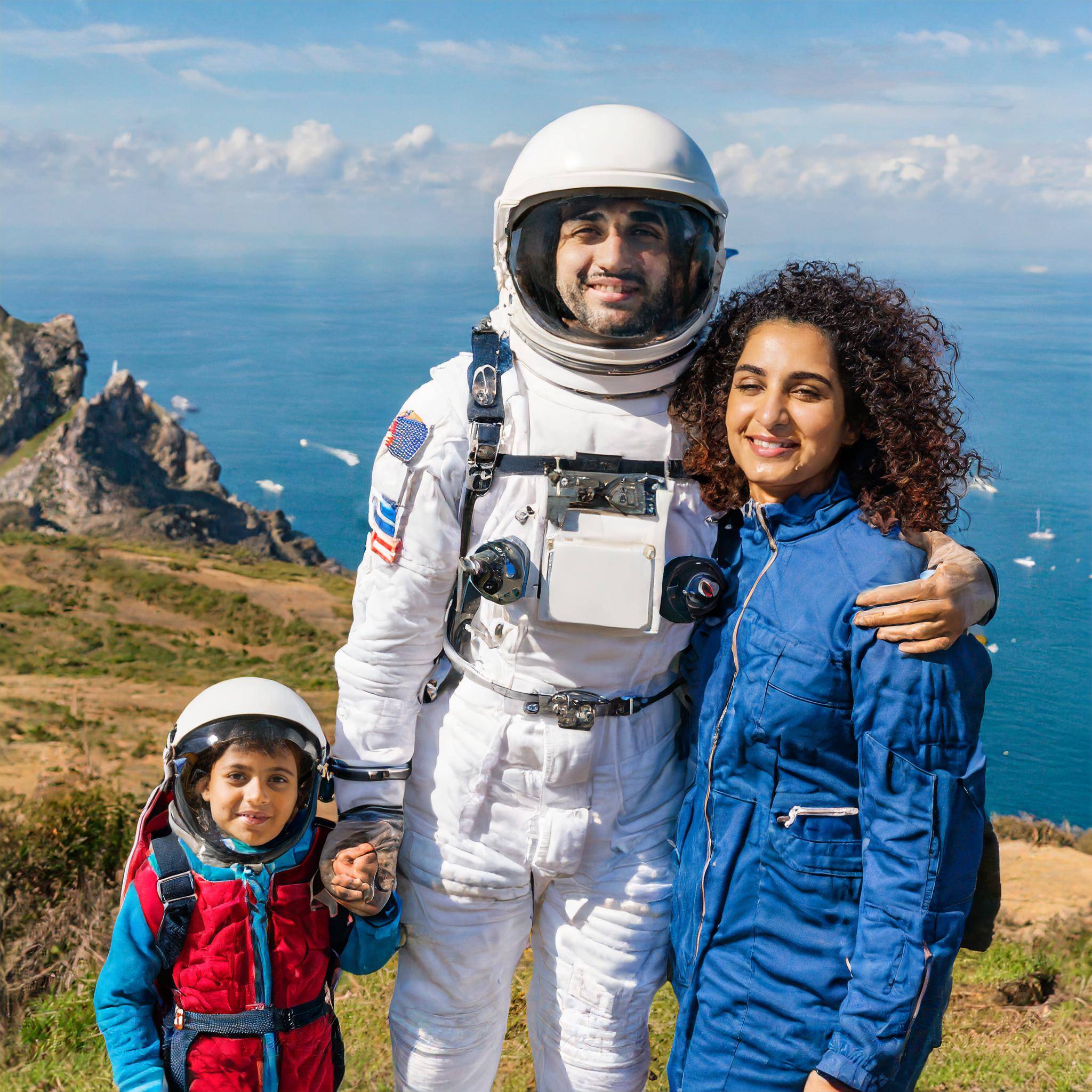}}}
\end{subfigure}
\hfill
\begin{subfigure}[t]{0.23\linewidth}  
\subcaption{}
\vtop{\vskip0pt\hbox{\includegraphics[width=\linewidth]{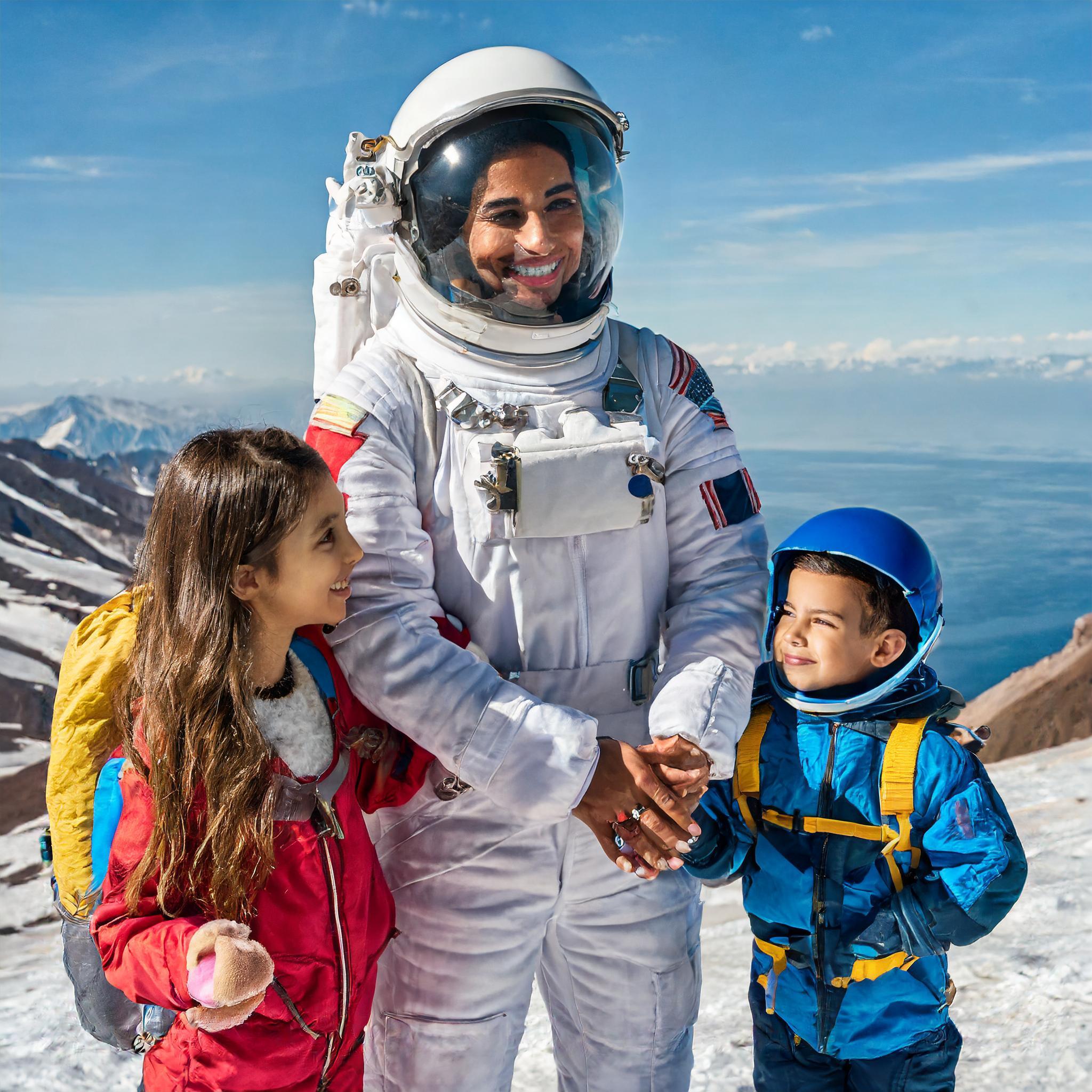}}}
\end{subfigure}
\hfill
\begin{subfigure}[t]{0.23\linewidth}  
\subcaption{}
\vtop{\vskip0pt\hbox{\includegraphics[width=\linewidth]{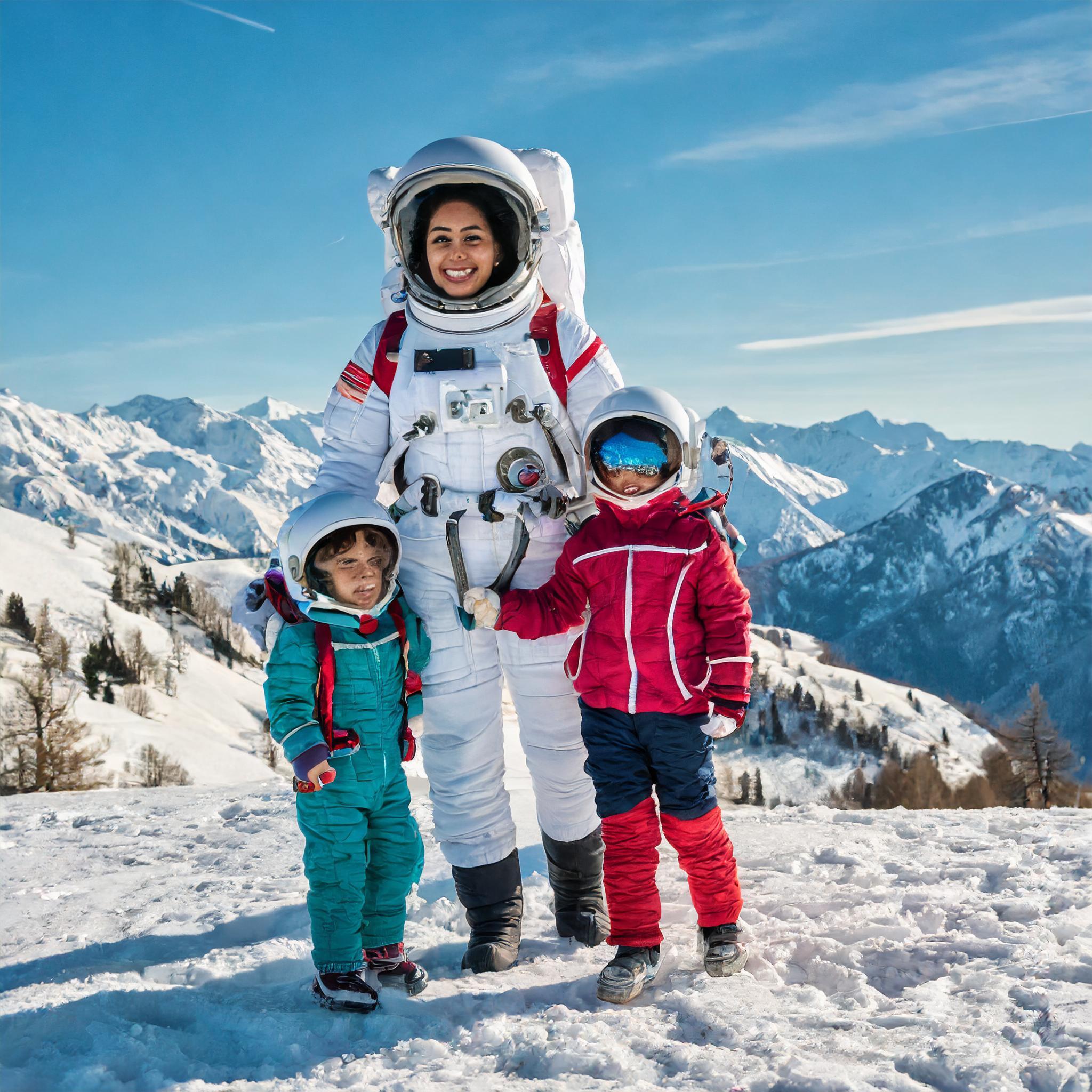}}}
\end{subfigure}

\end{minipage}
}
\vspace{-2mm}
\caption{\mybold{Example images generated by consistently photorealistic and consistently detectable prompts.} \normalfont{
\textbf{A.} Curated image generated with a consistently photorealistic prompt: ``American woman faculty portrait, not a close-up, blond." 
\textbf{B-D} Reprompted images generated with the same consistently photorealistic prompts. 
\textbf{E.} Curated image generated with a consistently detectable prompt: ``Persian woman astronaut in astronaut clothes, family photo with husband and two toddlers, high resolution, realistic." 
\textbf{F-H} Reprompted images of the same consistently detectable prompts.}}
\label{fig:goodandbadprompt}
\Description{Two example images where A shows a portrait image of an American woman faculty with few visible artifacts and B shows a Persian woman and her child and husband in a space suit with noticeable artifacts in all of their faces.}
\end{figure*}

\clearpage
\onecolumn
\section{Future Work on Videos}
\begin{figure*}[ht]
\centering
\resizebox{1.0\textwidth}{!}{ 
\begin{minipage}{\textwidth} 
\captionsetup{justification=raggedright, singlelinecheck=false, skip=2pt}

\begin{subfigure}[t]{0.33\linewidth}  
\subcaption{}
\vtop{\vskip0pt\hbox{\includegraphics[width=\linewidth]{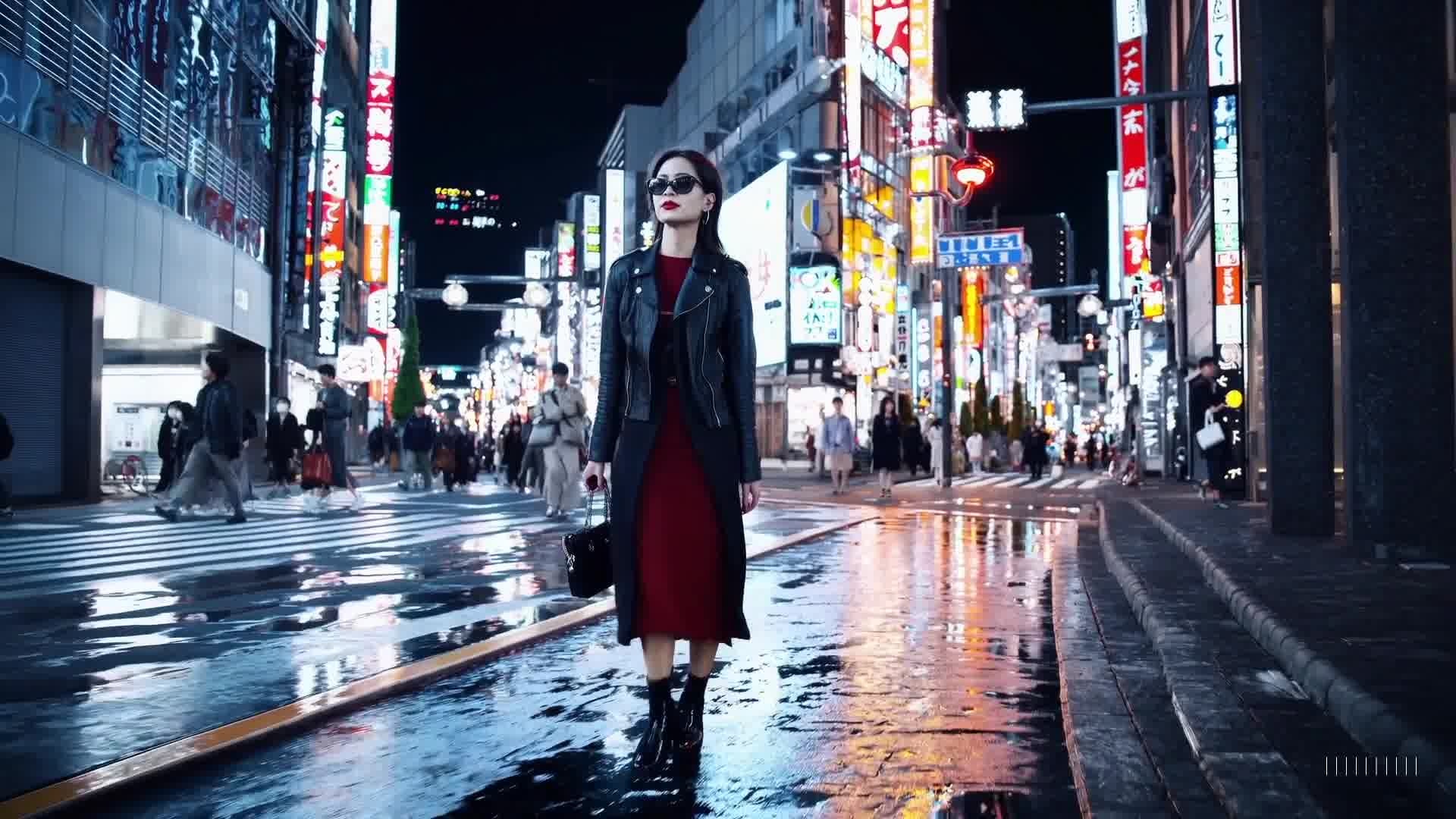}}}
\end{subfigure}
\hfill
\begin{subfigure}[t]{0.33\linewidth}  
\subcaption{}
\vtop{\vskip0pt\hbox{\includegraphics[width=\linewidth]{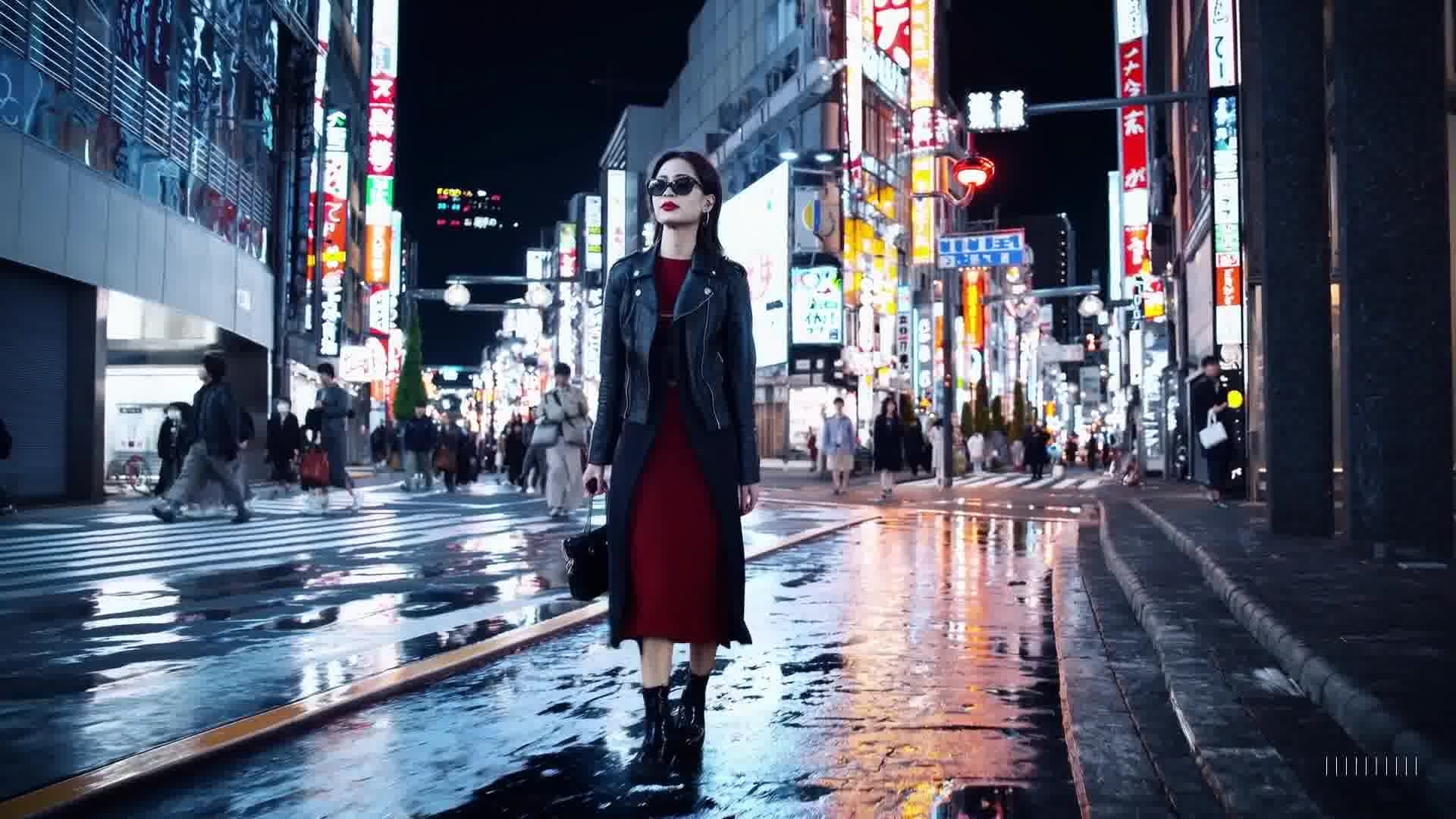}}}
\end{subfigure}
\hfill
\begin{subfigure}[t]{0.33\linewidth}  
\subcaption{}
\vtop{\vskip0pt\hbox{\includegraphics[width=\linewidth]{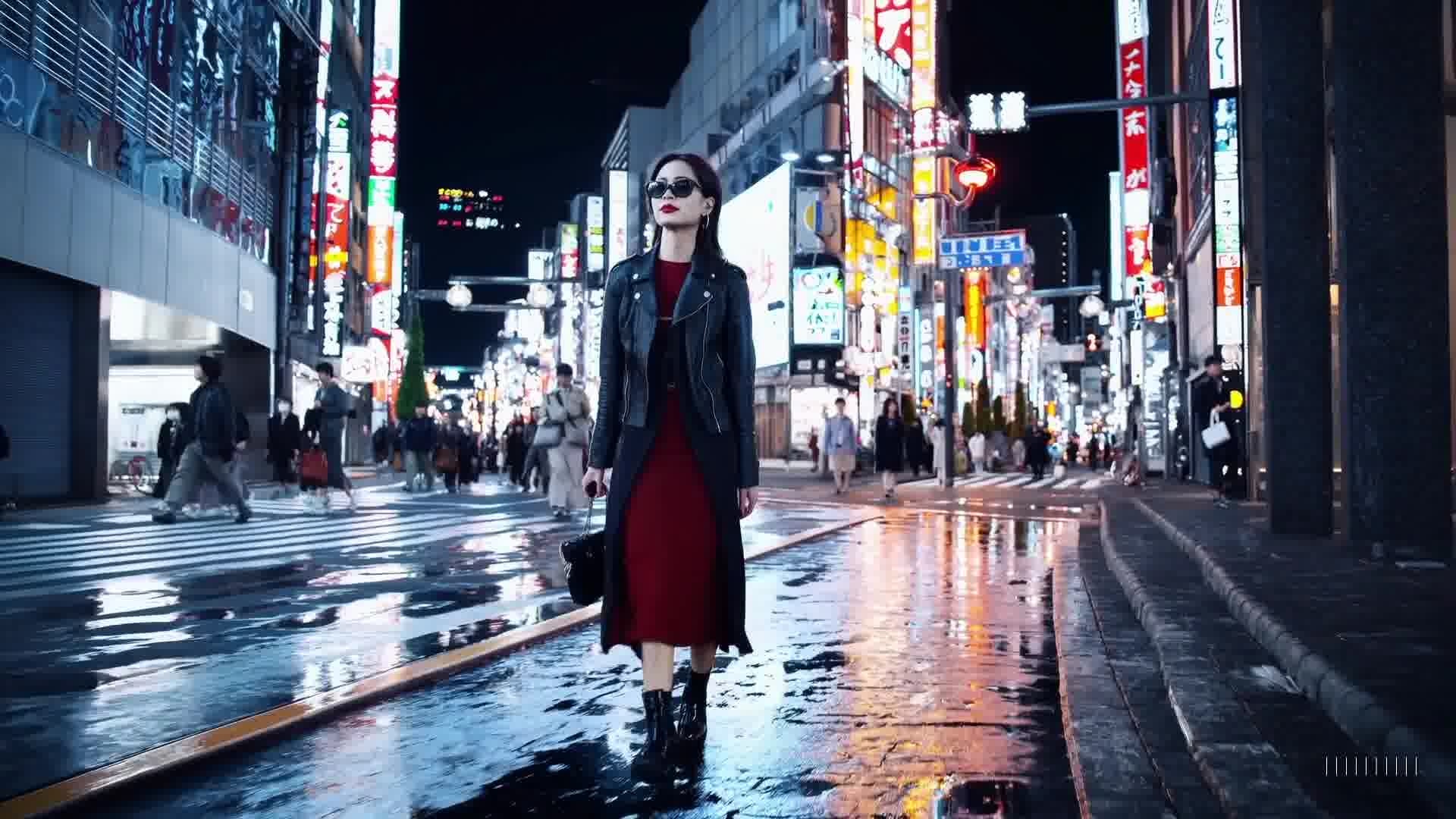}}}
\end{subfigure}

\vspace{10pt} 

\begin{subfigure}[t]{0.33\linewidth}  
\subcaption{}
\vtop{\vskip0pt\hbox{\includegraphics[width=\linewidth]{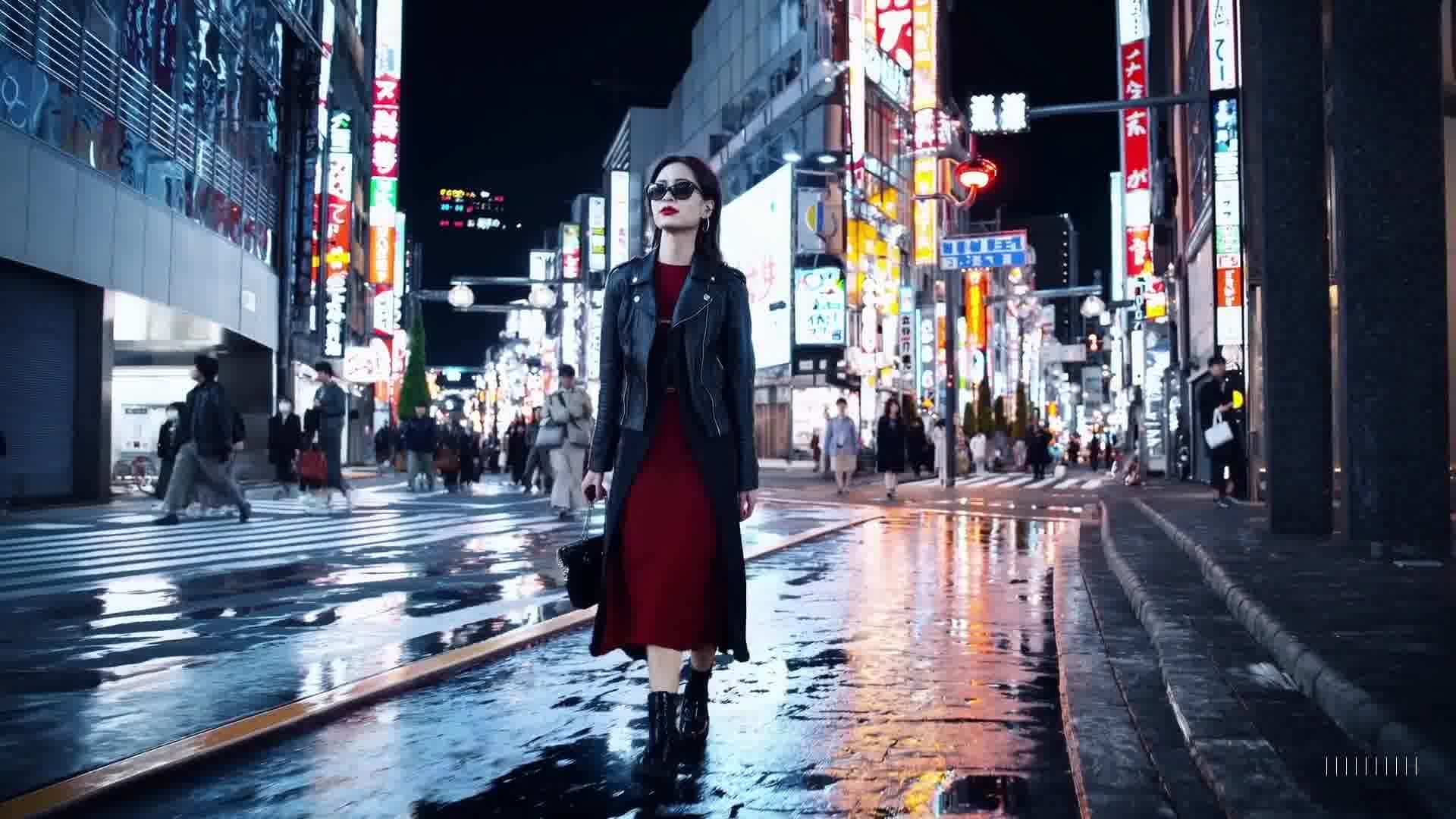}}}
\end{subfigure}
\hfill
\begin{subfigure}[t]{0.33\linewidth}  
\subcaption{}
\vtop{\vskip0pt\hbox{\includegraphics[width=\linewidth]{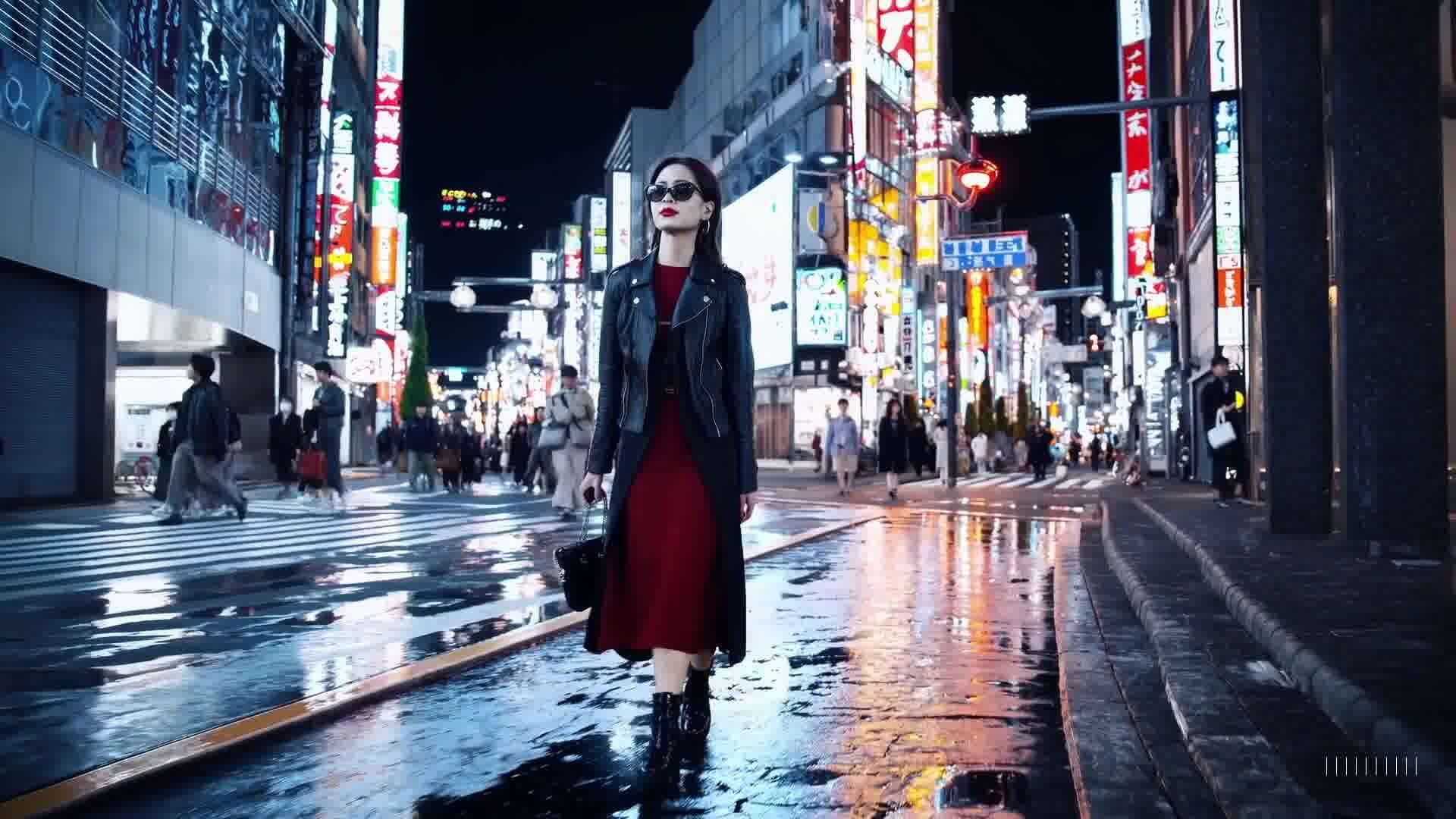}}}
\end{subfigure}
\hfill
\begin{subfigure}[t]{0.33\linewidth}  
\subcaption{}
\vtop{\vskip0pt\hbox{\includegraphics[width=\linewidth]{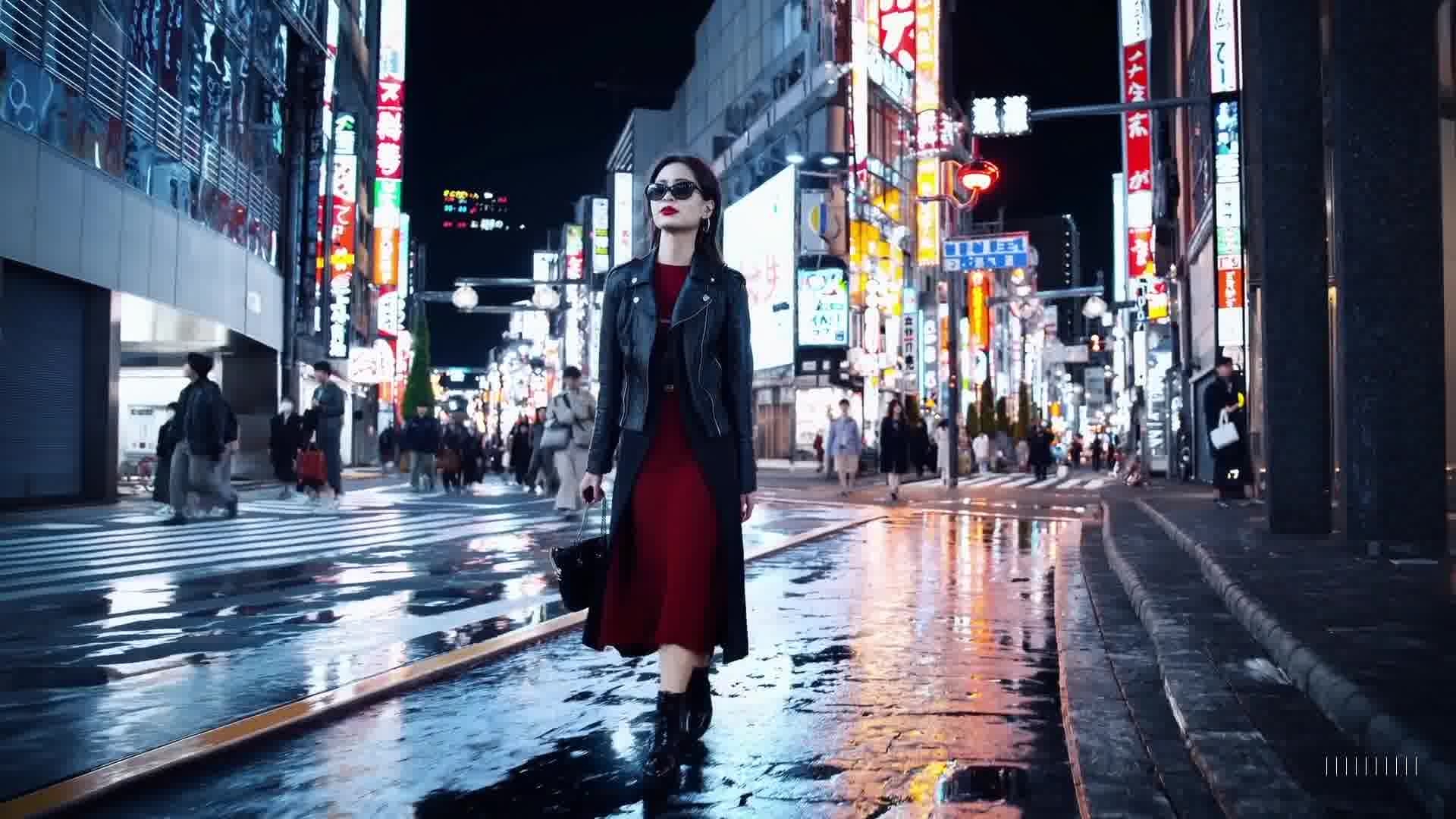}}}
\end{subfigure}

\vspace{10pt} 

\begin{subfigure}[t]{0.33\linewidth}  
\subcaption{}
\vtop{\vskip0pt\hbox{\includegraphics[width=\linewidth]{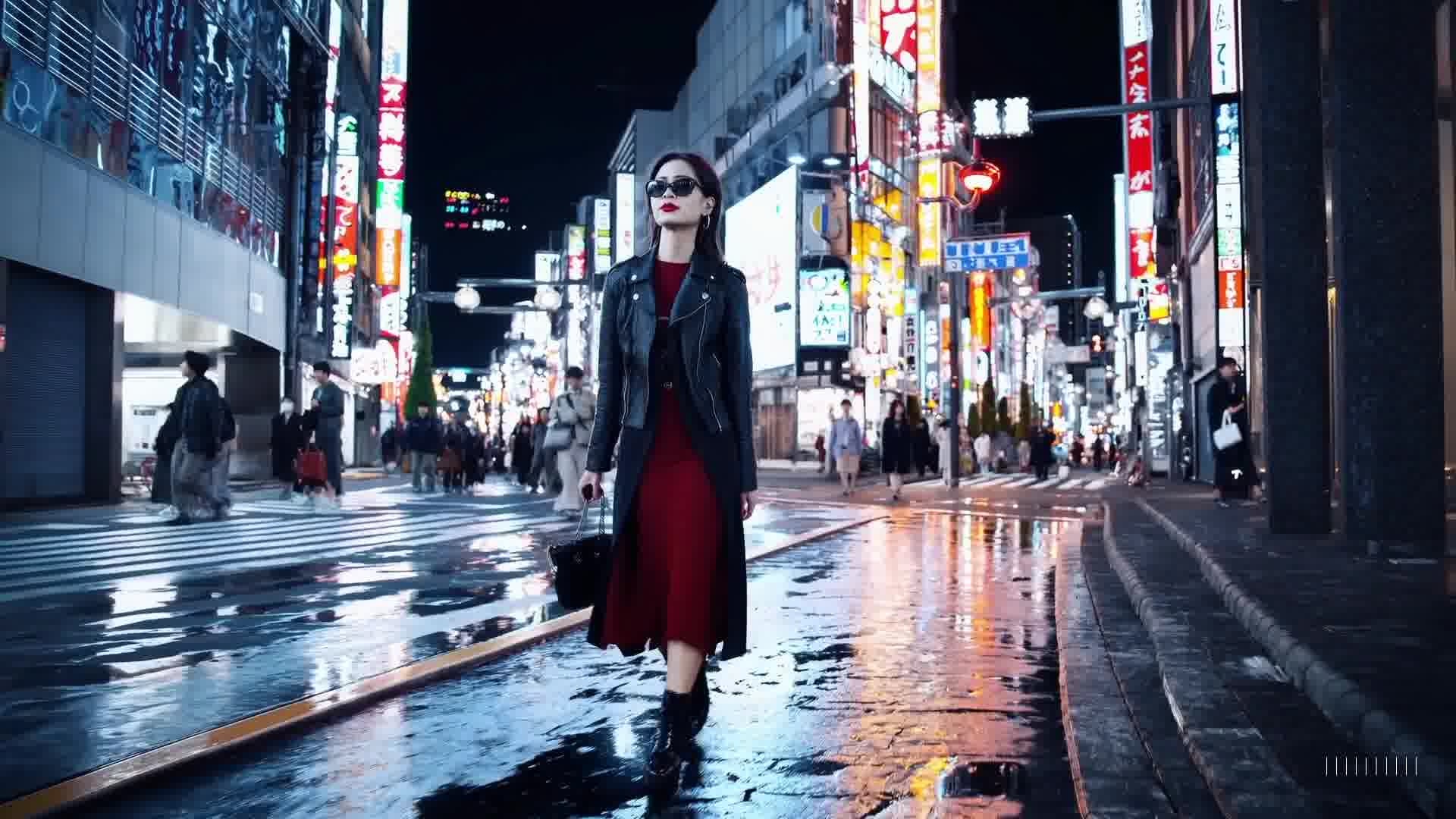}}}
\end{subfigure}
\hfill
\begin{subfigure}[t]{0.33\linewidth}  
\subcaption{}
\vtop{\vskip0pt\hbox{\includegraphics[width=\linewidth]{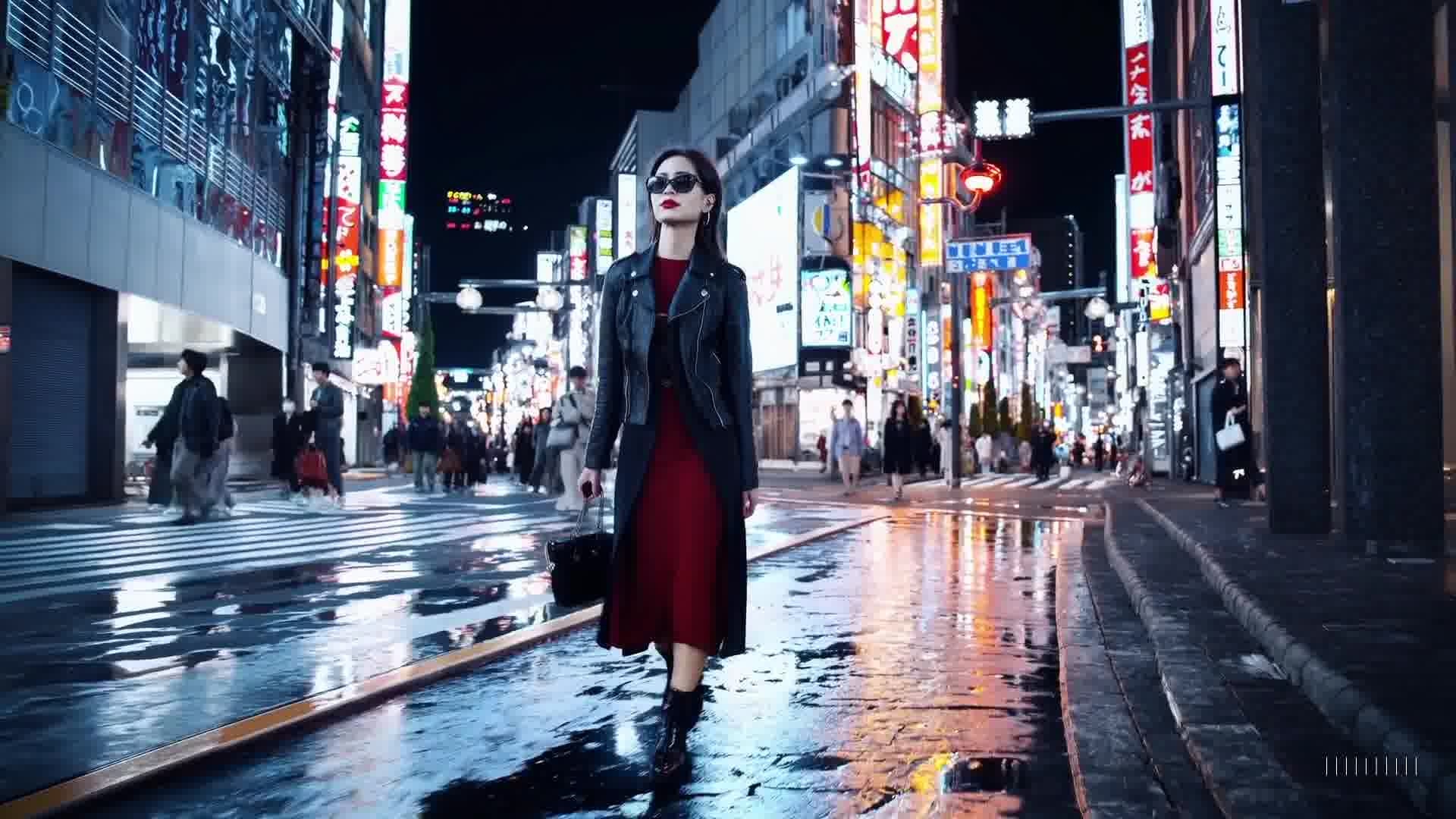}}}
\end{subfigure}
\hfill
\begin{subfigure}[t]{0.33\linewidth}  
\subcaption{}
\vtop{\vskip0pt\hbox{\includegraphics[width=\linewidth]{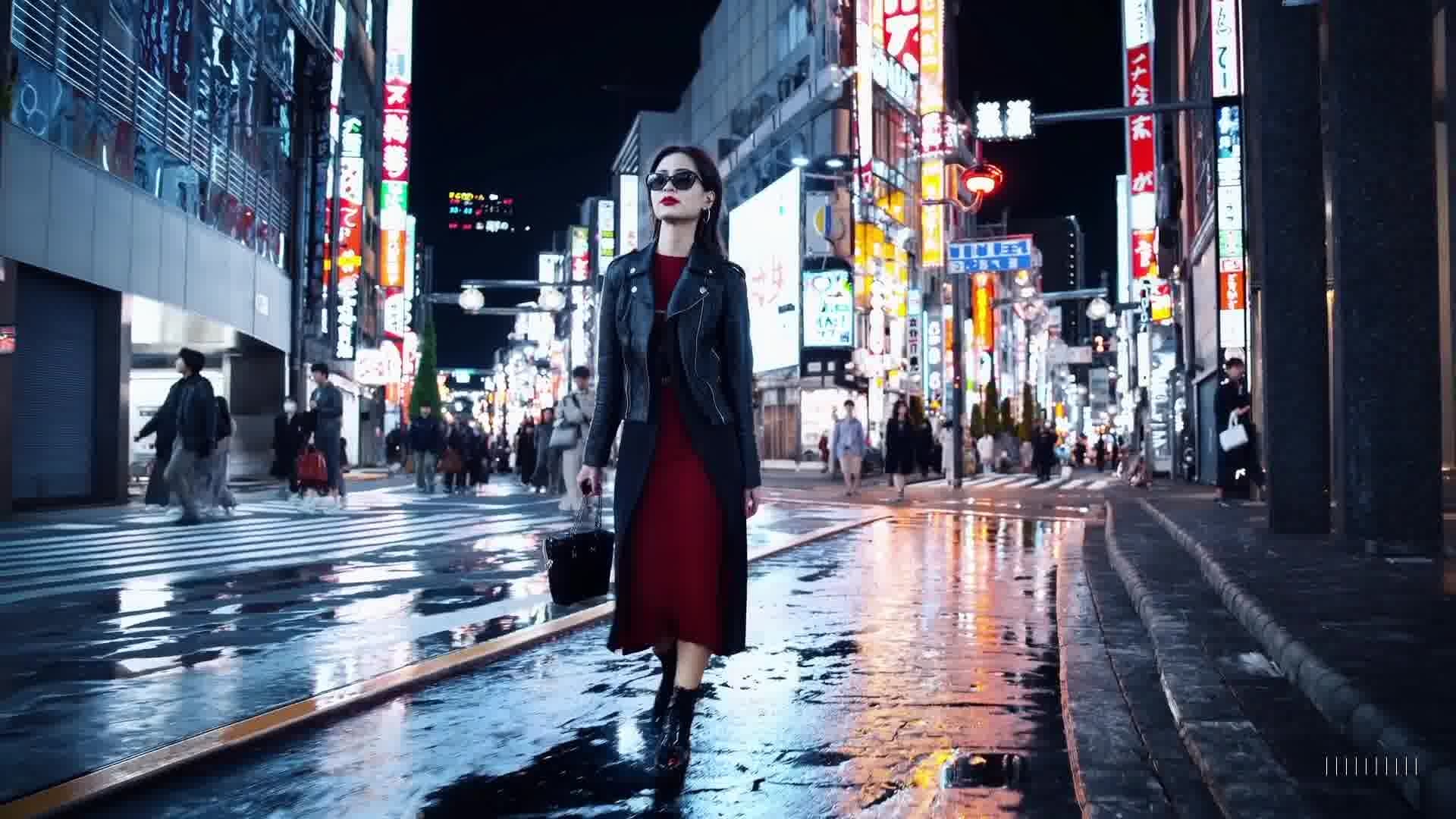}}}
\end{subfigure}

\vspace{10pt} 


\begin{subfigure}[t]{0.33\linewidth}  
\subcaption{}
\vtop{\vskip0pt\hbox{\includegraphics[width=\linewidth]{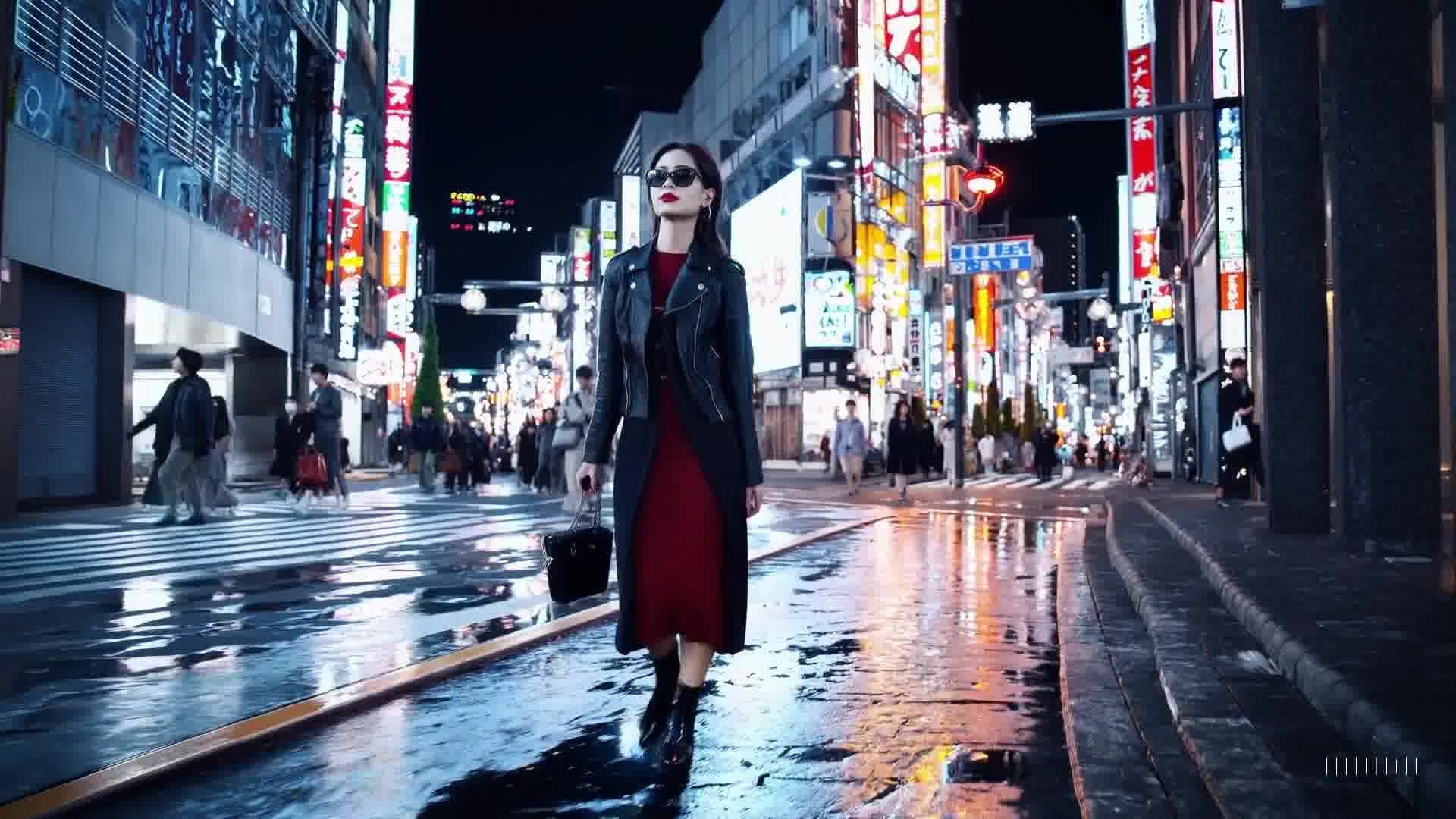}}}
\end{subfigure}
\hfill
\begin{subfigure}[t]{0.33\linewidth}  
\subcaption{}
\vtop{\vskip0pt\hbox{\includegraphics[width=\linewidth]{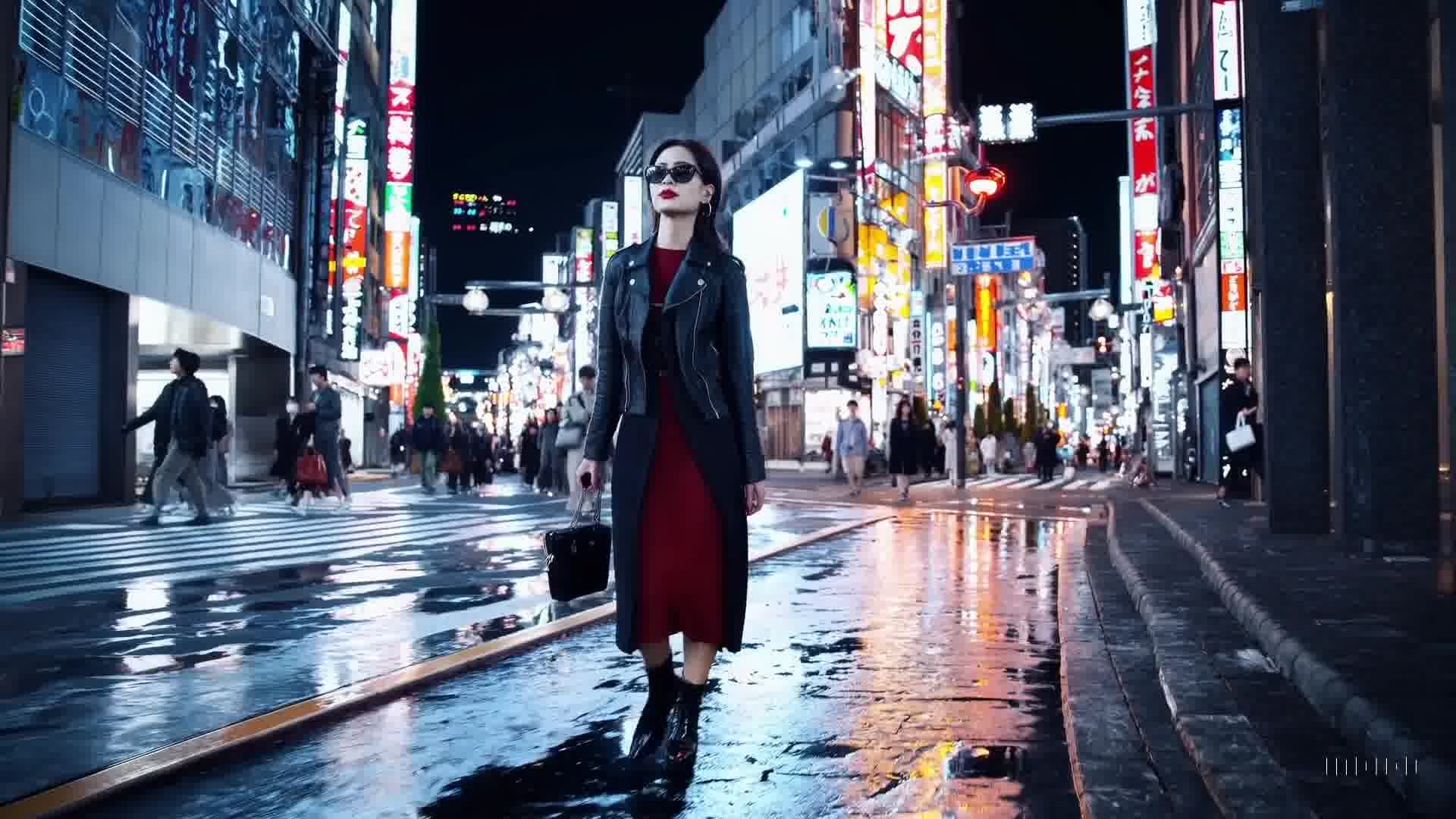}}}
\end{subfigure}
\hfill
\begin{subfigure}[t]{0.33\linewidth}  
\subcaption{}
\vtop{\vskip0pt\hbox{\includegraphics[width=\linewidth]{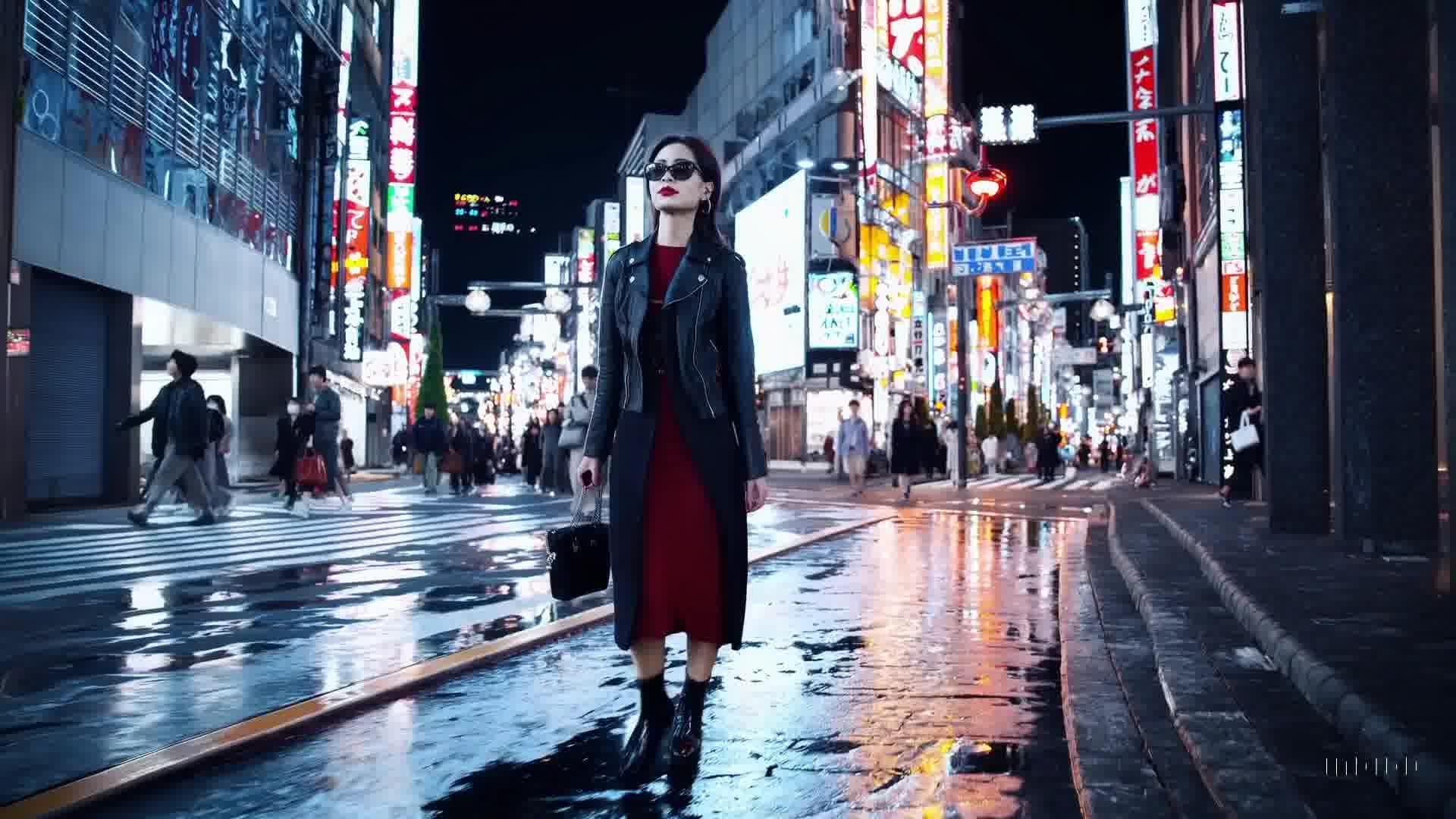}}}
\end{subfigure}

\end{minipage}
}
\vspace{-2mm}
\caption{\mybold{Example frames from an AI-generated video with a temporal anatomical implausibility.} \normalfont{
9 frames from a video generated by OpenAI's Sora diffusion-transformer model where the subject's right leg morphs into the left leg somewhere between E and J. Each frame is separated by 1/10 of a second. This particular artifact fits into the anatomical implausibility category of the taxonomy, but it's different from any anatomical plausibility seen in diffusion model-generated images. In particular, this implausibility has a temporal element: the transition from A to L involves the subject's right leg becoming her left in a split second, which does not fit with what we know about human anatomy.}}
\label{fig:sora}
\Description{9 frames from a video generated by OpenAI's Sora diffusion-transformer model where the subject's right morphs into the left leg somewhere between E and J. Each frame is separated by 1/10 of a second. This particular artifact fits into the anatomical implausibility category of the taxonomy, but it's different than any anatomical plausibility in diffusion model-generated images. In particular, this implausibility has a temporal element: the transition from A to L involves the subject's right leg becoming her left in a split second, which does not fit with what we know about human anatomy.}
\end{figure*}

\end{document}